\documentclass[vecphys]{svmult}

\newcommand{\beq}{\begin{equation}}
\newcommand{\eeq}{\end{equation}}
\newcommand{\bea}{\begin{eqnarray}}
\newcommand{\eea}{\end{eqnarray}}
\newcommand{\cir}{{\buildrel \circ \over =}}
\newcommand{\sgn}{\epsilon}
\newcommand{\eo}{{}^4{\buildrel \circ \over E}}

\usepackage{amssymb}

\usepackage{makeidx}         
\usepackage{graphicx}        
\usepackage{multicol}        
\usepackage[bottom]{footmisc}

\makeindex

\begin{document}

\title*{From Clock Synchronization to Dark Matter as a Relativistic Inertial Effect}

\titlerunning{From Clock Synchronization}

\author{Luca Lusanna}

\institute{Sezione INFN di Firenze, Polo Scientifico, Via Sansone 1,
50019 Sesto Fiorentino (FI), Italy \texttt{lusanna@fi.infn.it}
\hfill\break
 Lecture at the Black Objects in Supergravity School BOSS2011,
Frascati, 9-13 May 2011}

%
%
\maketitle


\section*{Contents}
\hfill\break

 ${}{}$1.  {\bf Introduction}..................................................
 .................................. 3
\bigskip

 2. {\bf Relativistic
 Metrology}...................................................................11
\bigskip

 3. {\bf Clock Synchronization and Global Non-Inertial Frames in
${}{}{}$ Minkowski
Space-Time}...........................................................................14
\medskip

{}{}{}{}{}{}3.1 {\bf 3+1 Splittings of Minkowski Spacetime and Radar
\hfill\break {}{}{}{}{}{}{}{}{}4-Coordinates}.........
.................................................................................
14

{}{}{}{}{}{}3.2 {\bf Global Non-Inertial Frames in Minkowski
Spacetime}.............16

{}{}{}{}{}{}3.3 {\bf Congruences of Timelike Observers Associated
with a 3+1
{}{}{}{}{}{}{}{}{}Splitting}......................................................................................................19

{}{}{}{}{}{}3.4 {\bf Parametrized Minkowski Theories}...
..........................................19

{}{}{}{}{}{}3.5 {\bf The Instant Form of Dynamics in the Inertial
Rest Frames {}{}{}{}{}{}{}{}{}and the Problem of the Relativistic
Center of Mass} ........................22

{}{}{}{}{}{}3.6 {\bf The Description of Isolated Systems in the Rest
Frame and {}{}{}{}{}{}{}{}{}their Poincar\'e
Generators}......................................................................24

\bigskip

4. {\bf Implications for Relativistic Mechanics and Classical Field
{}{}{}Theory in Special Relativity and the Multi-Temporal
{}{}{}Quantization
Approach}....................................................................................................28
\medskip

{}{}{}{}{}{}4.1 {\bf Relativistic Atomic Physics}...............
..........................................28

{}{}{}{}{}{}4.2 {\bf Relativistic Kinetic Theory and Relativistic
Micro-Canonical
{}{}{}{}{}{}{}{}{}Ensemble}....................................................................................................30

{}{}{}{}{}{}4.3 {\bf Relativistic Quantum Mechanics and Relativistic
{}{}{}{}{}{}{}{}{}Entanglement}.......................................
......................................................................31

{}{}{}{}{}{}4.4 {\bf Multitemporal Quantization in Non-Inertial
Frames}................32

{}{}{}{}{}{}4.5 {\bf Open
Problem}..................................................................................33

\bigskip

5. {\bf Non-Inertial Frames in Asymptotically Minkowskian Einstein
{}{}{}Spacetimes and ADM Tetrad
Gravity}.....................................................34

{}{}{}{}{}{}5.1 {\bf The Parametrization of Tetrads for ADM Tetrad
{}{}{}Gravity}...................................................................................................................36

{}{}{}{}{}{}5.2 {\bf The ADM Phase Space and the ADM Hamilton
{}{}{}Equations}...............................................................................................................39

\bigskip

6. {\bf The York Canonical Basis and the Inertial and Tidal Degrees
{}{}{}of Freedom of the Gravitational
Field}...........................................................43

\medskip

{}{}{}{}{}{}6.1 {\bf The York Canonical Basis}.....
..........................................................43

{}{}{}{}{}{}6.2 {\bf 3-Orthogonal Schwinger Time Gauges and Hamilton
{}{}{}Equations}...........................................................................................................47

{}{}{}{}{}{}6.3 {\bf The congruence of Eulerian Observers and the
Non-Hamiltonian {}{}{}First-Order ADM Equations of Cosmological
Spacetimes}....................51

\bigskip
7. {\bf Post-Minkowskian Linearization in Non-Harmonic 3-Orthogonal
{}{}{}Gauges and Post-Minkowskian Gravitational
Waves}.............................54
\bigskip

8. {\bf Post-Minkowskian Hamilton Equations for Particles, their
{}{}{}Post-Newtonian Limit and Dark Matter as a Relativistic
{}{}{}Inertial
Effect}...................................................................................................................63
\medskip

{}{}{}{}{}{}8.1 {\bf The Center-of-Mass Problem in General
Relativity and in the {}{}{}HPM
Linearization}.......................................................................................65

{}{}{}{}{}{}8.2 {\bf The Post-Newtonian Expansion at all orders in
the Slow {}{}{}Motion
Limit}........................................................................................................67

{}{}{}{}{}{}8.3 {\bf The HPM Binaries at the 1PN
Order}..............................................68

{}{}{}{}{}{}8.4 {\bf From the Three Signatures for Dark Matter
Reinterpreted as {}{}{}Relativistic Inertial Effects Induced by the
York Time to the {}{}{}Need of a PM
ICRS}..................................................................................................69

\bigskip

9. {\bf Dark Energy and Other Open
Problems}...........................................75

\bigskip

{\bf
References}..........................................................................
.......................79

\vfill\eject

\section{Introduction}
\label{sec:1}

One of the main open problems in astrophysics is the dominance of
{\it dark} entities, the dark matter and the dark energy, in the
existing description of the universe given by the standard
$\Lambda$CDM cosmological model \cite{1} based on the cosmological
principle (homogeneity and isotropy of the space-time), which
selects the class of Friedmann-Robertson-Walker (FWR) space-times.
After the transition from quantum cosmology to classical
astrophysics, with the Heisenberg cut roughly located  at a suitable
cosmic time ($\approx 10^5$ years after the big bang) and at the
recombination surface identified by the cosmic microwave background
(CMB), one has a description of the universe in which the known
forms of baryonic matter and radiation contribute only with a few
percents of the global budget. One has a great variety of models
trying to explain the composition of the universe in accelerated
expansion (based on data on high red-shift supernovae, galaxy
clusters and CMB): WIMPS (mainly super-symmetric particles), $f(R)$
modifications of Einstein gravity (with a modified Newton
potential), MOND (with a modification of Newton law),... for dark
matter; cosmological constant, string theory, back-reaction (spatial
averages, non-linearity of Einstein equations), inhomogeneous
space-times (Lemaitre-Tolman-Bondi, Szekeres), scalar fields
(quintessence, k-essence, phantom), fluids (Chaplygin fluid), ....
for dark energy.

\bigskip

Most of these developments rest on a description based on a family
of FRW space-times with nearly flat 3-spaces (as required by CMB
data) as the reference space-times where to interpret the
astronomical data (luminosity, light spectrum, angles) on the
2-dimensional sky vault. Therefore, the starting point is the
extension of the standards of relativistic metrology near the Earth
and in the Solar System to astronomy: to reconstruct a 4-dimensional
space-time one needs new standards of time and length like the
cosmic time and the luminosity distance (or any other astrometric
definition, see Ref.\cite{2}) allowing to define an International
Celestial Reference System (ICRS) \cite{3}, namely a 4-coordinate
system describing a 3-universe evolving in time, where the
astronomical data have to be dynamically interpreted according to
Einstein gravity or some of its extensions.

\bigskip

The aim of this Lecture is to suggest a new viewpoint on the origin
of dark matter, and maybe also of dark energy, starting from a
re-reading of the general covariance of Einstein general relativity
(GR), which could be also applied to every generally covariant
extension of this theory if needed. It is an extended version of the
review paper \cite{4}. In this Introduction one will delineate the
framework of our approach and then in the subsequent Sections one
will give more details of the various topics.

\bigskip

The gauge group of the Lagrangian formulation of Einstein GR, the
diffeomorphism group, implies that the 4-coordinates of the
space-time are {\it gauge variables}. As a consequence, the search
of GR observables is restricted to {\it 4-scalars} and at the
theoretical level one tries to describe gravitational dynamical
properties in term of them. However, inside the Solar System the
experimental localization of macroscopic classical objects is
unavoidably done by choosing some {\it convention} for the local
4-coordinates of space-time. Atomic physicists, NASA engineers and
astronomers have chosen a series of reference frames and standards
of time and length suitable for the existing technology \cite{5}.
These conventions determine certain Post-Minkowskian (PM)
4-coordinate systems of an asymptotically Minkowskian space-time, in
which the instantaneous 3-spaces are not strictly Euclidean. Then
these reference frames are seen as a local approximation of a
reference frame in ICRS, where however the space-time has become a
cosmological FWR one, which is only conformally asymptotically
Minkowskian at spatial infinity. A search of a consistent patching
of the 4-coordinates from inside the Solar System to the rest of the
universe will start when the data from the future GAIA mission
\cite{6} for the cartography of the Milky Way will be available.
This will allow a PM definition of a Galactic Reference System
containing at leat our galaxy. Let us remark that notwithstanding
the FRW instantaneous 3-spaces are not strictly Euclidean, all the
books on galaxy dynamics describe the galaxies by means of Kepler
theory in Galilei space-time.

\bigskip

This state of affairs requires to revisit Einstein GR to see whether
it is possible to identify which components of the 4-metric tensor
are connected with the gauge freedom in the choice of the
4-coordinates and which ones describe the dynamical degrees of
freedom of the gravitational field. Since this cannot be done at the
Lagrangian level, one must restrict himself to the class of globally
hyperbolic, asymptotically flat space-times allowing a Hamiltonian
description starting from the description of Einstein GR in terms of
the ADM action \cite{7} instead than in terms of the
Einstein-Hilbert one. In canonical ADM gravity one can use Dirac
theory of constraints \cite{8} to describe the Hamiltonian gauge
group, whose generators are the first-class constraints of the
model. The basic tool of this approach is the possibility to find
so-called Shanmugadhasan canonical transformations \cite{9}, which
identify special canonical bases adapted to the first-class
constraints (and also to the second-class ones when present). In
these special canonical bases the vanishing of certain momenta (or
of certain configurational coordinates) corresponds to the vanishing
of well defined Abelianized combinations of the first-class
constraints (Abelianized because the new constraints have exactly
zero Poisson brackets even if the original constraints were not in
strong involution). As a consequence, the variables conjugate to
these Abelianized constraints are inertial Hamiltonian gauge
variables describing the Hamiltonian gauge freedom. The remaining
2+2 conjugate variables describe the dynamical tidal degrees of
freedom of the gravitational field (the two polarizations of
gravitational waves in the linearized theory). If one would be able
to include all the constraints in the Shanmugadhasan canonical
basis, these 2+2 variables would be the {\it Dirac observables} of
the gravitational field, invariant under the Hamiltonian gauge
transformations. However such Dirac observables are not known: one
only has statements bout their existence \cite{10}. Moreover, in
general they are not 4-scalar observables. The problem of the
connection between the 4-diffeomorphism group and the Hamiltonian
gauge group was studied in Ref.\cite{11} by means of the inverse
Legendre transformation and of the notion of dynamical symmetry. The
conclusion is that on the space of solutions of Einstein equations
there is an overlap of the two types of observables: there should
exists special Shanmugadhasan canonical bases in which the 2+2 Dirac
observables become 4-scalars when restricted to the space of
solutions of the Einstein equations. In any case the identification
of the inertial gauge components of the 4-metric is what is needed
to make a fixation of 4-coordinates as required by relativistic
metrology.

\bigskip

Another problem is that asymptotically flat space-times have the SPI
group of asymptotic symmetries (direction-dependent asymptotic
Killing symmetries) \cite{12} and this is an obstruction to the
existence of asymptotic Lorentz generators for the gravitational
field \cite{13}. However if one restricts the class of space-times
to those {\it not containing super-translations} \cite{14}, then the
SPI group reduces to the asymptotic ADM Poincar\'e group \cite{15}:
these space-times are {\it asymptotically Minkowskian}, they contain
an asymptotic Minkowski 4-metric (to be used as an {\it asymptotic
background} at spatial infinity in the linearization of the theory)
and they have asymptotic inertial observers at spatial infinity
whose spatial axes may be identified by means of the fixed stars of
star catalogues \footnote{The fixed stars can be considered as an
empirical definition of spatial infinity of the observable
universe.}. Moreover, in the limit of vanishing Newton constant ($ G
= 0$) the asymptotic ADM Poincar\'e generators become the generators
of the special relativistic Poincar\'e group describing the matter
present in the space-time. This is an important condition for the
inclusion into GR of the classical version of the standard model of
particle physics, whose properties are all connected with the
representations of this group in the inertial frames of Minkowski
space-time. In absence of matter a sub-class of these space-times is
the (singularity-free) family of Chrstodoulou-Klainermann solutions
of Einstein equations \cite{16} (they are near to Minkowski
space-time in a norm sense and contain gravitational waves).

\bigskip

Moreover, in this restricted class of space-times the canonical
Hamiltonian is the ADM energy \cite{17}, so that there is no {\it
frozen picture} like in the "spatially compact space-times without
boundaries" used in loop quantum gravity \footnote{In these
space-times the canonical Hamiltonian vanishes and the Dirac
Hamiltonian is  a combination of first-class constraints, so that it
only generates Hamiltonian gauge transformations. In the reduced
phase space, quotient with respect the Hamilonian gauge group, the
reduced Hamiltonian is zero and one has a {\it frozen picture} of
dynamics. This class of space-times fits well with Machian ideas (no
boundary conditions) and with interpretations in which there is no
physical time like the one in Ref.\cite{18}. However, it is not
clear how to include in this framework the standard model of
particle physics.}.

\bigskip

To take into account the fermion fields present in the standard
particle model one must extend ADM gravity  to ADM tetrad gravity .
Since our class of space-times admits orthonormal tetrads and a
spinor structure \cite{19}, the extension can be done by simply
replacing the 4-metric in the ADM action with its expression in
terms of tetrad fields, considered as the basic 16 configurational
variables substituting the 10 metric fields.

\bigskip

To study ADM tetrad gravity the preliminary problem is to choose a
coordinatization of the space-time compatible with relativistic
metrology. This requires a definition of global non-inertial frames,
because the equivalence principle forbids the existence of global
inertial frames in GR. Due to the Lorentz signature of the
space-time this is a non-trivial task already in special relativity
(SR): there is no notion of instantaneous 3-space, because the only
intrinsic structure is the conformal one, i.e. the light-cone as the
locus of incoming and outgoing radiation. The existing
coordinatizations, like either Fermi or Riemann-normal coordinates,
hold only locally They are based on the {\it 1+3 point of view}, in
which only the world-line of a time-like observer is given. In each
point of the world-line the observer 4-velocity determines an
orthogonal 3-dimensional space-like tangent hyper-plane, which is
identified with an instantaneous 3-space. However, these tangent
planes intersect at a certain distance from the world-line (the
so-called acceleration length depending upon the 4-acceleration of
the observer \cite{20}), where 4-coordinates of the Fermi type
develop a coordinate singularity. Another type of coordinate
singularity is developed in rigidly rotating coordinate systems at a
distance $r$ from the rotation axis where $\omega\, r = c$ ($\omega$
is the angular velocity and $c$ the two-way velocity of light). This
is the so-called "horizon problem of the rotating disk": a time-like
4-velocity becomes a null vector at $\omega\, r = c$, like it
happens on the horizon of a black-hole. See Ref.\cite{21} for a
classification of the possible pathologies of non-inertial frames
and on how to avoid them.

\bigskip

In this Lecture one will review the way out from these problems
based on the {\it 3+1 point of view} in which, besides the
world-line of a time-like observer, one gives a global nice
foliation of the space-time with instantaneous 3-spaces. Then a
metrology-oriented notion of 4-coordinates, the so-called {\it radar
4-coordinates} first introduced by Bondi \cite{22}, is introduced in
these global non-inertial frames. One will give the conditions for a
foliation to be nice, i.e. for the absence of pathologies like the
ones of the rotating disk and of the Fermi coordinates.

\bigskip

Let us remark that the theory of global non-inertial frames is also
needed to speak of predictability in a (either classical or quantum)
theory in which the basic equations of motion are partial
differential equations (PDE). To be able to use the existence and
unicity theorem for the solutions of PDE's, one needs a well-posed
Cauchy problem, whose prerequisite is a sound definition of an
instantaneous 3-space (i.e. of a clock synchronization convention)
where the Cauchy data are given.  To give the data on a space-like
surface  is {\it not factual}, but with the data on the backward
light-cone of an observer it is not yet possible to demonstrate the
theorem. However, also the 1+3 point of view is non factual, because
it requires the knowledge of a world-line from the whole past to all
the future.

\bigskip

A Section of this Lecture will be devoted to the developments in
relativistic particle mechanics made possible by the 3+1 point of
view in SR \cite{23}, \cite{21}, \cite{24}. By means of {\it
parametrized Minkowski theories} \cite{23}, \cite{21}, one can get
the description of arbitrary isolated systems (particles, strings,
fluids, fields) admitting a Lagrangian formulation in arbitrary
non-inertial frames with the transition among non-inertial frames
described as a "gauge transformation" (general covariance under the
frame-preserving diffeomorphisms of Ref.\cite{25}). Moreover this
framework allows us to define the {\it inertial and non-inertial
rest frames} of the isolated systems, where to develop the
rest-frame instant form of the dynamics and to build the explicit
form of the Lorentz boosts for interacting systems. This makes
possible to study the problem of the relativistic center of mass
\cite{26}, relativistic bound states \cite{27,28,29}, relativistic
kinetic theory and relativistic micro-canonical ensemble \cite{30}
and various other systems \cite{31,32}. Moreover a Wigner-covariant
relativistic quantum mechanics \cite{33}, with a solution of all the
known problems introduced by SR, has been developed after some
preliminary work done in Ref.\cite{34}. This will allow us to study
relativistic entanglement.

\bigskip

After this digression in SR one defines global non-inertial frames
with radar 4-coordinates in the asymptotically Minkowskian
space-times of GR \footnote{While in SR Minkowski space-time is an
absolute notion, unifying the absolute notions of time and 3-space
of the non-relativistic Galilei space-time, in GR there is no
absolute notion: space-time becomes dynamical \cite{35} with its
metric structure satisfying Einstein equations.} and one gives the
parametrization of the tetrads and of the 4-metric in them. The
absence of super-translations implies that these non-inertial frames
are non-inertial rest frames of the 3-universe. Starting from the
ADM action for tetrad gravity one defines the Hamiltonian formalism
in a phase space containing 16 configurational field variables and
16 conjugate moments. One identifies the 14 first-class constraints
of the system and one finds that the canonical Hamiltonian is the
weak ADM energy (it is given as a volume integral over 3-space). The
existence of these 14 first-class constraints implies that 14
components of the tetrads (or of the conjugate momenta) are
Hamiltonian gauge variables describing the {\it inertial} aspects of
the gravitational field (6 of these inertial variables describe the
extra gauge freedom in the choice of the tetrads and in their
transport along world-lines). Therefore there are only 2+2 degrees
of freedom for the description of the {\it tidal} dynamical aspects
of the gravitational field. The asymptotic ADM Poincar\'e generators
can be evaluated explicitly. Till now the type of matter studied in
this framework consists of the electro-magnetic field and of N
charged scalar particles, whose signs of the energy and electric
charges are Grassmann-valued to regularize both the gravitational
and electro-magnetic self-energies (it is both a ultraviolet and an
infrared regularization),
\bigskip

Then it will be shown that there is a Shanmugadhasan canonical
transformation \cite{36} (implementing the so-called York map
\cite{37} and diagonalizing the York-Lichnerowics approach
\cite{38}) to a so-called York canonical basis adapted to 10 of the
14 first-class constraints. Only the super-Hamiltonian and
super-momentum constraints, whose general solution is not known, are
not included in the basis, but it is clarified which variables are
to be determined by their solution. Among the inertial gauge
variables there is the York time \cite{39}  ${}^3K$, i.e. the trace
of the extrinsic curvature of the 3-spaces as 3-manifolds embedded
into the space-time. It is the only gauge variable which is a
momentum in the York canonical basis \footnote{Instead in Yang-Mills
theory all the gauge variables are configurational.}: this is due to
the Lorentz signature of space-time, because the York time and three
other inertial gauge variables can be used as 4-coordinates of the
space-time (see Ref.\cite{35} for this topic and for its relevance
in the solution of the hole argument). Therefore  an identification
of the inertial gauge variables to be fixed to get a 4-coordinate
system in relativistic metrology was found. In the first paper of
Ref.\cite{40} there is the expression of the Hamilton equations for
all the variables of the York canonical basis.

\bigskip

An important remark is that in the framework of the York canonical
basis the natural family of gauges is not the harmonic one, but the
family of 3-orthogonal Schwinger time gauges in which the 3-metric
in the 3-spaces is diagonal.

\bigskip

Both in SR and GR an admissible 3+1 splitting of space-time has two
associated congruences of time-like observers \cite{21},
geometrically defined and not to be confused with the congruence of
the world-lines of fluid elements, when relativistic fluids are
added as matter in GR \cite{41,42,43}. One of the two congruences,
with zero vorticity, is the congruence of the Eulerian observers,
whose 4-velocity field is the field of unit normals to the 3-spaces.
This congruence allows us to re-express the non-vanishing momenta of
the York canonical basis in terms of the expansion ($\theta = -
{}^3K$) and of the shear of the Eulerian observers. This allows us
to compare the Hamilton equations of ADM canonical gravity with the
usual first-order non-Hamiltonian ADM equations deducible from
Einstein equations given a 3+1 splitting of space-time but without
using the Hamiltonian formalism. As a consequence, one can extend
our Hamiltonian identification of the inertial and tidal variables
of the gravitational field to the Lagrangian framework and use it in
the cosmological (conformally asymptotically flat) space-times: in
them it is not possible to formulate the Hamiltonian formalism but
the standard ADM equations are well defined. The time inertial gauge
variable needed for relativistic metrology is now the expansion of
the Eulerian observers of the given 3+1 splitting of the globally
hyperbolic cosmological space-time.
\bigskip

The next step (see the second paper of Ref.\cite{40}) is the
definition of a PM linearization of ADM tetrad gravity in the family
of 3-orthogonal Schwinger time gauges in which one chooses
3-coordinates diagonalizing the 3-metric in the 3-spaces and  an
arbitrarily given numerical function for the York time ${}^3K$. The
cosmic time $\tau_{cosm}$ has to be chosen so that the 3-spaces
$\tau_{cosm} = const.$ have an extrinsic curvature with the given
value of ${}^3K$. This PM linearization uses the asymptotic
Minkowski 4-metric as an {\it asymptotic background}, so that one
never splits the 4-metric with respect to a fixed Minkowski metric
in the bulk like in the standard approach to gravitational waves. A
ultraviolet cutoff on the matter is needed.

\bigskip

This leads to a PM formulation of gravitational waves in
non-harmonic 3-orthogonal gauges. All the constraints can be solved,
an explicit expression of the PM 4-metric can be given and the
explicit form of the Hamilton equations for the tidal degrees of
freedom of the gravitational field and the matter can be obtained.
It is non-trivial to show that all the standard results about
gravitational waves in harmonic gauges \cite{44} can be reproduced
in the 3-orthogonal gauges with the help of the formalism of
Ref.\cite{45}. As shown in the third paper of Ref.\cite{40} (where
the matter is restricted only to scalar particles), all the 4- and
3- curvature tensors of the space-time can be explicitly evaluated
and the time-like and null geodesics can be studied. It is also
possible to evaluate the red-shift of light rays and the luminosity
distance finding their dependence on the York time and verify the
old Hubble red-shift-distance law (see ref.\cite{46}), which becomes
the usual Hubble law (a velocity-distance relation) when one uses
the standard cosmological model. In the Solar System the results in
the 3-orthogonal gauges are compatible with the ones in the harmonic
gauges used in relativistic metrology \cite{5}.

\bigskip

The main important result or this lecture are the PM Hamilton
equations and the implied PM second-order  equations of motion for
the particles. Their PN limit identifies the Newton forces acting on
the particles at the lowest order augmented with 1PN forces
compatible with the known results on binaries in harmonic gauges
\cite{47}. However, there are extra 0.5PN forces, depending linearly
on the non-local York time ${}^3{\cal K} = {1\over {\triangle}}\,
{}^3K$ ($\triangle$ is the asymptotic Laplacian of the 3-space),
representing either a friction or an anti-friction force according
to the sign of ${}^3{\cal K}$. These 0.5PN forces are our main
result, because their effect can be re-interpreted as an extra {\it
effective} (time-, position- and velocity-dependent) {\it
contribution to the inertial mass} of the particles in the equations
used in the three main signatures for the existence of dark matter:
the rotation curves of spiral galaxies \cite{48} and the masses of
clusters of galaxies from the virial theorem \cite{49} and from weak
gravitational lensing \cite{50}, \cite{49}.

\bigskip

While gravitational and inertial masses are equal in Einstein GR,
the PM limit, followed by the PN one, shows that the non-Euclidean
nature of the 3-spaces  implies a breaking of the Newtonian equality
of the two types of masses, which holds only in the absolute
Euclidean 3-space of Galilei space-time.

\bigskip

As a consequence the data on dark matter can be re-read as a partial
fixation of the non-local York time ${}^3{\cal K}$. However to fix
the York time ${}^3K = \triangle\, {}^3{\cal K}$ one needs a global
information on ${}^3{\cal K}$ on the whole 3-space, in particular in
the voids among galaxy clusters.
\bigskip

Therefore one has an indication that (at least part of) dark matter
could be re-absorbed in a PM extension of the conventions in the
existing ICRS, such that the 3-spaces $\tau_{ICRS} = const.$
determined by a suitable ICRS time have a York time ${}^3K$ such
that the derived non-local York time reproduces the data for the
signatures of dark matter.

\bigskip

In the Conclusions it will be suggested that also the open problem
of dark energy could be rephrased as the determination of a suitable
York time in inhomogeneous cosmological space-times. Therefore there
is the possibility of an understanding of the "dark" aspects of the
universe in terms of relativistic metrology.

\vfill\eject

\section{Relativistic Metrology}
\label{sec:2}

As shown in Ref. \cite{51} modern relativistic metrology is not only
deeply rooted in Maxwell theory and its quantization but is also
beginning to take into account GR.
\bigskip

The basic metrological conventions on the Earth surface are:\medskip

a) An atomic clock as a standard of time. The fundamental conceptual
time scale is the {\it SI atomic second}: it is the duration of 9
192 631 770 periods of the radiation corresponding to the transition
between the two hyperfine levels of the ground state of the cesium
133 atom. This definition refers to a cesium atom at rest at a
temperature of $0^o$. However the practical time standard is the
International Atomic Time (TAI), which is defined as a suitable
weighted average of the SI time kept by over 200 atomic clocks in
about 70 national laboratories worldwide. Time scales connected with
TAI are the GPS Station Time and the Universal Time (UC) based on
Earth's rotation \footnote{It is based on Very Long Baseline
Interferometry (VLBI) observations of distant quasars, on Lunar
Laser Ranging (LLR) and on determination of GPS satellite orbits.}.
All the other existing time scales inside the Solar System are
connected to this standard by fixed conventions.\hfill\medskip

b) The 2-way velocity of light (only one clock is involved in its
definition), fixed to the value $c = 299\, 792\, 458 m\, s^{-1}$, in
place of the standard of length \footnote{The {\it meter} is the
length of the path traveled by light in vacuum during a time
interval of $1/c$ of a second.}. To measure the 3-distance between
two objects in an inertial frame one sends a ray of light from the
first object, to which is associated an atomic clock, to the second
one, where it is reflected and then reabsorbed by the first object.
The measure of the flight time and the 2-way velocity of light
determine the 3-distance between the objects. This convention is
compatible with the Euclidean 3-space of inertial frames in
Minkowski space-time. When the technology will allow one to measure
the deviations from Euclidean 3-space implied by PN gravity one will
need a modified convention taking into account a general
relativistic notion of length.

\bigskip

Given these standards one can think to the Global Positioning System
(GPS) as a local standard of space-time. To define GPS one needs a
conventional reference frame centered on a given time-like observer.
Inside the Solar System one has well defined conventions for the
following reference frames:\hfill\break

A) The description of satellites around the Earth is done by means
of NASA coordinates  either in the International Terrestrial
Reference System (ITRS; it is a frame fixed on the Earth surface) or
in the Geocentric Celestial Reference System (GCRS)  centered on the
world-line of the Earth center (see Ref.\cite{5}). Both of them use
a {\it geocentric coordinate time} $t_G$ connected to TAI.
\medskip

B) The description of planets and other objects in the Solar System
uses the Barycentric Celestial Reference System (BCRS), centered in
the barycenter of the Solar System (see Ref.\cite{5}). It uses a
{\it barycentric coordinate time} $t_B$ connected to $t_G$ and TAI.
\bigskip

While ITRS is essentially realized as a non-relativistic
non-inertial frame  in Galilei space-time, BCRS is defined as a {\it
quasi-inertial frame}, {\it non-rotating} with respect to some
selected fixed stars, in Minkowski space-time with nearly-Euclidean
3-spaces (one ignores the perturbations induced from the Milky Way).
It can also be considered as a PM space-time with 3-spaces having a
very small extrinsic curvature of order $c^{-2}$. GCRS is obtained
from BCRS by taking into account Earth's rotation around the Sun
with a suitable Lorentz boost with corrections from PN gravity
\footnote{See Ref.\cite{52} for possible gravitational anomalies
inside the Solar System.}. By taking into account the extension of
the geoid and Earth revolution around its axis one goes from the
quasi-Minkowskian GCRS to the quasi-Galilean ITRS.

\bigskip

New problems emerge by going outside the Solar System. In astronomy
the positions of stars and galaxies are determined from the data
(luminosity, light spectrum, angles) on the sky, i.e. on a
2-dimensional spherical surface around the Earth with the relations
between it and the observatory on the Earth  done with GPS. To get a
description of stars and galaxies as living in a 4-dimensional
space-time one introduces the International Celestial Reference
System ICRS (see Refs. \cite{3}). Its time scale is a "second"
connected to  GPS, TAI and SI and therefore to $t_G$ and $t_B$. ICRS
has the origin in the solar system barycenter, which is considered
as quasi-inertial observer carrying a quasi-inertial (essentially
non-relativistic) reference frame with rectangular 3-coordinates in
a {\it nearly Galilei space-time} whose 3-spaces are nearly
Euclidean. The directions of the spatial axes are effectively
defined by the adopted coordinates of 212 extragalactic radio
sources observed by VLBI . These radio sources (quasars and AGN,
active galactic nuclei) are assumed to have no observable intrinsic
angular momentum. Thus, the ICRS is a {\it space-fixed} system, more
precisely a {\it kinematically non-rotating} system, which  provides
the orientation of BCRS.

\medskip

In astronomy the unit of length is the {\it astronomical unit AU},
approximately equal to the mean Earth-Sun distance. Measurements of
the relative positions of planets in the Solar System are done by
radar: one measures the time taken for light to be reflected from an
object using the conventional value of the velocity of light $c$.
Both for objects inside the Solar System and for the nearest stars
one measure the distance with the {\it trigonometric parallax} by
using the propagation of light and its velocity $c$ in inertial
frames. One measures the difference (the inclination angle) in the
apparent position of an object viewed along two different lines of
sight at two different times and then uses Euclidean geometry to
evaluate the distance. The used unit in astronomy is the parsec,
which is 3.26 light-years or $3.26\, 10^{16}$ meters.
\medskip

This convention cannot be used for more distant either galactic or
extra-galactic objects. New notions like standard candles, dynamical
parallax, spectroscopic parallax, luminosity distance,..... are
needed \cite{2}. These notions involve both aspects of light
propagation in curved space-times and cosmological assumptions like
the Hubble law (velocity-red-shift linear relation).

\bigskip

However if one takes into account the description of the universe
given by cosmology, the actual cosmological space-time cannot be a
nearly Galilei space-time but it must be a cosmological space-time
with some theoretical {\it cosmic time}. In the standard
cosmological model \cite{1} it is a homogeneous and isotropic FRW
space-time whose instantaneous 3-spaces have nearly vanishing
internal 3-curvature, so that they may locally be replaced with
Euclidean 3-spaces as it is done in galactic dynamics. However they
have a time-dependent {\it conformal factor} (it is one in Galilei
space-time) responsible for the Hubble constant regulating the
expansion of the universe. As a consequence the transition from the
astronomical ICRS to an astrophysical description taking into
account cosmology is far from being understood.

\medskip

What is still lacking is a PM extension of the celestial frame such
that the PM BCRS frame is its restriction to the solar system inside
our galaxy. In particular one needs the definition of a coordinate
time $t_{ICRS}$ connected to $t_B$ such that the 3-spaces $t_{ICRS}
= const.$ have a very small internal 3-curvature and a suitable
extrinsic curvature as sub-manifolds of the space-time connected
with the Hubble constant. In this way this astronomical PM ICRS
would be consistent with the FRW cosmological space-times used in
astrophysics except for the conformal factor determining the
accelerated expansion of the universe and creating problems in the
metrological use of fixed stars.
\medskip

Hopefully at least an PM extension of ICRS including our galaxy
(with the definition of a galactic coordinate system) will be
achieved with the ESA GAIA mission devoted to the cartography of the
Milky Way \cite{6}.

\vfill\eject

\section{Clock Synchronization and Global Non-Inertial Frames in
Minkowski Space-Time}
 \label{sec:3}

Since in the Minkowski space-time of SR time is not absolute, there
is no intrinsic notion of 3-space and of synchronization of clocks:
both of them have to be defined with some convention. As a
consequence the {\it 1-way velocity of light} from one observer A to
an observer B has a meaning only after a choice of a convention for
synchronizing the clock in A with the one in B. Therefore the
crucial quantity in special relativity is the {\it 2-way (or round
trip) velocity of light $c$} involving only one clock. It is this
velocity (a kind of mean velocity) which is isotropic and constant
in SR and replaces the standard of length in relativistic metrology.
\medskip

Einstein convention for the synchronization of clocks in Minkowski
space-time uses the 2-way velocity of light to identify the
Euclidean 3-spaces of the inertial frames centered on an inertial
observer A by means of only its clock. The inertial observer A sends
a ray of light at $x^o_i$ towards the (in general accelerated)
observer B; the ray is reflected towards A at a point P of B
world-line and then reabsorbed by A at $x^o_f$; by convention P is
synchronous with the mid-point between emission and absorption on
A's world-line, i.e. $x^o_P = x^o_i + {1\over 2}\, (x^o_f - x^o_i) =
{1\over 2}\, (x^o_i + x^o_f)$. This convention selects the Euclidean
instantaneous 3-spaces $x^o = ct = const.$ of the inertial frames
centered on A. Only in this case the one-way velocity of light
between A and B coincides with the two-way one, $c$. However, as
said in the Introduction, if the observer A is accelerated, the
convention can breaks down due the possible appearance of coordinate
singularities.

\bigskip

As a consequence, a theory of global non-inertial frames in
Minkowski space-time has to be developed in a metrology-oriented way
to overcame the pathologies of the 1+3 point of view. This has been
done in the papers of Ref.\cite{21} based on the {\it 3+1 point of
view} and on the use of observer-dependent Lorentz scalar radar
4-coordinates. This theory and its implications for the description
of isolated systems in SR will be reviewed in this Section.

\subsection{3+1 Splittings of Minkowski Spacetime and Radar
4-Coordinates}

Assume that the world-line $x^{\mu}(\tau)$ of an arbitrary time-like
observer carrying a standard atomic clock is given: $\tau$ is an
arbitrary monotonically increasing function of the proper time of
this clock. Then one gives an admissible 3+1 splitting of Minkowski
space-time, namely a nice foliation with space-like instantaneous
3-spaces $\Sigma_{\tau}$. It is the mathematical idealization of a
protocol for clock synchronization: all the clocks in the points of
$\Sigma_{\tau}$ sign the same time of the atomic clock of the
observer \footnote{It is the non-factual idealization required by
the Cauchy problem generalizing the existing protocols for building
coordinate system inside the future light-cone of a time-like
observer.}. On each 3-space $\Sigma_{\tau}$ one chooses curvilinear
3-coordinates $\sigma^r$ having the observer as origin. These are
the Lorentz-scalar and observer-dependent {\it radar 4-coordinates}
$\sigma^A = (\tau; \sigma^r)$.

\bigskip

If $x^{\mu} \mapsto \sigma^A(x)$ is the coordinate transformation
from the Cartesian 4-coordinates $x^{\mu}$ of a reference inertial
observer to radar coordinates, its inverse $\sigma^A \mapsto x^{\mu}
= z^{\mu}(\tau ,\sigma^r)$ defines the {\it embedding} functions
$z^{\mu}(\tau ,\sigma^r)$ describing the 3-spaces $\Sigma_{\tau}$ as
embedded 3-manifold into Minkowski space-time. The induced 4-metric
on $\Sigma_{\tau}$ is the following functional of the embedding
${}^4g_{AB}(\tau ,\sigma^r) = [z^{\mu}_A\, \eta_{\mu\nu}\,
z^{\nu}_B](\tau ,\sigma^r)$, where $z^{\mu}_A = \partial\,
z^{\mu}/\partial\, \sigma^A$ and ${}^4\eta_{\mu\nu} = \sgn\, (+---)$
is the flat metric \footnote{$\sgn = \pm 1$ according to either the
particle physics $\sgn = 1$ or the general relativity $\sgn = - 1$
convention.}. While the 4-vectors $z^{\mu}_r(\tau ,\sigma^u)$ are
tangent to $\Sigma_{\tau}$, so that the unit normal $l^{\mu}(\tau
,\sigma^u)$ is proportional to $\epsilon^{\mu}{}_{\alpha
\beta\gamma}\, [z^{\alpha}_1\, z^{\beta}_2\, z^{\gamma}_3](\tau
,\sigma^u)$, one has $z^{\mu}_{\tau}(\tau ,\sigma^r) = [N\, l^{\mu}
+ N^r\, z^{\mu}_r](\tau ,\sigma^r)$  with $N(\tau ,\sigma^r) =
\sgn\, [z^{\mu}_{\tau}\, l_{\mu}](\tau ,\sigma^r) = 1 + n(\tau,
\sigma^r)$ and $N_r(\tau ,\sigma^r) = - \sgn\, g_{\tau r}(\tau
,\sigma^r)$ being the lapse and shift functions.\bigskip

As a consequence, the components of the 4-metric ${}^4g_{AB}(\tau
,\sigma^r )$ have the following expression

\bea
 \sgn\, {}^4g_{\tau\tau} &=& N^2 - N_r\, N^r,\qquad
  - \sgn\, {}^4g_{\tau r} = N_r = {}^3g_{rs}\, N^s,\nonumber \\
 {}^3g_{rs} &=& - \sgn\, {}^4g_{rs} = \sum_{a=1}^3\, {}^3e_{(a)r}\,
 {}^3e_{(a)s} = \nonumber \\
 &=& {\tilde \phi}^{2/3}\, \sum_{a=1}^3\,
 e^{2\, \sum_{\bar b =1}^2\, \gamma_{\bar ba}\, R_{\bar b}}\,
 V_{ra}(\theta^i)\, V_{sa}(\theta^i),
 \label{1}
 \eea

\noindent where ${}^3e_{(a)r}(\tau ,\sigma^u)$ are cotriads on
$\Sigma_{\tau}$, ${\tilde \phi}^2(\tau ,\sigma^r) = det\,
{}^3g_{rs}(\tau ,\sigma^r)$ is the 3-volume element on
$\Sigma_{\tau}$, $\lambda_a(\tau ,\sigma^r) = [{\tilde \phi}^{1/3}\,
e^{\sum_{\bar b =1}^2\, \gamma_{\bar ba}\, R_{\bar b}}](\tau
,\sigma^r)$ are the positive eigenvalues of the 3-metric
($\gamma_{\bar aa}$ are suitable numerical constants) and
$V(\theta^i(\tau ,\sigma^r))$ are diagonalizing rotation matrices
depending on three Euler angles.\medskip

Therefore starting from the {\it four} independent embedding
functions $z^{\mu}(\tau, \sigma^r)$ one obtains the {\it ten}
components ${}^4g_{AB}$ of the 4-metric (or the quantities $N$,
$N_r$, $\tilde \phi$, $R_{\bar a}$, $\theta^i$), which play the role
of the {\it inertial potentials} generating the relativistic
apparent forces in the non-inertial frame. It can be shown \cite{21}
that the usual non-relativistic Newtonian inertial potentials are
hidden in the functions $N$, $N_r$ and $\theta^i$. The extrinsic
curvature tensor ${}^3K_{rs}(\tau, \sigma^u) = [{1\over {2\, N}}\,
(N_{r|s} + N_{s|r} - \partial_{\tau}\, {}^3g_{rs})](\tau,
\sigma^u)$, describing the {\it shape} of the instantaneous 3-spaces
of the non-inertial frame as embedded 3-sub-manifolds of Minkowski
space-time, is a secondary inertial potential, functional of the ten
inertial potentials ${}^4g_{AB}$.

\bigskip

The foliation is nice and admissible if it satisfies the conditions:
\hfill\medskip

1) $N(\tau ,\sigma^r) > 0$ in every point of $\Sigma_{\tau}$ so that
the 3-spaces never intersect, avoiding the coordinate singularity of
Fermi coordinates;\hfill\medskip

2) $\sgn\, {}^4g_{\tau\tau}(\tau ,\sigma^r) > 0$, so to avoid the
coordinate singularity of the rotating disk, and with the
positive-definite 3-metric ${}^3g_{rs}(\tau ,\sigma^u) = - \sgn\,
{}^4g_{rs}(\tau ,\sigma^u)$ having three positive eigenvalues (these
are the M$\o$ller conditions \cite{53});\hfill\medskip

3) all the 3-spaces $\Sigma_{\tau}$ must tend to the same space-like
hyper-plane at spatial infinity with a unit  normal
$\epsilon^{\mu}_{\tau}$, which is the time-like 4-vector of a set of
asymptotic ortho-normal tetrads $\epsilon^{\mu}_A$. These tetrads
are carried by asymptotic inertial observers and the spatial axes
$\epsilon^{\mu}_r$ are identified by the fixed stars of star
catalogues. At spatial infinity the lapse function tends to $1$ and
the shift functions vanish.

\subsection{Global Non-Inertial Frames in Minkowski Spacetime}

By using the asymptotic tetrads $\epsilon^{\mu}_A$ one can give the
following parametrization of the embedding functions

\bea
 z^{\mu}(\tau, \sigma^r) &=& x^{\mu}(\tau) + \epsilon^{\mu}_A\,
 F^A(\tau, \sigma^r),\qquad  F^A(\tau, 0) = 0,\nonumber \\
 &&{}\nonumber \\
 x^{\mu}(\tau) &=& x^{\mu}_o + \epsilon^{\mu}_A\, f^A(\tau),
 \label{2}
 \eea

\noindent where $x^{\mu}(\tau)$ is the world-line of the observer.
The functions $f^A(\tau)$ determine the 4-velocity $u^{\mu}(\tau) =
{\dot x}^{\mu}(\tau)/ \sqrt{\sgn\, {\dot x}^2(\tau)}$ (${\dot
x}^{\mu}(\tau) = {{d x^{\mu}(\tau)}\over {d\tau}}$) and the
4-acceleration $a^{\mu}(\tau) = {{d u^{\mu}(\tau)}\over {d\tau}}$ of
the observer.
\bigskip

The M$\o$ller conditions are non-linear differential conditions on
the functions $f^A(\tau)$ and $F^A(\tau, \sigma^r)$, so that it is
very difficult to construct explicit examples of admissible 3+1
splittings. When these conditions are satisfied Eqs.(\ref{2})
describe a global non-inertial frame in Minkowski space-time.

\bigskip

Till now the solution of M$\o$ller conditions is known in the
following two cases in which the instantaneous 3-spaces are parallel
Euclidean space-like hyper-planes not equally spaced due to a linear
acceleration.\medskip

A)  {\it Rigid non-inertial reference frames with translational
acceleration}. An example are the following embeddings

\medskip

\bea
 z^{\mu}(\tau ,\sigma^u ) &=& x^{\mu}_o +
\epsilon^{\mu}_{\tau}\, f(\tau ) + \epsilon^{\mu}_r\,
\sigma^r,\nonumber \\
 &&{}\nonumber \\
 &&g_{\tau\tau}(\tau ,\sigma^u ) = \sgn\,
 \Big({{d f(\tau )}\over {d\tau}}\Big)^2,\quad g_{\tau r}(\tau ,\sigma^u )
 =0,\quad g_{rs}(\tau ,\sigma^u ) = -\sgn\, \delta_{rs}.\nonumber \\
 &&{}
 \label{3}
 \eea

\medskip

This is a foliation with parallel hyper-planes with normal $l^{\mu}
= \epsilon^{\mu}_{\tau} = const.$ and with the time-like observer
$x^{\mu}(\tau ) = x^{\mu}_o + \epsilon^{\mu}_{\tau}\, f(\tau )$ as
origin of the 3-coordinates. The hyper-planes have translational
acceleration ${\ddot x}^{\mu}(\tau ) = \epsilon^{\mu}_{\tau}\, \ddot
f(\tau )$, so that they are not uniformly distributed like in the
inertial case $f(\tau ) = \tau$.

\bigskip

B) {\it Differentially rotating non-inertial frames} without the
coordinate singularity of the rotating disk.  The embedding defining
this frames is

\begin{eqnarray*}
 z^{\mu}(\tau ,\sigma^u ) &=& x^{\mu}(\tau ) + \epsilon^{\mu}_r\,
R^r{}_s(\tau , \sigma )\, \sigma^s \, \rightarrow_{\sigma
\rightarrow \infty}\,
x^{\mu}(\tau) + \epsilon^{\mu}_r\, \sigma^r,\nonumber \\
 &&{}\nonumber \\
 R^r{}_s(\tau ,\sigma ) &=& R^r{}_s(\alpha_i(\tau,\sigma )) =
 R^r{}_s(F(\sigma )\, {\tilde \alpha}_i(\tau)),\nonumber \\
 &&{}\nonumber \\
 &&0 < F(\sigma ) < {1\over {A\, \sigma}},\qquad {{d\, F(\sigma
 )}\over {d\sigma}} \not= 0\,\, (Moller\,\, conditions),
 \end{eqnarray*}

\bea
 z^{\mu}_{\tau}(\tau ,\sigma^u) &=& {\dot x}^{\mu}(\tau ) -
 \epsilon^{\mu}_r\,  R^r{}_s(\tau
 ,\sigma )\, \delta^{sw}\, \epsilon_{wuv}\, \sigma^u\, {{\Omega^v(\tau
 ,\sigma )}\over c},\nonumber \\
  z^{\mu}_r(\tau ,\sigma^u) &=& \epsilon^{\mu}_k\, R^k{}_v(\tau
 ,\sigma )\, \Big(\delta^v_r + \Omega^v_{(r) u}(\tau ,\sigma )\,
 \sigma^u\Big),
 \label{4}
 \eea

\noindent where $\sigma = |\vec \sigma |$ and $R^r{}_s(\alpha_i(\tau
,\sigma ))$ is a rotation matrix satisfying the asymptotic
conditions $R^r{}_s(\tau , \sigma)\, {\rightarrow}_{\sigma
\rightarrow \infty} \delta^r_s$, $\partial_A\, R^r{}_s(\tau ,\sigma
)\, {\rightarrow}_{\sigma \rightarrow \infty}\, 0$, whose Euler
angles have the expression $\alpha_i(\tau ,\vec \sigma ) = F(\sigma
)\, {\tilde \alpha}_i(\tau )$, $i=1,2,3$. The unit normal is
$l^{\mu} = \epsilon^{\mu}_{\tau} = const.$ and the lapse function is
$1 + n(\tau ,\sigma^u) = \sgn\, \Big(z^{\mu}_{\tau}\,
l_{\mu}\Big)(\tau ,\sigma^u) = \sgn\, \epsilon^{\mu}_{\tau}\, {\dot
x}_{\mu}(\tau ) > 0$. In Eq.(\ref{4}) one uses the notations
$\Omega_{(r)}(\tau ,\sigma ) = R^{-1}(\tau ,\vec \sigma )\,
\partial_r\, R(\tau ,\sigma )$ and $\Big(R^{-1}(\tau ,\sigma )\,
\partial_{\tau}\, R(\tau ,\sigma )\Big)^u{}_v = \delta^{um}\,
\epsilon_{mvr}\, {{\Omega^r(\tau ,\sigma)}\over c}$, with
$\Omega^r(\tau ,\sigma ) = F(\sigma )\, \tilde \Omega (\tau ,\sigma
)$ ${\hat n}^r(\tau ,\sigma )$ \footnote{${\hat n}^r(\tau ,\sigma )$
defines the instantaneous rotation axis and $0 < \tilde \Omega (\tau
,\sigma ) < 2\, max\, \Big({\dot {\tilde \alpha}}(\tau ), {\dot
{\tilde \beta}}(\tau ), {\dot {\tilde \gamma}}(\tau )\Big)$.} being
the angular velocity. The angular velocity vanishes at spatial
infinity and has an upper bound proportional to the minimum of the
linear velocity $v_l(\tau ) = {\dot x}_{\mu}\, l^{\mu}$ orthogonal
to the space-like hyper-planes. When the rotation axis is fixed and
$\tilde \Omega (\tau ,\sigma ) = \omega = const.$, a simple choice
for the function $F(\sigma )$ is $F(\sigma ) = {1\over {1 +
{{\omega^2\, \sigma^2}\over {c^2}}}}$ \footnote{Nearly rigid
rotating systems, like a rotating disk of radius $\sigma_o$, can be
described by using a function $F(\sigma )$ approximating the step
function $\theta (\sigma - \sigma_o)$.}.\medskip

To evaluate the non-relativistic limit for $c \rightarrow \infty$,
where $\tau = c\, t$ with $t$ the absolute Newtonian time, one
chooses the gauge function $F(\sigma ) = {1\over {1 + {{\omega^2\,
\sigma^2}\over {c^2}}}}\, \rightarrow_{c \rightarrow \infty}\, 1 -
{{ \omega^2\, \sigma^2}\over {c^2}} + O(c^{-4})$. This implies that
the corrections to rigidly-rotating non-inertial frames coming from
M$\o$ller conditions are of order $O(c^{-2})$ and become important
at the distance from the rotation axis where the horizon problem for
rigid rotations appears.

\bigskip

As shown in the first paper in Refs.\cite{21},  {\it global rigid
rotations are forbidden in relativistic theories}, because, if one
uses the embedding $z^{\mu}(\tau ,\sigma^u)= x^{\mu}(\tau ) +
\epsilon^{\mu}_r\, R^r{}_s(\tau )\, \sigma^s$ describing a global
rigid rotation with angular velocity $\Omega^r = \Omega^r(\tau )$,
then the resulting $g_{\tau\tau}(\tau ,\sigma^u)$ violates M$\o$ller
conditions, because it vanishes at $\sigma = \sigma_R = {1\over
{\Omega (\tau )}}\, \Big[\sqrt{{\dot x}^2(\tau ) + [{\dot
x}_{\mu}(\tau )\, \epsilon^{\mu}_r\, R^r{}_s(\tau )\, (\hat \sigma
\times \hat \Omega (\tau ))^r]^2}$ $- {\dot x}_{\mu}(\tau )\,
\epsilon^{\mu}_r\, R^r{}_s(\tau )\, (\hat \sigma \times \hat \Omega
(\tau ))^r \Big]$ ( $\sigma^u = \sigma\, {\hat \sigma}^u$, $\Omega^r
= \Omega\, {\hat \Omega}^r$, ${\hat \sigma}^2 = {\hat \Omega}^2 =
1$). At this distance from the rotation axis the tangential
rotational velocity becomes equal to the velocity of light. This is
the {\it horizon problem} of the rotating disk (the horizon is often
named the {\it light cylinder}). Let us remark that even if in the
existing theory of rotating relativistic stars \cite{54} one uses
differential rotations, notwithstanding that in the study of the
magnetosphere of pulsars often the notion of light cylinder is still
used.

\bigskip

The search of admissible 3+1 splittings with non-Euclidean 3-spaces
is much more difficult. The simplest case is the following
parametrization of the embeddings (\ref{1}) in terms of Lorentz
matrices $\Lambda^A{}_B(\tau, \sigma)\, \rightarrow_{\sigma
\rightarrow \infty}\, \delta^A_B$ \footnote{It corresponds to the
{\it locality hypothesis} of Ref.\cite{20}, according to which at
each instant of time the detectors of an accelerated observer give
the same indications as the detectors of the instantaneously
comoving inertial observer.} with $\Lambda^A{}_B(\tau, 0)$ finite.
The Lorentz matrix is written in the form $\Lambda = {\cal B}\,
{\cal R}$ as the product of a boost ${\cal B}(\tau, \sigma)$ and a
rotation ${\cal R}(\tau, \sigma)$ like the one in Eq.(\ref{4})
(${\cal R}^{\tau}{}_{\tau} = 1$, ${\cal R}^{\tau}{}_r = 0$, ${\cal
R}^r{}_s = R^r{}_s$). The components of the boost are ${\cal
B}^{\tau}{}_{\tau}(\tau, \sigma) = \gamma(\tau, \sigma) = 1/ \sqrt{1
- {\vec \beta}^2(\tau, \sigma)}$, ${\cal B}^{\tau}{}_r(\tau, \sigma)
= \gamma(\tau, \sigma)\, \beta_r(\tau, \sigma)$, ${\cal
R}^r{}_s(\tau, \sigma) = \delta^r_s + {{\gamma\, \beta^r\,
\beta_s}\over {1 + \gamma}}(\tau, \sigma)$, with $\beta^r(\tau,
\sigma) = G(\sigma)\, \beta^r(\tau)$, where $\beta^r(\tau)$ is
defined by the 4-velocity of the observer $u^{\mu}(\tau) =
\epsilon^{\mu}_A\, \beta^A(\tau)/ \sqrt{1 - {\vec \beta}^2(\tau)}$,
$\beta^A(\tau) = (1; \beta^r(\tau))$. The M$\o$ller conditions are
restrictions on $G(\sigma)\, \rightarrow_{\sigma \rightarrow
\infty}\, 0$ with $G(0)$ finite, whose explicit form is still under
investigation.

\bigskip

See the second paper of Ref.\cite{21} for the description of the
electro-magnetic field and of phenomena like the Sagnac effect and
the Faraday rotation in this framework for non-inertial frames.
Moreover the embedding (\ref{4}) has been used in the first paper of
Ref.\cite{34} on quantum mechanics in non-inertial frames.

\subsection{Congruences of Timelike Observers Associated with a 3+1
Splitting}

Each admissible 3+1 splitting of space-time allows one to define two
associated congruences of time-like observers.
\medskip

i) The congruence of the Eulerian observers with the unit normal
$l^{\mu}(\tau, \sigma^r) = z^{\mu}_A(\tau, \sigma^r)\, l^A(\tau,
\sigma^r)$ to the 3-spaces embedded in Minkowski space-time as unit
4-velocity. The world-lines of these observers are the integral
curves of the unit normal and in general are not geodesics. In
adapted radar 4-coordinates the contro-variant orthonormal tetrads
carried by the Eulerian observers are $l^A(\tau, \sigma^r)$,
${}^4{\buildrel \circ \over {\bar E}}^A_{(a)}(\tau, \sigma^r) = (0;
{}^3e^u_{(a)}(\tau, \sigma^r))$, where ${}^3e^U_{(a)}(\tau,
\sigma^r)$ ($a=1,2,3$) are triads on the 3-space.\medskip

If ${}^4\nabla$ is the covariant derivative associated with the
4-metric ${}^4g_{AB}(\tau, \sigma^r)$ induced by a 3+1 splitting,
the equation

\beq
 {}^4\nabla_A\, \sgn\, l_B = \sgn\, l_A\, {}^3a_B + \sigma_{AB} +
 {1\over 3}\, \theta\, {}^3h_{AB} - \omega_{AB},\quad  ({}^3h_{AB} =
 {}^4g_{AB} - \sgn\, l_A\, l_B),
 \label{5}
 \eeq

\noindent defines the {\it acceleration} ${}^3a^A$ (${}^3a^A\, l_A =
0$), the {\it expansion} $\theta$, the {\it shear} $\sigma_{AB} =
\sigma_{BA}$ ($\sigma_{AB}\, l^B = 0$) and the {\it vorticity or
twist} $\omega_{AB} = - \omega_{BA}$ ($\omega_{AB}\, l^B = 0$) of
the Eulerian observers with $\omega_{AB} = 0$ since they are
surface-forming by construction. They will be useful in GR as shown
in Section 7.

\bigskip

ii) The skew congruence with unit 4-velocity $v^{\mu}(\tau,
\sigma^r) = z^{\mu}_A(\tau, \sigma^r)\, v^A(\tau, \sigma^r)$ (in
general it is not surface-forming, i.e. it has a non-vanishing
vorticity, like the one of a rotating disk). The observers of the
skew congruence have the world-lines (integral curves of the
4-velocity) defined by $\sigma^r = const.$ for every $\tau$, because
the unit 4-velocity tangent to the flux lines $x^{\mu}_{{\vec
\sigma}_o}(\tau) = z^{\mu}(\tau, \sigma^r_o)$ is $v^{\mu}_{{\vec
\sigma}_o}(\tau) = z^{\mu}_{\tau}(\tau, \sigma^r_o)/\sqrt{\sgn\,
{}^4g_{\tau\tau}(\tau, \sigma^r_o)}$ (there is no horizon problem
because it is everywhere time-like in admissible 3+1 splittings).
They carry contro-variant orthonormal tetrads, given in
Ref.\cite{41}, not adapted to the foliation, connected in each point
by a Lorentz transformation to the ones of the Eulerian observer
present in this point.

\subsection{Parametrized Minkowski Theories}

In the global non-inertial frames of Minkowski space-time it is
possible to describe isolated systems (particles, strings, fields,
fluids) admitting a Lagrangian formulation  by means of {\it
parametrized Minkowski theories} \cite{23,24}, \cite{21}.\bigskip

The existence of a Lagrangian, which can be coupled  to an external
gravitational field, makes possible the determination of the matter
energy-momentum tensor and of the ten conserved Poincar\'e
generators $P^{\mu}$ and $J^{\mu\nu}$ (assumed finite) of every
configuration of the isolated system.
\bigskip

First of all one must replace the matter variables of the isolated
system with new ones knowing the clock synchronization convention
defining the 3-spaces $\Sigma_{\tau}$. For instance a Klein-Gordon
field $\tilde \phi (x)$ will be replaced with $\phi(\tau ,\sigma^r)
= \tilde \phi (z(\tau ,\sigma^r))$; the same for every other field.
Instead for a relativistic particle with world-line $x^{\mu}(\tau )$
one must make a choice of its energy sign: then the positive- (or
negative-) energy particle will be described by 3-coordinates
$\eta^r(\tau )$ defined by the intersection of its world-line with
$\Sigma_{\tau}$: $x^{\mu}(\tau ) = z^{\mu}(\tau ,\eta^r(\tau ))$.
Differently from all the previous approaches to relativistic
mechanics, the dynamical configuration variables are the
3-coordinates $\eta^r(\tau)$ and not the world-lines $x^{\mu}(\tau)$
(to rebuild them in an arbitrary frame one needs the embedding
defining that frame). This fact eliminates the possibility to have
time-like excitations in the spectrum of relativistic bound states:
inside each 3-space only space-like correlations among the particles
are possible.\medskip

Then one replaces the external gravitational 4-metric in the coupled
Lagrangian with the 4-metric ${}^4g_{AB}(\tau ,\sigma^r)$, which is
a functional of the embedding defining an admissible 3+1 splitting
of Minkowski space-time, and the matter fields with the new ones
knowing the instantaneous 3-spaces $\Sigma_{\tau}$.

\bigskip

Parametrized Minkowski theories are defined by the resulting
Lagrangian depending on the given matter and on the embedding
$z^{\mu}(\tau ,\sigma^r)$. The resulting action is invariant under
tto take into account of fermion fieldshe {\it frame-preserving
diffeomorphisms} $\tau \mapsto \tau^{'}(\tau, \sigma^u)$, $\sigma^r
\mapsto \sigma^{' r}(\sigma^u)$ firstly introduced in Ref.\cite{25}.
As a consequence, there are four first-class constraints with
exactly vanishing Poisson brackets (an Abelianized analogue of the
super-Hamiltonian and super-momentum constraints of canonical
gravity) determining the momenta conjugated to the embeddings in
terms of the matter energy-momentum tensor. This implies that the
embeddings $z^{\mu}(\tau ,\sigma^r)$ are {\it gauge variables}, so
that {\it all the admissible non-inertial or inertial frames are
gauge equivalent}, namely physics does {\it not} depend on the clock
synchronization convention and on the choice of the 3-coordinates
$\sigma^r$: only the appearances of phenomena change by changing the
notion of instantaneous 3-space \footnote{In the first paper of
Ref.\cite{34} there is the definition of {\it parametrized Galilei
theories}, non relativistic limit of the parametrized Minkowski
theories. Also the inertial and non-inertial frames in Galilei
space-time are gauge equivalent in this formulation.}.\medskip

Even if the gauge group is formed by the frame-preserving
diffeomorphisms, the matter energy-momentum tensor allows the
determination of the ten conserved Poincar\'e generators $P^{\mu}$
and $J^{\mu\nu}$ (assumed finite) of every configuration of the
system (in non-inertial frames they are asymptotic generators at
spatial infinity like the ADM ones in GR).

\bigskip

As an example one may consider N free scalar particles with masses
$m_i$ and sign of the energy $\eta_i = \pm$, whose world-lines are
identified by the configurational variables $\eta^r_i(\tau)$:
$x^{\mu}_i(\tau) = z^{\mu}(\tau, \eta^r_i(\tau))$, $i=1,..,N$. In
parametrized Minkowski theories they are described by the following
action depending on the configurational variables $\eta^r_i(\tau)$
and $z^{\mu}(\tau, \sigma^r)$

\bea
  S &=&\int d\tau\, d^{3}\sigma \,{\cal L}(\tau ,\sigma^u) = \int d\tau\,
  L(\tau),  \nonumber \\
 &&\nonumber\\
 {\cal L}(\tau ,\sigma^u) &=& -\sum_{i=1}^{N}\, \delta^3(\sigma^u - \eta^u_i(\tau ))
 \nonumber \\
 &&m_ic\, \eta_i\, \sqrt{\sgn\, [{}^4g_{\tau \tau }(\tau ,\sigma^u) + 2\,
 {}^4g_{\tau r}(\tau ,\sigma^u)\, {\dot{\eta}}_i^r(\tau ) + {}^4g_{rs }(\tau
 ,\sigma^u)\, {\dot{\eta}}_i^r(\tau )\, {\dot{\eta}}_i^s(\tau
 )]}.\nonumber \\
 &&{}
 \label{6}
 \eea

The resulting canonical momenta $\kappa_{ir}(\tau) = {{\partial\,
L(\tau)}\over {\partial\, \eta^r_i}}$, $\rho_{\mu}(\tau, \sigma^u) =
- \sgn\, {{\partial\, {\cal L}(\tau, \sigma^u)}\over {\partial\,
z^{\mu}_{\tau}(\tau, \sigma^u))}}$ satisfy the Poisson brackets $\{
\eta_i^r(\tau), \kappa_{js}(\tau) \} = - \delta^r_s\, \delta_{ij}$,
$\{ z^{\mu}(\tau, \sigma^u), \rho_{\nu}(\tau, \sigma^{' u}) \} = -
\delta^{\mu}_{\nu}\, \delta^3(\sigma^u - \sigma^{' u})$. The
Poincar\'e generators and the energy-momentum tensor of this system
are ($h^{rs} = - \sgn\, \gamma^{rs}$ with $\gamma^{ru}\, {}^4g_{us}
= \delta^r_s$; $\gamma = - \sgn\, det\, {}^4g_{rs}$)

\begin{eqnarray*}
  P^{\mu } &=& \int d^{3}\sigma \rho ^{\mu }(\tau ,\sigma^u),
\qquad J^{\mu \nu } =  \int d^{3}\sigma (z^{\mu }\rho ^{\nu }-z^{\nu
}\rho ^{\mu })(\tau ,\sigma^u),\nonumber \\
 &&{}\nonumber \\
  T^{AB}(\tau ,\sigma^u ) &=& - {2\over {\sqrt{- det\, {}^4g_{CD}(\tau ,\sigma^u
 )}}}\, {{\delta\, S}\over {\delta\, {}^4g_{AB}(\tau ,\sigma^u
 )}},\qquad T^{\mu\nu} = z_A^{\mu}\, z_B^{\nu}\, T^{AB},
 \end{eqnarray*}

 \bea
  T_{\perp\perp}(\tau, \sigma^u) &=& \Big(l_{\mu}\, l_{\nu}\, T^{\mu\nu}\Big)(\tau, \sigma^u)
 =\sum_{i=1}^N\, {{\delta^3(\sigma^u - \eta^u_i(\tau ))}\over {\sqrt{\gamma(\tau,
 \sigma^u)}}}\nonumber \\
 && \eta_i\, \sqrt{m_i^2 c^2 + h^{rs}(\tau, \sigma^u)\,
 \kappa_{ir}(\tau)\, \kappa_{is}(\tau)},\nonumber \\
  T_{\perp r}(\tau, \sigma^u) &=& \Big(l_{\mu}\, z_{r\nu}\, T^{\mu\nu}\Big)(\tau, \sigma^u)
 =\sum_{i=1}^N\, {{\delta^3(\sigma^u - \eta^u_i(\tau ))}\over {\sqrt{\gamma(\tau,
 \sigma^u)}}}\, \kappa_{ir}(\tau),\nonumber \\
  T_{rs}(\tau, \sigma^u) &=& \Big(z_{r\mu}\, z_{s\nu}\, T^{\mu\nu}\Big)(\tau, \sigma^u)
 =\sum_{i=1}^N\, {{\delta^3(\sigma^u - \eta^u_i(\tau ))}\over {\sqrt{\gamma(\tau,
 \sigma^u)}}}\nonumber \\
 && \eta_i\, {{\kappa_{ir}(\tau)\, \kappa_{is}(\tau)}\over
{\sqrt{m_i^2 c^2 +
 h^{vw}(\tau, \sigma^u)\, \kappa_{iv}(\tau)\, \kappa_{iw}(\tau)}}}.
 \label{7}
 \eea

The four first-class constraints implying the gauge nature of the
embedding and the gauge equivalence of the description in different
non-inertial frames are

\beq
 \rho_{\mu}(\tau, \sigma^u) - \sqrt{\gamma(\tau,
 \sigma^u)}\, \Big[l_{\mu}\, T_{\perp\perp} - z_{r\mu}\, h^{rs}\,
 T_{\perp s}\Big](\tau, \sigma^u) \approx 0.
 \label{8}
 \eeq

The same description can be given for the Klein-Gordon \cite{55} and
Dirac  \cite{56} fields and for the electro-magnetic field
\cite{21}.
\bigskip

To describe the physics in a given admissible non-inertial frame
described by an embedding $z^{\mu}_F(\tau, \sigma^u)$ one must add
the gauge-fixings $z^{\mu}(\tau, \sigma^u) - z^{\mu}_F(\tau,
\sigma^u) \approx 0$.

\subsection{The Instant Form of Dynamics in the Inertial Rest Frames
and the Problem of the Relativistic Center of Mass}

If one restricts himself to inertial frames, one can define the {\it
inertial rest-frame instant form of dynamics for isolated systems}
by choosing the 3+1 splitting corresponding to the intrinsic
inertial rest frame of the isolated system centered on an inertial
observer: the instantaneous 3-spaces, named {\it Wigner 3-spaces}
due to the fact that the 3-vectors inside them are Wigner spin-1
3-vectors \cite{23,24}, are orthogonal to the conserved 4-momentum
$P^{\mu}$ (assumed time-like, $\sgn\, P^2 > 0$) of the
configuration.
\bigskip

In this framework one can  give the final solution to the old
problem of the relativistic extension of the Newtonian center of
mass of an isolated system. In its rest frame there are {\it only}
three notions of collective variables, which can be built by using
{\it only} the Poincar\'e generators :

the canonical non-covariant Newton-Wigner center of mass (or center
of spin) ${\tilde x}^{\mu}(\tau)$,

the non-canonical covariant Fokker-Pryce center of inertia
$Y^{\mu}(\tau)$

the non-canonical non-covariant M$\o$ller center of energy
$R^{\mu}(\tau)$.

While $Y^{\mu}(\tau )$ is a 4-vector, ${\tilde x}^{\mu}(\tau )$ and
$R^{\mu}(\tau )$ are not 4-vectors. All of them tend to the
Newtonian center of mass in the non-relativistic limit. Since the
Poincar\'e generators know the whole $\Sigma_{\tau}$, they and
therefore also these three collective variables are {\it non-local}
quantities: as a consequence they are {\it non measurable} with
local means \cite{21,26,28,33}.
\bigskip

If one centers the inertial rest frame on the world-line of the
Fokker-Planck center of inertia thought as an inertial observer,
then the corresponding embedding has the expression \cite{23,24,28}

\beq
 z_W^{\mu}(\tau, \vec \sigma) = Y^{\mu}(\tau) + \epsilon^{\mu}_r(\vec
 h)\, \sigma^r,
 \label{9}
 \eeq

\noindent where $Y^{\mu}(\tau)$ is the Fokker-Pryce
center-of-inertia 4-vector, $\vec h = \vec P/\sqrt{\sgn\, P^2}$ and
$\epsilon^{\mu}{}_{A=\nu}(\vec h) = L^{\mu}{}_{A = \nu}(P,
{\buildrel \circ \over P})$ are the columns of the standard Wigner
boost for time-like orbits sending $P^{\mu} = \sqrt{\sgn\, P^2}\,
(\sqrt{1 + {\vec h}^2}; \vec h)$ to ${\buildrel \circ \over P}^{\mu}
= \sqrt{\sgn\, P^2}\, (1; 0)$. Their expression is
$\epsilon^{\mu}_{\tau}(\vec h) = h^{\mu} = (\sqrt{1 + {\vec h}^2};
\vec h)$ and $\epsilon^{\mu}_r(\vec h) = \Big( h_r; \delta^i_r +
{{h^i\, h_r}\over {1 + \sqrt{1 + {\vec h}^2}}}\Big)$ as shown in
Appendix B of Ref.\cite{57}.

\medskip

As shown in Ref.\cite{26,28,29,33}, the three collective variables
can be expressed as known functions of the Lorentz-scalar rest time
$\tau = c\, T_s = h \cdot \tilde x = h \cdot Y = h \cdot R$, of
canonically conjugate Jacobi data (frozen Cauchy data) $\vec z =
Mc\, {\vec x}_{NW}(0)$ and $\vec h = \vec P/Mc$  \footnote{Their
Poisson brackets are$\{ z^i, h^j\} = \delta^{ij}$. ${\vec
x}_{NW}(\tau )$ is the standard Newton-Wigner non-covariant
3-position, classical counterpart of the corresponding position
operator; the use of $\vec z$ avoids to take into account the mass
spectrum of the isolated system in the description of the center of
mass.  The non-covariance of $\vec z$ under Poincar\'e
transformations $(a ,\Lambda )$ has the following form \cite{33,57}
$z^i\, \mapsto z^{{'}\, i} = \Big(\Lambda^i{}_j -
{{\Lambda^i{}_{\mu}\, h^{\mu}}\over {\Lambda^o{}_{\nu}\, h^{\nu}}}\,
\lambda^o{}_j\Big)\, z^j + \Big(\Lambda^i{}_{\mu} -
{{\Lambda^i{}_{\nu}\, h^{\nu}}\over {\Lambda^o{}_{\rho}\,
h^{\rho}}}\, \Lambda^o{}_{\mu}\Big)\, (\Lambda^{-1}\, a)^{\mu}$.},
of the invariant mass $Mc = \sqrt{\sgn\, P^2}$ of the system and of
its rest spin ${\vec {\bar S}}$.

\bigskip

While the world-line of the non-canonical covariant external
Fokker-Pryce 4-center of inertia is

\bea
 Y^{\mu}(\tau ) &=& z_W^{\mu}(\tau ,\vec 0) =
 \Big({\tilde x}^o(\tau ); \vec Y(\tau )\Big) =\nonumber \\
 &=& \Big(\sqrt{1 + {\vec h}^2}\, (\tau + {{\vec h \cdot \vec z}\over
 {Mc}});  {{\vec z}\over {Mc}} + (\tau + {{\vec h \cdot \vec z}\over
 {Mc}})\, \vec h + {{\vec S \times \vec h}\over {Mc\, (1 + \sqrt{1 +
 {\vec h}^2})}} \Big),\nonumber \\
 &&{}
 \label{10}
  \eea

\noindent  the pseudo-world-line of the canonical non-covariant
external 4-center of mass is (${\tilde {\vec \sigma}} = {{- \vec S
\times \vec h}\over {Mc\, (1 + \sqrt{1 + {\vec h}^2})}}$ from
Ref.\cite{26})

\bea
 {\tilde x}^{\mu}(\tau ) &=& \Big({\tilde x}^o(\tau ); {\tilde
 {\vec x}}(\tau )\Big) = z^{\mu}_W(\tau ,{\tilde {\vec \sigma}}) = Y^{\mu}(\tau ) +
 \Big(0; {{- \vec S \times \vec h}\over {Mc\, (1 + \sqrt{1 + {\vec
 h}^2})}}\Big)  =\nonumber \\
 &=& \Big(\sqrt{1 + {\vec h}^2}\, (\tau + {{\vec
 h \cdot \vec z}\over {Mc}}); {{\vec z}\over {Mc}} + (\tau + {{\vec h
 \cdot \vec z}\over {Mc}})\, \vec h\Big).
 \label{11}
  \eea

\bigskip

The world-lines of the positive-energy particles are parametrized by
the Wigner 3-vectors ${\vec \eta}_i(\tau)$, $i = 1,2, ..,N$, and are
given by

\beq
 x^{\mu}_i(\tau) = z^{\mu}_W(\tau, {\vec \eta}_i(\tau)) =
Y^{\mu}(\tau) + \epsilon^{\mu}_r(\tau)\, \eta^r_i(\tau).
 \label{12}
 \eeq

\medskip

The world-lines $x^{\mu}_i(\tau)$ of the particles are derived
(interaction-dependent) quantities. Also the standard particle
4-momenta are derived quantities, whose expression is
$p^{\mu}_i(\tau) = \epsilon^{\mu}_A(\vec h)\, \kappa_i^A(\tau) =
h^{\mu}\, \sqrt{m_i^2 c^2 + {\vec \kappa}_i^2(\tau)} -
\epsilon^{\mu}_r(\vec h)\, \kappa_{ir}(\tau)$ with $\sgn\, p_i^2 =
m_i^2 c^2$ in the free case.\medskip

In the case of interacting particles the reconstruction of the
world-lines requires a complex interaction-dependent procedure
delineated in Ref.\cite{29}, where there is also a comparison of the
present approach with the other formulations of relativistic
mechanics developed for the study of the problem of {\it
relativistic bound states}. See Ref.\cite{21} for the extension to
non-inertial frames.

In general the world-lines $x^{\mu}_i(\tau)$ do not satisfy
vanishing Poisson brackets (they are relativistic predictive
coordinates, see Ref.\cite{29}): already at the classical level a
{\it non-commutative structure} emerges due to the Lorentz signature
of the space-time \cite{33}.

\bigskip

In each Lorentz frame one has different pseudo-world-lines
describing $R^{\mu}$ and ${\tilde x}^{\mu}$: the canonical 4-center
of mass ${\tilde x}^{\mu}$ {\it lies in between} $Y^{\mu}$ and
$R^{\mu}$ in every (non rest)-frame. As discussed in Subsection IIF
of Ref.\cite{28}, this leads to the existence of the {\it M$\o$ller
non-covariance world-tube}, around the world-line $Y^{\mu}$ of the
covariant non-canonical Fokker-Pryce 4-center of inertia $Y^{\mu}$.
The {\it invariant radius} of the tube is $\rho =\sqrt{- \sgn\,
W^2}/p^2 = |\vec S|/\sqrt{\sgn\, P^2}$ where ($W^2 = - P^2\, {\vec
S}^2$ is the Pauli-Lubanski invariant when $\sgn\, P^2 > 0$). This
classical intrinsic radius is a non-local effect of Lorentz
signature absent in Euclidean spaces and delimits the non-covariance
effects (the pseudo-world-lines) of the canonical 4-center of mass
${\tilde x}^{\mu}$ \footnote{In the rest-frame the world-tube is a
cylinder: in each instantaneous 3-space there is a disk of possible
positions of the canonical 3-center of mass orthogonal to the spin.
In the non-relativistic limit the radius $\rho$ of the disk tends to
zero and one recovers the non-relativistic center of mass.}. They
are not detectable because the M$\o$ller radius is of the order of
the Compton wave-length: an attempt to test its interior would mean
to enter in the quantum regime of pair production. The M$\o$ller
radius $\rho$ is also a remnant of the energy conditions of general
relativity in flat Minkowski space-time \cite{23}.

\bigskip

Finally Eqs.(\ref{7}) can be used to extend the multipolar
expansions of Ref.\cite{58} to this framework for relativistic
isolated systems as it is shown in the third paper of
Refs.\cite{26}.

\subsection{The Description of Isolated Systems in the Rest Frame and their
Poincar\'e Generators }

In the inertial rest frame of an isolated system Eqs.(\ref{7}) are
the starting point to get the explicit form of its Poincar\'e
generators, in particular of the Lorentz boosts, which, differently
from the Galilei ones, are interaction dependent.

\bigskip

As shown in Ref.\cite{28}, every isolated system (i.e. a closed
universe) can be visualized as a decoupled non-covariant collective
(non-local) pseudo-particle (the {\it external} center of mass),
described by the frozen Jacobi data $\vec z$, $\vec h$, carrying a
{\it pole-dipole structure}, namely the invariant mass $M\, c$ (the
monopole) and the rest spin ${\vec {\bar S}}$ (the dipole) of the
system, and with an associated {\it external} realization  of the
Poincar\'e group \footnote{The last term in the Lorentz boosts
induces the Wigner rotation of the 3-vectors inside the Wigner
3-spaces.}:\medskip

\bea
 P^{\mu} &=& M\, c\, h^{\mu} = M\, c\, \Big(\sqrt{1 + {\vec
h}^2}; \vec h\Big),\nonumber \\
&&{}\nonumber \\
J^{ij} &=& z^i\, h^j - z^j\, h^i + \epsilon^{ijk}\, S^k,\qquad K^i =
J^{oi} = - \sqrt{1 + {\vec h}^2}\, z^i + {{(\vec S \times \vec
h)^i}\over {1 + \sqrt{1 + {\vec h}^2}}}.\nonumber \\
&&{}
 \label{13}
  \eea

\bigskip

The universal breaking of Lorentz covariance is connected to this
decoupled non-local collective variable and is irrelevant because
all the dynamics of the isolated system leaves inside the Wigner
3-spaces and is Wigner-covariant. The invariant mass and the rest
spin are built in terms of the Wigner-covariant variables of the
given isolated system (${\vec \eta}_i(\tau)$ and ${\vec
\kappa}_i(\tau)$) inside the Wigner 3-spaces \cite{21,26,28,33}.

\bigskip

In each Wigner 3-space $\Sigma_{\tau}$ there is a {\it unfaithful
internal} realization of the Poincar\'e algebra, whose generators
are built by using the energy-momentum tensor (\ref{7}) of the
isolated system. While the internal energy and angular momentum are
$Mc$ and ${\vec {\bar S}}$ respectively, the internal 3-momentum
vanishes: it is the rest frame condition. Also the internal
(interaction dependent) Lorentz boost  vanishes: this condition
identifies the covariant non-canonical Fokker-Pryce center of
inertia as the natural inertial observer origin of the 3-coordinates
$\sigma^r$ in each Wigner 3-space.

\bigskip

For N free particles  the internal Poincar\'e generators have the
following expression

\bea
 M\, c &=& {1\over c}\, {\cal E}_{(int)} = \sum_{i=1}^N\,
\sqrt{m_i^2\, c^2 + {\vec
\kappa}^2_i},\nonumber \\
{\vec {\cal P}}_{(int)} &=& \sum_{i=1}^N\, {\vec \kappa}_i \approx
0,\nonumber \\
\vec S &=& {\vec {\cal J}}_{(int)} = \sum_{i=1}^N\,
{\vec \eta}_i \times {\vec \kappa}_i,\nonumber \\
{\vec {\cal K}}_{(int)} &=& - \sum_{i=1}^N\, {\vec \eta}_i\,
\sqrt{m_i^2\, c^2 + {\vec \kappa}_i^2} \approx 0.
 \label{14}
  \eea

\bigskip

Since one is in an instant form of the dynamics, in the interacting
case only $M c$ and ${\vec {\cal K}}_{(int)}$ become interaction
dependent.
\bigskip

The three pairs of second-class (interaction dependent) constraints
${\vec {\cal P}}_{(int)} \approx 0$,  ${\vec {\cal K}}_{(int)}
\approx 0$, eliminate the {\it internal} 3-center of mass and its
conjugate momentum inside the Wigner 3-spaces \footnote{One can show
\cite{26,28} that one has ${\vec {\cal K}}_{(int)} = - M\, {\vec
R}_+$, where ${\vec R}_+$ is the internal M$\o$ller 3-center of
energy inside the Wigner 3-spaces. The rest frame condition ${\vec
{\cal P}}_{(int)} \approx 0$ implies ${\vec R}_+ \approx {\vec q}_+
\approx {\vec y}_+$, where ${\vec q}_+$ is the internal 3-center of
mass and ${\vec y}_+$ the internal Fokker-Pryce 3-center of
inertia.}: this avoids a double counting of the collective variables
(external and internal center of mass). As a consequence the
dynamics inside the Wigner 3-spaces is described in terms of
internal Wigner-covariant relative variables. In the case of N
relativistic particles one defines the following canonical
transformation \cite{33} (see Ref.\cite{26} for other variants) ($m
= \sum_{i=1}^N\, m_i$)

\bea
 {\vec \eta}_+ &=& \sum_{i=1}^N\, {{m_i}\over m}\, {\vec
\eta}_i,\qquad {\vec \kappa}_+ = {\vec {\cal P}}_{(int)} =
\sum_{i=1}^N\, {\vec \kappa}_i,\nonumber \\
{\vec \rho}_a &=& \sqrt{N}\, \sum_{i=1}^N\, \gamma_{ai}\, {\vec
\eta}_i,\qquad {\vec \pi}_a = {1\over {\sqrt{N}}}\, \sum_{i=1}^N\,
\Gamma_{ai}\,
{\vec \kappa}_i,\qquad a = 1,..,N-1,\nonumber \\
&&{}\nonumber \\
{\vec \eta}_i &=& {\vec \eta}_+ + {1\over {\sqrt{N}}}\,
\sum_{a-1}^{N-1}\, \Gamma_{ai}\, {\vec \rho}_a,\qquad {\vec
\kappa}_i = {{m_i}\over m}\, {\vec \kappa}_+ + \sqrt{N}\,
\sum_{a=1}^{N-1}\, \gamma_{ai}\, {\vec \pi}_a,
 \label{15}
  \eea

\noindent with the following canonicity conditions
\footnote{Eqs.(\ref{15}) describe a family of canonical
transformations, because the $\gamma_{ai}$'s depend on
${\frac{1}{2}}(N-1)(N-2)$ free independent parameters.}

\bea
 &&\sum_{i=1}^{N}\, \gamma _{ai} = 0,\qquad  \sum_{i=1}^{N}\,
\gamma _{ai}\, \gamma _{bi} = \delta _{ab},\qquad \sum_{a=1}^{N-1}\,
\gamma _{ai}\, \gamma _{aj} = \delta _{ij} - {\frac{1}{N}},
\nonumber \\
&&{}\nonumber \\
&&\Gamma_{ai} = \gamma_{ai} - \sum_{k=1}^N\, {\frac{{m_k}}{ m}}\,
\gamma_{ak},\qquad \gamma_{ai} = \Gamma_{ai} -
{\frac{1}{ N}}\, \sum_{k=1}^N\, \Gamma_{ak},\nonumber \\
&&{}\nonumber \\
&&\sum_{i=1}^N\, {\frac{{m_i}}{m}}\, \Gamma_{ai} = 0,\qquad
\sum_{i=1}^N\, \gamma_{ai}\, \Gamma_{bi} = \delta_{ab},\qquad
\sum_{a=1}^{N-1}\, \gamma_{ai}\, \Gamma_{aj} =  \delta_{ij} -
{\frac{{m_i}}{ m}}.\nonumber \\
&&{}
 \label{16}
  \eea

Since Eqs.(\ref{14}) imply ${\vec \kappa}_+(\tau) = {\vec {\cal
P}}_{(int)} \approx 0$ and ${\vec \eta}_+(\tau) \approx {\vec
f}_+({\vec \rho}_a(\tau), {\vec \pi}_a(\tau))$ due to ${\vec {\cal
K}}_{(int)} \approx 0$, the invariant mass $M c$ and the rest spin
${\vec {\bar S}}$ become functions only of the N-1 pairs of relative
canonical variables.
\medskip

As a consequence, Eqs.(\ref{10}), (\ref{12}) and (\ref{15}) imply
that the world-lines $x^{\mu}_i(\tau)$ can be expressed in terms of
the Jacobi data $\vec z$, $\vec h$, and of the relative variables
${\vec \rho}_a(\tau)$, ${\vec \pi}_a(\tau)$, $a=1,..,N-1$. See
Ref.\cite{55} for the collective and relative variables of the
Klein-Gordon field and the second paper in Ref.\cite{28} for such
variables for the electro-magnetic field in the radiation gauge. For
these systems one can give for the first time the explicit closed
form of the interaction-dependent Lorentz boosts.

\medskip

One finds that disregarding the unobservable external center of mass
all the dynamics is described only by relative variables: this is a
form of {\it weak relationism} without the heavy foundational
problem of approaches like the one in Ref.\cite{59}.

\bigskip

The non-relativistic limit of this description \cite{33} is Newton
mechanics with the Newton center of mass decoupled from the relative
variables and moreover after a canonical transformation to the
frozen Hamilton-Jacobi description of the center of mass.

\bigskip

An important remark  is that the internal space of relative
variables is independent from the reference inertial frame used for
the description of the isolated system. As shown in Ref.\cite{33},
the formalism is built in such a way that the Wigner rotation
induced on the relative variables by a Lorentz transformation
connecting two reference inertial frames is the identity, i.e. the
space of the relative variables in an {\it abstract internal space}
insensitive to Lorentz transformations carried by the external
center of mass (or in a more covariant description carried by the
Fokker-Pryce center-of-inertia 4-vector origin of the embedding
(\ref{9})).

\bigskip

Finally in Ref.\cite{21} there is the extension of the formalism  to
admissible {\it non-inertial rest frames}, where $P^{\mu}$ is
orthogonal to the asymptotic space-like hyper-planes to which the
instantaneous 3-spaces tend at spatial infinity. In these
non-inertial rest frames the internal Poincar\'e generators are
asymptotic (constant of the motion) symmetry generators like the
asymptotic ADM ones in the asymptotically Minkowskian space-times.

\vfill\eject

\section{Implications for Relativistic Mechanics and Classical Field
Theory in Special Relativity and the Multi-Temporal Quantization
Approach}
 \label{sec:4}

In the rest-frame instant form of the dynamics it has been possible
to find the explicit form of the internal Poincar\'e generators (in
particular of the interaction-dependent invariant mass and Lorentz
boosts) not only for the Klein-Gordon \cite{55} and Dirac
\cite{31,56} fields, but also for the electro-magnetic field in the
radiation gauge (the only one suitable for the Shanmugadhasan
canonical transformations of constraint theory \cite{60})
\cite{21,28}, for relativistic fluids \cite{41,42}, spinning
particles \cite{27,31} and for massless particles, the Nambu string
and the two-level atom \cite{32}.

\medskip

In this Section some other developments in SR will be reviewed.

\subsection{Relativistic Atomic Physics}

Standard atomic physics \cite{61} is a semi-relativistic treatment
of quantum electro-dynamics (QED) in which the matter fields are
approximated by scalar (or spinning) particles, the relevant
energies are below the threshold of pair production and the
electro-magnetic field is described in the Coulomb gauge at the
order $1/c$.

In Refs.\cite{27,28} a fully relativistic formulation of classical
atomic physics in the rest-frame instant form of dynamics was given
with the electro-magnetic field in the radiation gauge and with the
electric charges $Q_i$ of the positive-energy particles being
Grassmann-valued ($Q_i^2 = 0$, $Q_i\, Q_j = Q_j\, Q_i$ for $i\not=
j$) to regularize the electro-magnetic self-energies on the
world-lines of particles. In the language of QED this is both a
ultraviolet regularization (no loop contributions) and an infrared
one (no brehmstrahlung), so that only the one-photon exchange
diagram contributes and its static and non-static effects are
replaced by potentials in a formulation based on the Cauchy problem.
Therefore the starting point is a parametrized Minkowski theory with
N charged positive-energy particles mutually interacting with a
Coulomb potential and coupled to a dynamical transverse
electro-magnetic field described by the canonical variables ${\vec
A}_{\perp}(\tau, \sigma^r)$ and ${\vec \pi}_{\perp}(\tau, \sigma^r)
= {\vec E}_{\perp}(\tau, \sigma^r)$.

In the first paper of Ref.\cite{27} (the second paper is devoted to
spinning particles) it is shown that the use of the Lienard-Wiechert
solution (see the third paper in Ref.\cite{27}) with "no incoming
radiation field" allows one to arrive at a description of N charged
particles dressed with a Coulomb cloud and mutually interacting
through the Coulomb potential augmented with the full relativistic
Darwin potential. This happens independently from the choice of the
Green function (retarded, advanced, symmetric,..) due to the
Grassmann regularization. The quantization allows one to recover the
standard instantaneous approximation for relativistic bound states,
which till now had only been obtained starting from QED (either in
the instantaneous approximations of the Bethe-Salpeter equation or
in the quasi-potential approach). In the case of spinning particle
the relativistic Salpeter potential was identified.

Moreover in Ref.\cite{28} it is shown that by using the previous
results one can find a canonical transformation from the canonical
basis ${\vec \eta}_i(\tau)$, ${\vec \kappa}_i(\tau)$, ${\vec
A}_{\perp}(\tau, \sigma^r)$, ${\vec \pi}_{\perp}(\tau, \sigma^r)$,
in which the internal Poincar\'e generators have the expression in
the case N=2 ($\vec B = \vec \partial \times {\vec A}_{\perp}$,
$c(\vec \sigma) = - 1/4\pi\, |\vec \sigma|$)

\begin{eqnarray*}
 \mathcal{E}_{(int)} &=&  M\, c^2 =
 c\, \sum_{i=1}^{N}\, \sqrt{ m_{i}^{2}\, c^2 + \Big({\vec{
 \kappa}} _i(\tau ) - {\frac{{Q_i}}{c}}\, {\vec{A }}_{\perp }(\tau
 ,\vec{\eta} _i(\tau ))\Big)^2} +  \nonumber \\
 &+&\sum_{i\neq j}\, \frac{Q_{i}\, Q_{j}}{4\pi\, \mid
 \vec{\eta}_{i}(\tau ) - \vec{\eta} _{j}(\tau )\mid } +
 {\frac{1}{2}}\, \int d^{3}\sigma \, [{\vec{ \pi }}_{\perp }^{2} +
 {\vec{B}}^{2}](\tau ,\vec{\sigma}),\nonumber \\
 &&{}\nonumber \\
  \mathcal{\vec{P}}_{(int)} &=&
   \sum_{i=1}^N\, {\vec{\kappa}}_i(\tau ) + {\frac{1}{c}}\, \int
 d^{3}\sigma\, \lbrack {\vec{\pi}}_{\perp } \times
 {\vec{B}}](\tau ,\vec{\sigma}) \approx 0,  \nonumber \\
 &&{}  \nonumber \\
  {\bar S}^r &=&\sum_{i=1}^{N}\,\Big(\vec{\eta}_{i}(\tau )\times
 {\vec{\kappa}}_{i}(\tau ) \Big)^{r} + {\frac{1}{c}}\, \int
 d^{3}\sigma (\vec{\sigma}\times \,\Big([{ \vec{\pi}}_{\perp }{\
 \times }{\vec{B}}]\Big)^{r}(\tau ,\vec{\sigma}),
 \end{eqnarray*}

\bea
  \mathcal{K}_{(int)}^{r} &=& - \sum_{i=1}^{N}\, \eta^r_{i}(\tau )\, \sqrt{m_{i}^{2}\, c^2 +
 \Big({{\ \vec{\kappa} }}_{i}(\tau ) - {\frac{{Q_{i}}}{c}}\,
 {\vec{A}}_{\perp }(\tau ,{
 \ \ \vec{\eta}}_{i}(\tau))\Big)^2} +  \nonumber \\
 &+& {\frac{1}{c}}\, \sum_{i=1}^{N}\, \Big[\sum_{j\not=i}^{1..N}\,
 Q_{i}\, Q_{j}\, [{\ \frac{1}{{\ \triangle
 _{{\vec{\eta}}_{j}}}}}{\frac{{\partial }}{{
 \partial \eta _{j}^{r}}} }c({\vec{\eta}}_{i}(\tau ) - {\vec{\eta}}_{j}(\tau
 )) - \nonumber \\
 &-& \eta _{j}^{r}(\tau )\, c({\ \vec{\eta}}_{i}(\tau ) -
 {\vec{\eta}}_{j}(\tau ))] +  \nonumber \\
 &+& Q_{i}\, \int d^{3}\sigma\, {\pi}_{\perp }^{r}(\tau
 ,\vec{\sigma})\, c( \vec{\sigma} - {\ \vec{\eta}}_{i}(\tau ))\Big] -
 {\frac{1}{2c}}\, \int d^{3}\sigma\, \sigma ^{r}\,
 ({{\vec{\pi}}}_{\perp }^{2} + {{\vec{B} }} ^{2})(\tau
 ,\vec{\sigma}),
 \label{17}
 \eea

\noindent to a new canonical basis ${\hat {\vec \eta}}_i(\tau)$,
${\hat {\vec \kappa}}_i(\tau)$, ${\vec A}_{\perp rad}(\tau,
\sigma^r)$, ${\vec \pi}_{\perp rad}(\tau, \sigma^r)$ so that in the
rest frame there is a decoupled free radiation transverse field and
a system of charged particles mutually interacting with Coulomb plus
Darwin potential. See the first paper in Ref.\cite{27} for the
explicit form of the relativistic Darwin potential. The new internal
Poincar\'e generators in the N=2 case are

\begin{eqnarray*}
 \mathcal{E}_{(int)} &=&
 M\,c^{2} = c\,\sum_{i=1}^2\,\sqrt{m_{i}^{2}\,c^{2}+{\hat{\vec{
 \kappa}}}_{i}^{2}(\tau )} + {\frac{{Q_1\,Q_2}}{{4\pi \,|{\
 \hat{\vec{\eta}}}_1(\tau )-{\hat{\vec{\eta}}}_2(\tau )|}}}+
 \nonumber \\
 &+& V_{DARWIN}({\hat {\vec \kappa}}_1(\tau ), {\hat {\vec \kappa}}_2(\tau ),
 {\hat {\vec \eta}}_1(\tau ) - {\hat {\vec \eta}}_2(\tau )) +  \nonumber \\
 &&{}\nonumber \\
 &+& {\frac{1}{2}}\,\int d^{3}\sigma \,\Big({\vec{ \pi}}_{\perp
 rad}^{2}+{\vec{B}}_{rad}^{2}\Big)(\tau ,\vec{\sigma}) =
  \mathcal{E}_{matter} + \mathcal{E}_{rad},\nonumber \\
 &&{}\nonumber \\
  {\vec {\mathcal{P}}}_{(int)} &=&
  \sum_{i=1}^2\, {\hat {\vec \kappa}}_i(\tau ) + {\frac{1}{c}}\, \int
 d^3\sigma\, \Big({\vec \pi} _{\perp rad} \times {\vec B}_{rad}
 \Big)(\tau ,\vec \sigma ) =   {\vec {\mathcal{P}}}_{matter} + {\vec
 {\mathcal{P}}}_{rad}\approx 0,\nonumber \\
 &&{}\nonumber \\
  {\vec {\bar S}} &=&
 \sum_i\, {\hat {\vec \eta}}_i \times {\hat {\vec \kappa}}_i +
 {\frac{1}{c }}\, \int d^3\sigma\, \vec \sigma \times \Big({\vec
 \pi}_{\perp rad} \times {\vec B} _{rad}\Big)(\tau ,\vec \sigma ) =
 {\vec {\bar S}}_{matter} + {\vec {\bar S}}_{rad},
 \end{eqnarray*}

 \bea
  {\vec{\mathcal{K}}}_{(int)} &=& - \sum_{i=1}^2\, {\hat {\vec \eta}}_i\, \sqrt{m_i^2\, c^2 +
 {\hat {\vec \kappa}}_i^2} -  \nonumber \\
 &-& {\frac{1}{2}}\, {\frac{{Q_1\, Q_2}}{c}}\, \Big[{\hat {\vec
 \eta}}_1\, { \frac{{{\hat {\vec \kappa}}_1 \cdot
 \Big({\frac{1}{2}}\, {\frac{{\partial\, { \hat
 {\mathcal{K}}}_{12}({\hat {\vec \kappa}}_1, {\hat {\vec \kappa}}_2,
 { \hat {\vec \rho}}_{12})}}{{\partial\, {\hat {\vec \rho}}_{12}}}} -
 2\, {\vec A}_{\perp S2}({\hat {\vec \kappa}}_2, {\hat {\vec
 \rho}}_{12})\Big)}}{\sqrt{
 m_1^2\, c^2 + {\hat {\vec \kappa}}_1^2}}} +  \nonumber \\
 &+& {\hat {\vec \eta}}_2\, {\frac{{{\hat {\vec \kappa}}_2 \cdot
 \Big({\frac{1 }{2}}\, {\frac{{\partial\, {\hat
 {\mathcal{K}}}_{12}({\hat {\vec \kappa}}_1, {\hat {\vec \kappa}}_2,
 {\hat {\vec \rho}}_{12})}}{{\partial\, {\hat {\vec \rho}}_{12}}}} -
 2\, {\vec A}_{\perp S1}({\hat {\vec \kappa}}_1, {\hat {\vec
 \rho}}_{12})\Big)}}{\sqrt{m_2^2\, c^2 + {\hat {\vec \kappa}}_2^2}}}
 \Big] -\nonumber \\
 &-& {\frac{1}{2}}\, {\frac{{Q_1\, Q_2}}{c}}\, \Big(\sqrt{m_1^2\, c^2
 + { \hat {\vec \kappa}}_1^2}\, {\frac{{\partial}}{{\partial\, {\hat
 {\vec \kappa} }_1}}} + \sqrt{m_2^2\, c^2 + {\hat {\vec
 \kappa}}_2^2}\, {\frac{{\partial}}{{
 \partial\, {\hat {\vec \kappa}}_2}}} \Big)\, {\hat {\mathcal{K}}}_{12}({
 \hat {\vec \kappa}}_1, {\hat {\vec \kappa}}_2, {\hat {\vec
 \rho}}_{12}) -\nonumber \\
 &-& {\frac{{Q_1\, Q_2}}{{4\pi\, c}}}\, \int d^3\sigma\,
 \Big({\frac{{{\hat { \vec \pi}}_{\perp S1}(\vec \sigma - {\hat {\vec
 \eta}}_1, {\hat {\vec \kappa} }_1)}}{{|\vec \sigma - {\hat {\vec
 \eta}}_2|}}} + {\frac{{{\hat {\vec \pi}} _{\perp S2}(\vec \sigma -
 {\hat {\vec \eta}}_2, {\hat {\vec \kappa}}_2)}}{{
 |\vec \sigma - {\hat {\vec \eta}}_1|}}} \Big) -  \nonumber \\
 &-& {\frac{{Q_1\, Q_2}}{c}}\, \int d^3\sigma\, \vec \sigma\,\,
 \Big({\hat { \vec \pi}}_{\perp S1}(\vec \sigma - {\hat {\vec
 \eta}}_1, {\hat {\vec \kappa} }_1) \cdot {\hat {\vec \pi}}_{\perp
 S2}(\vec \sigma - {\hat {\vec \eta}}_2, {
 \hat {\vec \kappa}}_2) +  \nonumber \\
 &+& {\hat {\vec B}}_{S1}(\vec \sigma - {\hat {\vec \eta}}_1, {\hat
 {\vec \kappa}}_1) \cdot {\hat {\vec B}}_{ S2}(\vec \sigma - {\hat
 {\vec \eta}}_2, {\hat {\vec \kappa}}_2) \Big) -  \nonumber \\
 &&{}\nonumber \\
 &-& {\frac{1}{{2\, c}}}\, \int d^3\sigma\, \vec \sigma\,\,
 \Big({\vec \pi}^2_{\perp rad} + {\vec B}^2_{rad}\Big)(\tau ,\vec
 \sigma ) =
 {\vec {\mathcal{K}}}_{matter} + {\vec {\mathcal{K}}}_{rad}
 \approx 0.
 \label{18}
 \eea

\noindent The only restriction on the two decoupled systems is the
elimination of their overall internal 3-center of mass inside the
Wigner 3-spaces. Therefore, at the classical level there is a way
out from the Haag theorem forbidding the existence of the
interaction picture in QED, so that there is no unitary evolution
based on interpolating fields from the "in" states to the "out" ones
in scattering processes. While the extension of these results to the
non-inertial rest frame is done in Ref.\cite{21}, the quantization
of this framework is under investigation.
\bigskip

In the first paper of Ref.\cite{32} there is the formulation in the
rest-frame instant form of the relativistic quark model in the
radiation gauge for the SU(3) Yang-Mills fields with scalar quarks
having Grassmann-valued color charges. While in Eq.(101) of that
paper there is the rest-frame condition, in Eqs.(97)  there is the
invariant mass $M c^2$ for a quark-antiquark system. In it the
electro-magnetic Coulomb potential of Eq.(\ref{17}) is replaced with
a potential, given in Eq.(95), depending on the color transverse
vector potential through the Green function of the SU(3) covariant
divergence. The non-linearity of the problem does not allow to
evaluate a Lienard-Wiechert solution and to find the analogue of
Eqs.(\ref{18}) .

\subsection{Relativistic Kinetic Theory and Relativistic
Micro-Canonical Ensemble}

In the rest-frame instant form of dynamics it is also possible to
give a finally consistent treatment of relativistic kinetic theory
and relativistic statistical mechanics \cite{30}. In particular one
can give a definition of the relativistic micro-canonical ensemble
for an isolated system of N interacting particles with fixed
internal energy ${\cal E}$ and rest spin ${\vec {\cal S}}$ only in
terms of the internal Poincar\'e generators in the Wigner 3-spaces
by means of the partition function (V is the volume defined by the
function $\chi(V)$ vanishing outside it)

\bea
  \tilde Z({\cal E}, {\vec {\cal S}}, V, N) &=& {1\over {N!}}\,
  \int\, \prod^{1..N}_i\, d^3\eta_i\,
 \chi(V)\, \int\, \prod^{1..N}_j\, d^3\kappa_j\, \delta(M\, c^2 -
 {\cal E})\nonumber \\
 &&\delta^3({\vec {\bar S}} - {\vec {\cal S}})\, \delta^3({\vec {\cal P}}_{(int)})\,
 \delta^3({{{\vec {\cal K}}_{(int)}}\over {M c}}).
 \label{19}
 \eea

Also it extension to non-inertial rest frames can be given by using
the results of Ref.\cite{21} with the result that notwithstanding
the presence of long-range inertial forces one has still an
equilibrium distribution.

\subsection{Relativistic Quantum Mechanics and Relativistic
Entanglement}

A new formulation of {\it relativistic quantum mechanics} in the
Wigner 3-spaces of the inertial rest frame  is developed in
Ref.\cite{33} in absence of the electro-magnetic field. It englobes
all the known results about relativistic bound states (absence of
relative times) and avoids the causality problems of the Hegerfeldt
theorem \cite{62} (the instantaneous spreading of wave packets).
\medskip

In it one quantizes the frozen Jacobi data $\vec z$ and $\vec h$ of
the canonical non-covariant decoupled external center of mass and
the relative variables in the Wigner 3-spaces. Since the center of
mass is decoupled, its non-covariance is irrelevant: like for the
wave function of the universe, who will observe it?\medskip

The resulting Hilbert space  has the following tensor product
structure: $H = H_{com, HJ} \otimes H_{rel}$, where $H_{com, HJ}$ is
the Hilbert space of the external center of mass (in the
Hamilton-Jacobi formulation due to the use of frozen Jacobi data)
while $H_{rel}$ is the Hilbert space of the  relative variables in
the abstract internal space living in the Wigner 3-spaces. While at
the non-relativistic level this presentation of the Hilbert space is
unitarily equivalent to the tensor product of the Hilbert spaces
$H_i$ of the individual particles $ H = H_1 \otimes H_2 \otimes
...$, this is not true at the relativistic level.
\medskip

If one considers two scalar quantum particles with Klein-Gordon wave
functions belonging to Hilbert spaces ${\cal H}_{x^o_i}$, in the
tensor-product Hilbert space $({\cal H}_1)_{x^o_1} \otimes ({\cal
H}_2)_{x^o_2} \otimes ...$ there is no correlation among the times
of the particles (their clocks are not synchronized) so that in most
of the states there are some particles in the absolute future of the
others. As a consequence the two types of Hilbert spaces lead to
unitarily inequivalent descriptions and have different scalar
products (compare Refs. \cite{33} and \cite{57}).\medskip

As a consequence, at the relativistic level the zeroth postulate of
non-relativistic quantum mechanics does not hold: the Hilbert space
of composite systems is not the tensor product of the Hilbert spaces
of the sub-systems. Contrary to Einstein's notion of separability
(separate objects have their independent real states) \cite{63} one
gets a {\it kinematical spatial non-separability} induced by the
need of clock synchronization for eliminating the relative times and
to be able to formulate a well-posed relativistic Cauchy problem.
\medskip

Moreover one has the {\it non-locality} of the non-covariant
external center of mass which implies its {\it non-measurability}
with local instruments \footnote{In Ref. \cite{64} it was shown that
the quantum Newton-Wigner position should not be a self-adjoint
operator, but only a symmetric one, with an implication of bad
localization.}. While its conjugate momentum operator must be well
defined and self-adjoint, because its eigenvalues describe the
possible values for the total momentum of the isolated system (the
momentum basis is therefore a {\it preferred basis} in the Hilbert
space), it is not clear whether it is meaningful to define
center-of-mass wave packets.\medskip

These {\it non-locality} and {\it kinematical spatial
non-separability} are due to the Lorentz signature of Minkowski
space-time and this fact reduce the relevance of {\it quantum
non-locality} in the study of the foundational problems of quantum
mechanics \cite{63} which have to be rephrased in terms of relative
variables.

\medskip

The quantization defined in Ref.\cite{33} leads to a first
formulation of a theory for {\it relativistic entanglement}, which
is deeply different from the non-relativistic entanglement due to
these kinematical non-locality and spatial non-separability. To have
control on the Poincar\'e group one needs an isolated systems
containing all the relevant entities (for instance both Alice and
Bob) of the experiment under investigation and also the environment
when needed. One has to learn to reason in terms of relative
variables adapted to the experiment like molecular physicists do
when they look to the best system of Jacobi coordinates adapted to
the main chemical bonds in the given molecule. This theory has still
to be developed together with its extension to non-inertial rest
frames.

\subsection{Multitemporal Quantization in Non-Inertial Frames}

This quantization of relativistic mechanics can be extended to the
class of global non-inertial frames with space-like hyper-planes as
3-spaces and differentially rotating 3-coordinates defined in
Ref.\cite{21} by using the {\it multi-temporal quantization}
approach developed in Ref.\cite{65}.\medskip

As shown in Ref.\cite{34}, in this type of quantization one
quantizes only the 3-coordinates $\eta^r_i(\tau)$ of the particles
and {\it not} the inertial effects (like the Coriolis and
centrifugal ones): they remain c-numbers describing the appearances
of phenomena. The known results in atomic and nuclear physics are
reproduced.

\subsection{Open Problem}

The main open problem in SR is the quantization of {\it fields} in
non-inertial frames  due to the no-go theorem of Ref.\cite{66}
showing the existence of obstructions to the unitary evolution of a
massive quantum Klein-Gordon field between two space-like surfaces
of Minkowski space-time. It turns out that the Bogoljubov
transformation connecting the creation and destruction operators on
the two surfaces is not of the Hilbert-Schmidt type, i.e. that the
Tomonaga-Schwinger approach in general is not unitary. One must
reformulate the problem using the nice foliations of the admissible
3+1 splittings of Minkowski space-time and to try to identify all
the 3+1 splittings allowing unitary evolution. This will be a
prerequisite to any attempt to quantize canonical gravity taking
into account the equivalence principle (global inertial frames do
not exist) with the further problem that in general the Fourier
transform does not exist in Einstein space-times.

\vfill\eject

\section{Non-Inertial Frames in Asymptotically Minkowskian Einstein
Space-Times and ADM Tetrad Gravity}
 \label{sec:5}

After this description of SR induced by the metrology-oriented
problem of clock synchronization, one has to face the same problems
in the globally hyperbolic, topologically trivial, asymptotically
Minkowskian space-times without super-translations \footnote{These
space-times must also be without Killing symmetries, because,
otherwise, at the Hamiltonian level one should introduce complicated
sets of extra Dirac constraints for each existing Killing vector.}
of GR. As shown in the first paper of Ref.\cite{17}, in the chosen
class of space-times the ten {\it strong} asymptotic ADM Poincar\'e
generators $P^A_{ADM}$, $J^{AB}_{ADM}$ (they are fluxes through a
2-surface at spatial infinity) are well defined functionals of the
4-metric fixed by the boundary conditions at spatial infinity.
\medskip

While in SR Minkowski space-time is an absolute notion, in Einstein
GR also the space-time is a dynamical object \cite{35} and the
gravitational field is described by the metric structure of the
space-time, namely by the ten dynamical fields ${}^4g_{\mu\nu}(x)$
($x^{\mu}$ are world 4-coordinates). The 4-metric
${}^4g_{\mu\nu}(x)$ tends in a suitable way to the flat Minkowski
4-metric ${}^4\eta_{\mu\nu}$ at spatial infinity \cite{17}: having
an {\it asymptotic Minkowskian background} the usual splitting of
the 4-metric in the bulk in a background plus perturbations in the
weak field limit can be avoided as shown in Section VII.

\medskip

The ten dynamical fields ${}^4g_{\mu\nu}(x)$ are not only a
(pre)potential for the gravitational field (like the
electro-magnetic and Yang-Mills fields are the potentials for
electro-magnetic and non-Abelian forces) but also determines the
{\it chrono-geometrical structure of space-time} through the line
element $ds^2 = {}^4g_{\mu\nu}\, dx^{\mu}\, dx^{\nu}$. Therefore the
4-metric teaches relativistic causality to the other fields: it says
to massless particles like photons and gluons which are the allowed
world-lines in each point of space-time. This basic property is lost
in every quantum field theory approach to gravity with a fixed
background 4-metric \footnote{The ACES mission of ESA \cite{67} will
give the first precision measurement of the gravitational red-shift
of the geoid, namely of the $1/c^2$ deformation of Minkowski
light-cone caused by the geo-potential. In every quantum field
theory approach to gravity, where the definition of the Fock space
requires the use of the Fourier transform on a fixed background
space-time with a fixed light-cone, this is a non-perturbative
effect requiring the re-summation of the perturbative expansion.}.

\bigskip

In these space-times one can define global non-inertial frames by
using the same admissible 3+1 splittings, centered on a time-like
observer, and the observer-dependent radar 4-coordinates $\sigma^A =
(\tau; \sigma^r)$ employed in SR. This  will allow to separate the
{\it inertial} (gauge) degrees of freedom of the gravitational field
(playing the role of inertial potentials) from the dynamical {\it
tidal} ones at the Hamiltonian level.

\medskip

In GR the dynamical fields are the components ${}^4g_{\mu\nu}(x)$ of
the 4-metric and not the  embeddings $x^{\mu} = z^{\mu}(\tau,
\sigma^r)$ defining the admissible 3+1 splittings of space-time like
in  the parametrized Minkowski theories of SR. Now the gradients
$z^{\mu}_A(\tau, \sigma^r)$ of the embeddings give the transition
coefficients from radar to world 4-coordinates, so that the
components ${}^4g_{AB}(\tau, \sigma^r) = z^{\mu}_A(\tau, \sigma^r)\,
z^{\nu}_B(\tau, \sigma^r)\, {}^4g_{\mu\nu}(z(\tau, \sigma^r))$ of
the 4-metric will be the dynamical fields in the ADM action. Like in
SR the 4-vectors $z^{\mu}_A(\tau, \sigma^r)$, tangent to the
3-spaces $\Sigma_{\tau}$, are used to define the unit normal
$l^{\mu}(\tau, \sigma^r) = z^{\mu}_A(\tau, \sigma^r)\, l^A(\tau,
\sigma^r)$ to $\Sigma_{\tau}$, while the 4-vector
$z^{\mu}_{\tau}(\tau, \sigma^r)$ has the lapse function as component
along the unit normal and the shift functions as components along
the tangent vectors.

\medskip

Since the world-line of the time-like observer can be chosen as the
origin of a set of the spatial world coordinates, i.e.
$x^{\mu}(\tau) = (x^o(\tau); 0)$, it turns out that with this choice
the space-like surfaces of constant coordinate time $x^o(\tau) =
const.$ coincide with the dynamical instantaneous 3-spaces
$\Sigma_{\tau}$ with $\tau = const.$. By using asymptotic flat
tetrads $\epsilon^{\mu}_A = \delta^{\mu}_o\, \delta^{\tau}_A +
\delta^{\mu}_i\, \delta^i_A$ (with $\epsilon^A_{\mu}$ denoting the
inverse flat cotetrads) and by choosing a coordinate world time
$x^o(\tau) = x^o_o + \epsilon^o_{\tau}\, \tau = x^o_o + \tau$, one
gets the following preferred embedding corresponding to these given
world 4-coordinates

\beq
 x^{\mu} = z^{\mu}(\tau, \sigma^r) = x^{\mu}(\tau) +
 \epsilon^{\mu}_r\, \sigma^r = \delta^{\mu}_o\, x^o_o +
 \epsilon^{\mu}_A\, \sigma^A.
 \label{20}
 \eeq

\noindent This choice implies $z^{\mu}_A(\tau, \sigma^r) =
\epsilon^{\mu}_A$ and ${}^4g_{\mu\nu}(x = z(\tau, \sigma^r)) =
\epsilon^A_{\mu}\, \epsilon_{\nu}^B\, {}^4g_{AB}(\tau, \sigma^r)$.

\bigskip

As shown in Ref.\cite{35}, the dynamical nature of space-time
implies that each solution (i.e. an Einstein 4-geometry) of
Einstein's equations (or of the associated ADM Hamilton equations)
dynamically selects a preferred 3+1 splitting of the space-time,
namely in GR the instantaneous 3-spaces  are dynamically determined
in the chosen world coordinate system. Eq.(\ref{20}) can be used to
describe this 3+1 splitting and then by means of 4-diffeomorphisms
the solution can be written in an arbitrary world 4-coordinate
system in general not adapted to the dynamical 3+1 splitting. This
gives rise to the 4-geometry corresponding to the given solution.

\bigskip

To define the canonical formalism the Einstein-Hilbert action for
metric gravity (depending on the second derivative of the metric)
must be replaced with the ADM action (the two actions differ for a
surface tern at spatial infinity). As shown in the first paper of
Refs.\cite{17}, the Legendre transform and the definition of a
consistent canonical Hamiltonian require the introduction of the
DeWitt surface term at spatial infinity: the final canonical
Hamiltonian turns out to be the {\it strong} ADM energy (a flux
through a 2-surface at spatial infinity), which is equal to the {\it
weak} ADM energy (expressed as a volume integral over the 3-space)
plus constraints. Therefore there is not a frozen picture but an
evolution generated by a Dirac Hamiltonian equal to the weak ADM
energy plus a linear combination of the first class constraints.
Also the other strong ADM Poincar\'e generators are replaced by
their weakly equivalent weak form ${\hat P}^A_{ADM}$, ${\hat
J}^{AB}_{ADM}$.\medskip

In the first paper of Ref.\cite{17} it is also shown that the
boundary conditions on the 4-metric required by the absence of
super-translations imply that the only admissible 3+1 splittings of
space-time (i.e. the allowed global non-inertial frames) are the
{\it non-inertial rest frames}:  their 3-spaces are asymptotically
orthogonal to the weak ADM 4-momentum. Therefore one gets ${\hat
P}^r_{ADM} \approx 0$ as the rest-frame condition of the 3-universe
with a mass and a rest spin fixed by the boundary conditions. Like
in SR the 3-universe can be visualized as a decoupled non-covariant
(non-measurable) external relativistic center of mass plus an
internal non-inertial rest-frame 3-space containing only relative
variables (see the first paper in Ref.\cite{40}).

\subsection{The Parametrization of Tetrads for ADM Tetrad Gravity}

To take into account the coupling of fermions to the gravitational
field metric gravity has to be replaced with tetrad gravity. This
can be achieved by decomposing the 4-metric on cotetrad fields (by
convention a sum on repeated indices is assumed)

\beq
 {}^4g_{AB}(\tau, \sigma^r) = E_A^{(\alpha)}(\tau, \sigma^r)\,
 {}^4\eta_{(\alpha)(\beta)}\, E^{(\beta)}_B(\tau, \sigma^r),
 \label{21}
 \eeq

\noindent by putting this expression into the ADM action and by
considering the resulting action, a functional of the 16 fields
$E^{(\alpha)}_A(\tau, \sigma^r)$, as the action for ADM tetrad
gravity. In Eq.(\ref{21}) $(\alpha)$ are flat indices and the
cotetrad fields $E^{(\alpha)}_A$ are the inverse of the tetrad
fields $E^A_{(\alpha)}$, which are connected to the world tetrad
fields by $E^{\mu}_{(\alpha)}(x) = z^{\mu}_A(\tau, \sigma^r)\,
E^A_{(\alpha)}(z(\tau, \sigma^r))$ by the embedding of
Eq.(\ref{20}).

\medskip

This leads to an interpretation of gravity based on a congruence of
time-like observers endowed with orthonormal tetrads: in each point
of space-time the time-like axis is the  unit 4-velocity of the
observer, while the spatial axes are a (gauge) convention for
observer's gyroscopes. This framework was developed in the second
and third paper of  Refs.\cite{17}.
\medskip

Even if the action of ADM tetrad gravity depends upon 16 fields, the
counting of the physical degrees of freedom of the gravitational
field does not change, because this action is invariant not only
under the group of 4-difeomorphisms but also under the O(3,1) gauge
group of the Newman-Penrose approach \cite{68} (the extra gauge
freedom acting on the tetrads in the tangent space of each point of
space-time).

\medskip

The cotetrads $E^{(\alpha)}_A(\tau, \sigma^r)$ are the new
configuration variables. They are connected to cotetrads
$\eo^{(\alpha )}_A(\tau, \sigma^r)$ adapted to the 3+1 splitting of
space-time, namely such that the inverse adapted time-like tetrad
$\eo_{(o)}^A(\tau, \sigma^r)$ is the unit normal to the 3-space
$\Sigma_{\tau}$, by a standard Wigner boosts for time-like
Poincar\'e orbits with parameters $\varphi_{(a)}(\tau, \sigma^r)$,
$a=1,2,3$

\bea
 E_A^{\alpha)} &=& L^{(\alpha)}{}_{(\beta)}( \varphi_{(a)})\,
 {\buildrel o\over E}_A^{(\beta)}, \qquad {}^4g_{AB} = \eo^{(\alpha
 )}_A\, {}^4\eta_{(\alpha )(\beta )}\, \eo^{(\beta )}_B,\nonumber \\
 &&{}\nonumber \\
 L^{(\alpha )}{}_{(\beta )}(\varphi_{(a)}) &{\buildrel {def}\over
 =}& L^{(\alpha )}{}_{(\beta )}(V(z(\sigma ));\,\, {\buildrel \circ
 \over V}) = \delta^{(\alpha )}_{(\beta )} + 2 \sgn\, V^{(\alpha
 )}(z(\sigma ))\, {\buildrel \circ \over V}_{(\beta )} -\nonumber \\
 &-&\sgn\, {{(V^{(\alpha )}(z(\sigma )) + {\buildrel \circ \over
 V}^{(\alpha )})\, (V_{(\beta )}(z(\sigma )) + {\buildrel \circ \over
 V}_{(\beta )})}\over {1 + V^{(o)}(z(\sigma ))}}.
 \label{22}
 \eea

\noindent In each tangent plane to a point of $\Sigma_{\tau}$ this
point-dependent standard Wigner boost sends the unit future-pointing
time-like vector ${\buildrel o\over V}^{(\alpha )} = (1; 0)$ into
the unit time-like vector $V^{(\alpha )} = {}^4E^{(\alpha )}_A\, l^A
= \Big(\sqrt{1 + \sum_a\, \varphi^2_{(a)}}; \varphi^{(a)} = - \sgn\,
\varphi_{(a)}\Big)$. As a consequence, the flat indices $(a)$ of the
adapted tetrads and cotetrads and of the triads and cotriads on
$\Sigma_{\tau}$ transform as Wigner spin-1 indices under
point-dependent SO(3) Wigner rotations $R_{(a)(b)}(V(z(\sigma
));\,\, \Lambda (z(\sigma ))\, )$ associated with Lorentz
transformations $\Lambda^{(\alpha )}{}_{(\beta )}(z)$ in the tangent
plane to the space-time in the given point of $\Sigma_{\tau}$.
Instead the index $(o)$ of the adapted tetrads and cotetrads is a
local Lorentz scalar index.

\bigskip

The adapted tetrads and cotetrads   have the expression

\bea
 \eo^A_{(o)} &=& {1\over {1 + n}}\, (1; - \sum_a\, n_{(a)}\,
 {}^3e^r_{(a)}) = l^A,\qquad \eo^A_{(a)} = (0; {}^3e^r_{(a)}), \nonumber \\
 &&{}\nonumber  \\
 \eo^{(o)}_A &=& (1 + n)\, (1; \vec 0) = \sgn\, l_A,\qquad \eo^{(a)}_A
= (n_{(a)}; {}^3e_{(a)r}),
 \label{23}
 \eea

\noindent where ${}^3e^r_{(a)}$ and ${}^3e_{(a)r}$ are triads and
cotriads on $\Sigma_{\tau}$ and $n_{(a)} = n_r\, {}^3e^r_{(a)} =
n^r\, {}^3e_{(a)r}$ \footnote{Since one uses the positive-definite
3-metric $\delta_{(a)(b)} $, one will use only lower flat spatial
indices. Therefore for the cotriads one uses the notation
${}^3e^{(a)}_r\,\, {\buildrel {def}\over =}\, {}^3e_{(a)r}$ with
$\delta_{(a)(b)} = {}^3e^r_{(a)}\, {}^3e_{(b)r}$.} are adapted shift
functions. In Eqs.(\ref{23}) $N(\tau, \vec \sigma) = 1 + n(\tau,
\vec \sigma) > 0$, with $n(\tau ,\vec \sigma)$ vanishing at spatial
infinity (absence of super-translations), so that $N(\tau, \vec
\sigma)\, d\tau$ is positive from $\Sigma_{\tau}$ to $\Sigma_{\tau +
d\tau}$, is the lapse function; $N^r(\tau, \vec \sigma) = n^r(\tau,
\vec \sigma)$, vanishing at spatial infinity (absence of
super-translations), are the shift functions.

\bigskip

The adapted tetrads $\eo^A_{(a)}$ are defined modulo SO(3) rotations
$\eo^A_{(a)} = \sum_b\, R_{(a)(b)}(\alpha_{(e)})\, {}^4{\buildrel
\circ \over {\bar E}}^A_{(b)}$, ${}^3e^r_{(a)} = \sum_b\,
R_{(a)(b)}(\alpha_{(e)})\, {}^3{\bar e}^r_{(b)}$, where
$\alpha_{(a)}(\tau ,\vec \sigma )$ are three point-dependent Euler
angles. After having chosen an arbitrary point-dependent origin
$\alpha_{(a)}(\tau ,\vec \sigma ) = 0$, one arrives at the following
adapted tetrads and cotetrads [${\bar n}_{(a)} = \sum_b\, n_{(b)}\,
R_{(b)(a)}(\alpha_{(e)})\,$, $\sum_a\, n_{(a)}\, {}^3e^r_{(a)} =
\sum_a\, {\bar n}_{(a)}\,
 {}^3{\bar e}^r_{(a)}$]

\bea
 {}^4{\buildrel \circ \over {\bar E}}^A_{(o)}
 &=& \eo^A_{(o)} = {1\over {1 + n}}\, (1; - \sum_a\, {\bar n}_{(a)}\,
 {}^3{\bar e}^r_{(a)}) = l^A,\qquad {}^4{\buildrel \circ \over
 {\bar E}}^A_{(a)} = (0; {}^3{\bar e}^r_{(a)}), \nonumber \\
 &&{}\nonumber  \\
 {}^4{\buildrel \circ \over {\bar E}}^{(o)}_A
 &=& \eo^{(o)}_A = (1 + n)\, (1; \vec 0) = \sgn\, l_A,\qquad
 {}^4{\buildrel \circ \over {\bar E}}^{(a)}_A
= ({\bar n}_{(a)}; {}^3{\bar e}_{(a)r}),
 \label{24}
 \eea

\noindent which one will use as a reference standard.\medskip

The expression for the general tetrad

\bea
 {}^4E^A_{(\alpha )} &=& \eo^A_{(\beta )}\, L^{(\beta )}{}_{(\alpha
 )}(\varphi_{(a)}) = {}^4{\buildrel \circ \over {\bar E}}^A_{(o)}\,
 L^{(o)}{}_{(\alpha )}(\varphi_{(c)}) +\nonumber \\
 &+& \sum_{ab}\, {}^4{\buildrel \circ \over
 {\bar E}}^A_{(b)}\, R^T_{(b)(a)}(\alpha_{(c)})\,
 L^{(a)}{}_{(\alpha )}(\varphi_{(c)}),
 \label{25}
 \eea
\medskip

\noindent shows that every point-dependent Lorentz transformation
 $\Lambda$ in the tangent planes may be parametrized with the
 (Wigner) boost parameters $\varphi_{(a)}$ and the Euler angles
 $\alpha_{(a)}$, being the product $\Lambda = R\, L$ of a rotation
 and a boost.

\bigskip

The future-oriented unit normal to $\Sigma_{\tau}$ and the projector
on $\Sigma_{\tau}$ are $l_A = \sgn\, (1 + n)\, \Big(1;\, 0\Big)$,
${}^4g^{AB}\, l_A\, l_B = \sgn $, $l^A = \sgn\, (1 + n)\,
{}^4g^{A\tau} = {1\over {1 + n}}\, \Big(1;\, - n^r\Big) = {1\over {1
+ n}}\, \Big(1;\, - \sum_a\, {\bar n}_{(a)}\, {}^3{\bar
e}_{(a)}^r\Big)$, ${}^3h^B_A = \delta^B_A - \sgn\, l_A\, l^B$.

\bigskip

The 4-metric has the following expression

 \bea
 {}^4g_{\tau\tau} &=& \sgn\, [(1 + n)^2 - {}^3g^{rs}\, n_r\,
 n_s] = \sgn\, [(1 + n)^2 - \sum_a\, {\bar n}^2_{(a)}],\nonumber \\
 {}^4g_{\tau r} &=& - \sgn\, n_r = -\sgn\, \sum_a\, {\bar n}_{(a)}\,
 {}^3{\bar e}_{(a)r},\nonumber \\
  {}^4g_{rs} &=& -\sgn\, {}^3g_{rs} = - \sgn\, \sum_a\, {}^3e_{(a)r}\, {}^3e_{(a)s}
  = - \sgn\, \sum_a\, {}^3{\bar e}_{(a)r}\, {}^3{\bar e}_{(a)s},\nonumber \\
 &&{}\nonumber \\
 {}^4g^{\tau\tau} &=& {{\sgn}\over {(1 + n)^2}},\qquad
  {}^4g^{\tau r} = -\sgn\, {{n^r}\over {(1 + n)^2}} = -\sgn\, {{\sum_a\, {}^3{\bar e}^r_{(a)}\,
 {\bar n}_{(a)}}\over {(1 + n)^2}},\nonumber \\
 {}^4g^{rs} &=& -\sgn\, ({}^3g^{rs} - {{n^r\, n^s}\over
 {(1 + n)^2}}) = -\sgn\, \sum_{ab}\, {}^3{\bar e}^r_{(a)}\, {}^3{\bar e}^s_{(b)}\, (\delta_{(a)(b)} -
 {{{\bar n}_{(a)}\, {\bar n}_{(b)}}\over {(1 + n)^2}}),\nonumber \\
 &&{}\nonumber \\
 &&\sqrt{- g } = \sqrt{|{}^4g|} = {{\sqrt{{}^3g}}\over {\sqrt{\sgn\,
 {}^4g^{\tau\tau}}}} = \sqrt{\gamma}\, (1 + n) = {}^3e\, (1 +
 n),\nonumber \\
 && {}^3g = \gamma = ({}^3e)^2,\quad {}^3e = det\, {}^3e_{(a)r}.
 \label{26}
 \eea

\bigskip

The 3-metric ${}^3g_{rs}$ has signature $(+++)$, so that one may put
all the flat 3-indices {\it down}. One has ${}^3g^{ru}\, {}^3g_{us}
= \delta^r_s$.

\subsection{The ADM Phase Space and the ADM Hamilton Equations}

The given parametrization of the cotetrad fields leads to rewrite
the action of ADM tetrad gravity in terms of the following 16 fields
as configuration variables: three boost parameters
$\varphi_{(a)}(\tau, \sigma^u)$; the lapse $N(\tau, \sigma^u) = 1 +
n(\tau, \sigma^u)$ and shift $n_{(a)}(\tau, \sigma^u)$ functions;
the nine components of cotriad fields ${}^3e_{(a)r}(\tau, \sigma^u)$
on the 3-spaces $\Sigma_{\tau}$. As shown in the second and third
paper of Ref.\cite{17}, in Ref.\cite{36} and in the first paper of
Ref.\cite{40}, the ADM action for the gravitational field has the
expression

\bea
 S_{grav} &=& {{c^3}\over {16\pi\, G}}\, \int d\tau d^3\sigma \,
 \Big[ (1 + n)\, {}^3e\, \epsilon_{(a)(b)(c)}\, {}^3e^r_{(a)}\,
 {}^3e^s_{(b)}\, {}^3\Omega_{rs(c)}+\nonumber \\
 &+&{{{}^3e}\over
 {2 (1 + n)}} ({}^3G_o^{-1})_{(a)(b)(c)(d)} {}^3e^r_{(b)}(n_{(a) |
 r}- \partial_{\tau}\, {}^3e_{(a)r})\nonumber \\
 && {}^3e^s_{(d)}(n_{(c) |
 s}-\partial_{\tau} \, {}^3e_{(c) \ s})\Big]  (\tau , \sigma^u ).
 \label{27}
 \eea

\noindent In it ${}^3\Omega_{rs(a)} = \partial_r\, {}^3\omega_{s(a)}
-
\partial_s\, {}^3\omega_{r(a)} - \epsilon_{(a)(b)(c)}\,
{}^3\omega_{r(b)}\, {}^3\omega_{s(c)}$ is the field strength
associated with the 3-spin connection ${}^3\omega_{r(a)} = {1\over
2}\, \epsilon_{(a)(b)(c)}\, \Big[{}^3e^u_{(b)}\, (\partial_r\,
{}^3e_{(c)u} - \partial_u\, {}^3e_{(c)r}) + {1\over 2}\,
{}^3e^u_{(b)}\, {}^3e^v_{(c)}\, {}^3e_{(d)r}\, (\partial_v\,
{}^3e_{(d)u} - \partial_u\, {}^3e_{(d)v})\Big]$ and
$({}^3G_o^{-1})_{(a)(b)(c)(d)}=\delta_{(a)(c)}\delta_{(b)(d)}+
\delta_{(a)(d)}\delta_{(b)(c)}-2\delta_{(a)(b)}\delta_{(c)(d)}$ is
the flat (with lower indices) inverse  of the flat Wheeler-DeWitt
super-metric ${}^3G_{o(a)(b)(c)(d)}\, = \delta_{(a)(c)}\,
\delta_{(b)(d)} + \delta_{(a)(d)}\, \delta_{(b)(c)} -
\delta_{(a)(b)}\, \delta_{(c)(d)}$, ${}^3G_{o(a)(b)(e)(f)}\,
{}^3G^{-1}_{o(e)(f)(c)(d)} = 2\, (\delta_{(a)(c)}\, \delta_{(b)(d)}
+ \delta_{(a)(d)}\, \delta_{(b)(c)})$.

\bigskip

The canonical momenta $\pi_{\varphi_{(a)}}(\tau, \sigma^u)$,
$\pi_n(\tau, \sigma^u)$, $\pi_{n_{(a)}}(\tau, \sigma^u)$,
${}^3\pi^r_{(a)}(\tau, \sigma^u)$, conjugate to the configuration
variables satisfy 14 first-class constraints: the ten primary
constraints (the last   three constraints generate rotations on
quantities with  flat indices $(a)$ like the cotriads)

\bea
 &&\pi_{\varphi_{(a)}}(\tau , \sigma^u )\,
 \approx\, 0,\qquad \pi_n(\tau , \sigma^u )\, \approx\, 0,\qquad
 \pi_{n_{(a)}}(\tau , \sigma^u
 )\, \approx\, 0,\nonumber \\
 &&{}^3M_{(a)}(\tau ,  \sigma^u )
 = \epsilon_{(a)(b)(c)}\, {}^3e_{(b)r} (\tau , \sigma^u )\,
 {}^3\pi^r_{(c)}(\tau , \sigma^u )\, \approx\, 0,
 \label{28}
 \eea

\noindent and the secondary super-Hamiltonian and super-momentum
constraints

\bea
  {\cal H}(\tau , \sigma^u )&=&   \Big[{{c^3}\over {16\pi\,
 G}}\,\,\, {}^3e\,\, \epsilon_{(a)(b)(c)} \, {}^3e^r_{(a)}\,
 {}^3e^s_{(b)}\,{}^3\Omega_{rs(c)} -\nonumber \\
 &-&{{2\pi\, G}\over {c^3\, {}^3e}}\,\,\,
 {}^3G_{o(a)(b)(c)(d)}\, {}^3e_{(a)r}\, {}^3\pi^r_{(b)}\,
 {}^3e_{(c)s}\, {}^3\pi^s_{(d)}\Big] (\tau , \sigma^u
 ) +  {\cal M}(\tau , \sigma^u )\approx 0,\nonumber \\
 &&{}\nonumber \\
 {\cal H}_{(a)}(\tau , \sigma^u )&=&\Big[\partial_r\,
 {}^3\pi^r_{(a)} -\epsilon_{(a)(b)(c)}\, {}^3\omega_{r(b)}\, {}^3
 \pi^r_{(c)} + {}^3e^r_{(a)} {\cal M}_r\Big](\tau , \sigma^u
 )\approx 0.
 \label{29}
 \eea

The functions ${\cal M}(\tau, \sigma^u)$ and ${\cal M}_r(\tau,
\sigma^u)$ describe the matter present in the space-time: ${\cal
M}(\tau, \sigma^u)$ is the (matter- and metric-dependent) internal
mass density, while ${\cal M}_r(\tau, \sigma^u)$ is the universal
(metric-independent) internal momentum density. If the action of
matter is added to Eq.(\ref{27}), one can evaluate the
energy-momentum tensor $T^{AB}(\tau, \sigma^u) = - \Big[ {2\over
{\sqrt{-{}^4g}}}\, {{\delta S_{matter}}\over {\delta\,
{}^4g_{AB}}}\Big](\tau ,\vec \sigma )$ of the matter \footnote{The
Hamilton equations imply ${}^4\nabla_A\, T^{AB} \equiv 0$ in accord
with Einstein's equations and the Bianchi identity.}  and determine
these functions from the following parametrization

 \bea
 T^{\tau\tau}(\tau ,\sigma^u ) &=& {{{\cal M}(\tau ,\sigma^u )}\over
 {[{}^3e\, (1 + n)^2](\tau , \sigma^u )}},\nonumber \\
 T^{\tau r}(\tau , \sigma^u ) &=& {{{}^3e^r_{(a)}\, \Big[
 (1 + n)\, {}^3e^s_{(a)}\, {\cal M}_s - n_{(a)}\,
 {\cal M}\Big]}\over {{}^3e\, (1 + n)^2}}(\tau , \sigma^u ).
 \label{30}
 \eea

\bigskip

The extrinsic curvature tensor of the 3-spaces $\Sigma_{\tau}$ as
3-manifolds embedded into the space-time has the following
expression in terms of the barred cotriads of Eq.(\ref{24}) and
their conjugate barred momenta

\bea
  {}^3K_{rs} &=& - {{4\pi\, G}\over {c^3\,\, {}^3{\bar e}}}\,
  \sum_{abu}\, \Big[\Big({}^3{\bar
 e}_{(a)r}\, {}^3{\bar e}_{(b)s} + {}^3{\bar e}_{(a)s}\, {}^3{\bar
 e}_{(b)r}\Big)\, {}^3{\bar e}_{(a)u}\, {\bar \pi}^u_{(b)}
 -\nonumber \\
 &-& {}^3{\bar e}_{(a)r}\, {}^3{\bar e}_{(a)s}\, {}^3{\bar e}_{(b)u}\,
 {\bar \pi}^u_{(b)}\Big].
 \label{31}
 \eea

\bigskip

Therefore the basis of canonical variables for this formulation of
tetrad gravity, naturally adapted to 7 of the 14 first-class
constraints, is

\beq
 \begin{minipage}[t]{3cm}
\begin{tabular}{|l|l|l|l|} \hline
$\varphi_{(a)}$ & $n$ & $n_{(a)}$ & ${}^3e_{(a)r}$ \\ \hline $
\pi_{\varphi_{(a)}}\, \approx 0$ & $\pi_n\, \approx 0$ &
$\pi_{n_{(a)}}\, \approx 0 $ & ${}^3{ \pi}^r_{(a)}$
\\ \hline
\end{tabular}
\end{minipage}
 \label{32}
 \eeq

The behavior of these fields at spatial infinity (compatible with
the absence of super-translations) is given in Eqs.(5.5) of the
third paper in Refs.\cite{17}; in particular for the cotriads one
has ${}^3e_{(a)r}(\tau, \sigma^r) \rightarrow_{r\, \rightarrow\,
\infty}\, \Big(1 + {{const.}\over {2 r}}\Big)\, \delta_{ar} + O(r^{-
3/2})$ ($r = \sqrt{\sum_r\, (\sigma^r)^2}$).

\bigskip

From the action (\ref{29}), after having added the matter action,
one can obtain the standard non-Hamiltonian ADM equations ($|r$
denotes the 3-covariant derivative in the 3-space $\Sigma_{\tau}$
with 3-metric ${}^3g_{rs}$; ${}^3R_{rs}$ is the 3-Ricci tensor of
$\Sigma_{\tau}$)

\bea
  \partial_{\tau}\, {}^3g_{rs} &\cir& n_{r|s} + n_{s|r} - 2\,
 (1 + n)\, {}^3K_{rs},\nonumber \\
 &&{}\nonumber \\
 \partial_{\tau}\, {}^3K_{rs} &\cir& (1 + n)\, \Big({}^3R_{rs} + {}^3K\, {}^3K_{rs} -
 2\, {}^3K_{ru}\, {}^3K^u{}_s\Big) -\nonumber \\
 &-& n_{|s|r} + n^u{}_{|s}\, {}^3K_{ur} + n^u{}_{|r}\, {}^3K_{us} +
 n^u\, {}^3K_{rs|u},
 \label{33}
 \eea

\noindent with the quantities appearing in these equations
re-expressed in terms of the configurational variables of
Eq.(\ref{32}).

\bigskip

Instead at the Hamiltonian level one can get the Hamilton equations
for all the variables of the canonical basis (\ref{32}), as shown in
the first paper of Ref.\cite{40}, by using the Dirac Hamiltonian. As
shown in Refs.\cite{17}, the Dirac Hamiltonian has the form (if the
matter contains the electro-magnetic field there are extra terms
with the electro-magnetic first-class constraints)

\bea
  H_D&=& {1\over c}\, {\hat E}_{ADM} + \int d^3\sigma\, \Big[ n\,
 {\cal H} - n_{(a)}\, {\cal H}_{(a)} \Big](\tau , \sigma^u )
 +\nonumber \\
 &+&\int d^3\sigma\, \Big[\lambda_n\, \pi_n + \lambda_{
 n_{(a)}}\, \pi_{n_{(a)}} + \lambda_{\varphi_{(a)}}\,  \pi_{
 \varphi_{(a)}} + \mu_{(a)}\, {}^3M_{(a)}\Big](\tau , \sigma^u ),
 \label{34}
 \eea

\noindent where ${\hat E}_{ADM}$ is the weak ADM energy and the
$\lambda$'s are arbitrary Dirac multipliers.

\bigskip

See Eqs. (2.22), (3.43)and (3.47) of the first paper of
Ref.\cite{40} for the expression of the ten weak asymptotic ADM
Poincar\'e generators ${\hat E}_{ADM}$, ${\hat P}^r_{ADM}$, ${\hat
J}^r_{adm}$, ${\hat {\cal K}}^r_{ADM}$. Since one is in a {\it
non-inertial rest frame} (due to the absence of super-translations),
one has the rest-frame conditions ${\hat P}^r_{ADM} \approx 0$ like
in SR. Then one has to add the conditions ${\hat {\cal K}}^r_{ADM}
\approx 0$ to eliminate the internal 3-center of mass of the
3-universe like in SR \cite{21}. Therefore the 3-universe can be
seen as a decoupled external canonical non-covariant center of mass
carrying a pole-dipole structure: the invariant mass $M c = {1\over
c}\, {\hat E}_{ADM}$ and the rest spin ${\hat J}^{rs}_{ADM}$. This
view is in accord with an old suggestions of Dirac \cite{8}.

\bigskip

In Ref.\cite{40} the study of ADM canonical tetrad gravity was done
with the following type of matter: N charged scalar particles
(described by the canonical variables $\eta^r_i(\tau)$,
$\kappa_{ir}(\tau)$) and the electro-magnetic field in the
non-covariant radiation gauge (described by the canonical variables
$A^r_{\perp}(\tau, \sigma^u)$, $\pi^r_{\perp}(\tau, \sigma^u)$ as
shown in Ref.\cite{21}). The particles (described by an action like
the one in Eq.(\ref{6})) have not only Grassmann-valued electric
charges $Q_i$ ($Q^2_i = 0$, $Q_i\, Q_j = Q_j\, Q_i$ for $i\not= j$)
to regularize the electro-magnetic self-energies, but also
Grassmann-valued signs of the energy ($\eta^2_i = 0$, $\eta_i\,
\eta_j = \eta_j\, \eta_i$ for $i\not= j$) to regularize the
gravitational self-energies \footnote{Both quantities are
two-valued. The elementary electric charges are $Q = \pm e$, with
$e$ the electron charge. Analogously the sign of the energy of a
particle is a topological two-valued number (the two branches of the
mass-shell hyperboloid). The formal quantization of these Grassmann
variables gives two-level fermionic oscillators. At the classical
level the self-energies make the classical equations of motion
ill-defined on the world-lines of the particles. The ultraviolet and
infrared Grassmann regularization allows to cure this problem and to
get consistent solution of regularized equations of motion. See
Refs.\cite{28} for the electro-magnetic case.}.

Instead in Ref.\cite{41} the matter is a perfect fluid described by
the action of Ref.\cite{43} re-expressed in the 3+1 point of view in
Refs.\cite{42}.

\medskip

In the case of N particles the functions ${\cal M}$ and ${\cal M}_r$
have the expression (see Ref.\cite{40} for their form in presence of
the electro-magnetic field)

\bea
  {\cal M}(\tau , \sigma^u )&=& \sum_{i=1}^N\, \delta^3(
 \sigma^u , \eta^u_i(\tau ))\,
 \eta_i\, \sqrt{m_i^2\, c^2 + {}^3e^r_{(a)}(\tau , \sigma^u)\,
  \kappa_{ir}(\tau )\, {}^3e^s_{(a)}(\tau , \sigma^u)\,
  \kappa_{is}(\tau ) },\nonumber \\
 {\cal M}_r(\tau , \sigma^u )&=& \sum_{i=1}^N\, \eta_i\,
 \kappa_{ir}(\tau ),
  \label{35}
  \eea

\noindent while in the case of dust \cite{41}, described by
canonical coordinates $\alpha^i(\tau, \sigma^u)$, $\Pi_i(\tau,
\sigma^u)$, $i=1,2,3$, they have the expression

\bea
  {\cal M}(\tau, \sigma^u) &=& \sqrt{\mu^2\, [det\,
  (\partial_s\, \alpha^j)]^2 + {\tilde \phi}^{-2/3}\,
 \sum_{arsij}\, Q_a^{-2}\, V_{ra}\, V_{sa}\, \partial_r\, \alpha^i\,
 \partial_s\, \alpha^j\, \Pi_i\, \Pi_j}(\tau, \sigma^u),\nonumber \\
 {\cal M}_r(\tau, \sigma^u) &=&  \sum_i\,  \partial_r\, \alpha^i(\tau,
 \sigma^u)\, \Pi_i(\tau, \sigma^u).
 \label{36}
 \eea

\vfill\eject

 \section{The York Canonical Basis and the
 Inertial and Tidal Degrees of Freedom of the Gravitational Field}
  \label{sec:6}

The presence of 14 first-class constraints in the phase space having
the 32 fields of Eq.(\ref{32}) as a canonical basis implies that
there are 14 gauge variables describing {\it inertial effects} and 2
canonical pairs of physical degrees of freedom describing the {\it
tidal effects} of the gravitational field (namely gravitational
waves in the weak field limit). To disentangle the inertial effects
from the tidal ones one needs a canonical transformation to a new
canonical basis adapted to all the ten primary constraints
(\ref{28}) and containing the barred variables defined in
Eq.(\ref{24}). This is the topic of this Section.

\subsection{The York Canonical Basis}

A canonical transformation adapted to the ten primary constraints
(\ref{28}) was found in Ref.\cite{36}. It implements the York map of
Ref.\cite{37} in the cases in which the 3-metric ${}^3g_{rs}$ has
three distinct eigenvalues and diagonalizes the York-Lichnerowicz
approach (see Ref.\cite{38} for a review).
\medskip

As said before Eq.(\ref{24}), one can decompose the cotriads on
$\Sigma_{\tau}$ in the product of a rotation matrix, belonging to
the subgroup SO(3) of the tetrad gauge group and depending on three
Euler angles $\alpha_{(a)}(\tau, \sigma^r)$, and of barred cotriads
depending only on six independent fields. The canonical
transformation Abelianizes the constraints ${}^3M_{(a)}(\tau,
\sigma^u) \approx 0$ of Eqs.(\ref{28}), satisfying $\{
{}^3M_{(a)}(\tau , \sigma^u ),{}^3M_{(b)} (\tau , \sigma^{' u})\}\,
=\, \epsilon_{(a)(b)(c)}\, {}^3M_{(c)} (\tau , \sigma^u )
\delta^3(\sigma^u ,\sigma^{' u})$, and replaces them with the
vanishing of the three momenta $\pi^{(\alpha)}_{(a)}(\tau, \sigma^r)
\approx 0$ conjugate to the Euler angles.

\medskip

The new canonical basis, named York canonical basis, is ($a=1,2,3$;
$\bar a =1,2$)

\bea
 &&\begin{minipage}[t]{10 cm}
 \begin{tabular}{|ll|ll|l|l|l|} \hline
 $\varphi_{(a)}$ & $\alpha_{(a)}$ & $n$ & ${\bar n}_{(a)}$ &
 $\theta^r$ & $\tilde \phi$ & $R_{\bar a}$\\ \hline
 $\pi_{\varphi_{(a)}} \approx0$ &
 $\pi^{(\alpha)}_{(a)} \approx 0$ & $\pi_n \approx 0$ & $\pi_{{\bar n}_{(a)}} \approx 0$
 & $\pi^{(\theta )}_r$ & $\pi_{\tilde \phi} = {{c^3}\over {12\pi G}}\, {}^3K$ & $\Pi_{\bar a}$ \\
 \hline
 \end{tabular}
 \end{minipage}.\nonumber \\
 &&{}
 \label{37}
 \eea

In it the cotriads and the components of the 4-metric have the
following expression

\bea
 &&{}^3e_{(a)r} = \sum_b\, R_{(a)(b)}(\alpha_{(c)})\, {}^3{\bar e}_{(b)r} =
 \sum_b\, R_{(a)(b)}(\alpha_{(c)})\, V_{rb}(\theta^i)\,
 {\tilde \phi}^{1/3}\, e^{\sum_{\bar a}^{1,2}\, \gamma_{\bar aa}\, R_{\bar a}},\nonumber \\
 &&{}\nonumber \\
 &&{}^4g_{\tau\tau} = \sgn\, [(1 + n)^2 - \sum_a\, {\bar n}^2_{(a)}],
 \nonumber \\
 && {}^4g_{\tau r} = - \sgn\, \sum_a\,  n_{(a)}\, {}^3e_{(a)r}
 = - \sgn\, \sum_a\, {\bar n}_{(a)}\, {}^3{\bar e}_{(a)r},\qquad
 \tilde \phi = \phi^6 = \sqrt{det\, {}^3g_{rs}},\nonumber \\
 &&{}^4g_{rs} = - \sgn\, {}^3g_{rs} = - \sgn\, {\tilde \phi}^{2/3}\,
 \sum_a\, V_{ra}(\theta^i)\, V_{sa}(\theta^i)\, Q^2_a,\qquad
 Q_a = e^{ \sum_{\bar a}^{1,2}\, \gamma_{\bar aa}\, R_{\bar
 a}},
 \label{38}
 \eea

The set of numerical parameters $\gamma_{\bar aa}$ appearing in
$Q_a$ satisfies \cite{17} $\sum_u\, \gamma_{\bar au} = 0$, $\sum_u\,
\gamma_{\bar a u}\, \gamma_{\bar b u} = \delta_{\bar a\bar b}$,
$\sum_{\bar a}\, \gamma_{\bar au}\, \gamma_{\bar av} = \delta_{uv} -
{1\over 3}$. Each solution of these equations defines a different
York canonical basis.

\medskip

This canonical basis has been found  due to the fact that the
3-metric $ {}^3g_{rs}$ is a real symmetric $3 \times 3$ matrix,
which may be diagonalized with an {\it orthogonal} matrix
$V(\theta^r)$, $V^{-1} = V^T$ ($\sum_u\, V_{ua}\, V_{ub} =
\delta_{ab}$, $\sum_a\, V_{ua}\, V_{va} = \delta_{uv}$, $\sum_{uv}\,
\epsilon_{wuv}\, V_{ua}\, V_{vb} = \sum_c\, \epsilon_{abc}\,
V_{cw}$), $det\, V = 1$, depending on three parameters $\theta^i$
($i=1,2,3$) \footnote{Due to the positive signature of the 3-metric,
one defines the matrix $V$ with the following indices: $V_{ru}$.
Since the choice of Shanmugadhasan canonical bases breaks manifest
covariance, one will use the notation $V_{ua} = \sum_v\, V_{uv}\,
\delta_{v(a)}$ instead of $V_{u(a)}$.}, whose conjugate momenta
$\Pi_i^{(\theta)}$ are to be determined as solutions of the
super-momentum constraints. If one chooses these three gauge
parameters to be Euler angles ${\hat \theta}^i(\tau, \vec \sigma)$,
one gets a description of the 3-coordinate systems on
$\Sigma_{\tau}$ from a local point of view, because they give the
orientation of the tangents to the three 3-coordinate lines through
each point. However, for the calculations (see Refs.\cite{40}) it is
more convenient to choose the three gauge parameters as first kind
coordinates $\theta^i(\tau, \vec \sigma)$ ($- \infty < \theta^i < +
\infty$) on the O(3) group manifold, so that by definition one has
$V_{ru}(\theta^i) = \Big(e^{- \sum_i\, {\hat T}_i\,
\theta^i}\Big)_{ru}$, where $({\hat T}_i)_{ru} = \epsilon_{rui}$ are
the generators of the o(3) Lie algebra in the adjoint
representation, and the Euler angles may be expressed as ${\hat
\theta}^i = f^i(\theta^n)$. The Cartan matrix has the form
$A(\theta^n) = {{1 - e^{- \sum_i\, {\hat T}_i\, \theta^i} }\over
{\sum_i\, {\hat T}_i\, \theta^i}}$ and satisfies $A_{ri}(\theta^n)\,
\theta^i = \delta_{ri}\, \theta^i$; $B(\theta^i) =
A^{-1}(\theta^i)$.\medskip

From now on for the sake of notational simplicity the symbol $V$
will mean $V(\theta^i)$.

\bigskip

The extrinsic curvature tensor of the 3-space $\Sigma_{\tau}$ has
the expression

\bea
  {}^3K_{rs}(\tau, \sigma^u) &=&  - {{4\pi\, G}\over {c^3}}\, \Big[ {\tilde \phi}^{-1/3}\,
 \Big(\sum_a\, Q^2_a\, V_{ra}\, V_{sa}\, [2\, \sum_{\bar b}\, \gamma_{\bar ba}\,
 \Pi_{\bar b} -  \tilde \phi\, \pi_{\tilde \phi}] +\nonumber \\
 &+& \sum_{ab}\, Q_a\, Q_b\, (V_{ra}\, V_{sb} +
 V_{rb}\, V_{sa})\, \sum_{twi}\, {{\epsilon_{abt}\,
 V_{wt}\, B_{iw}\, \pi_i^{(\theta )}}\over {
 Q_b\, Q^{-1}_a  - Q_a\, Q^{-1}_b}} \Big) \Big](\tau, \sigma^u).
 \nonumber \\
 &&{}
 \label{39}
 \eea

\medskip

This canonical transformation realizes a {\it York map} because the
gauge variable $\pi_{\tilde \phi}$ (describing the freedom in the
choice of the trace of the extrinsic curvature of the instantaneous
3-spaces $\Sigma_{\tau}$) is proportional to {\it York internal
extrinsic time} ${}^3K$. It is the only gauge variable among the
momenta: this is a reflex of the Lorentz signature of space-time,
because $\pi_{\tilde \phi}$ and $\theta^n$ can be used as a set of
4-coordinates for the space-time \cite{35}. The York time describes
the effect of gauge transformations producing a deformation of the
shape of the 3-space along the 4-normal to the 3-space as a
3-sub-manifold of space-time.\medskip

Its conjugate variable, to be determined by the super-Hamiltonian
constraint (interpreted as the Lichnerowicz equation), is $\tilde
\phi = \phi^6 = {}^3\bar e = \sqrt{det\, {}^3g_{rs}}$, which is
proportional to {\it Misner's internal intrinsic time}; moreover
$\tilde \phi$ is the {\it 3-volume density} on $\Sigma_{\tau}$: $V_R
= \int_R d^3\sigma\, \tilde \phi$, $R \subset \Sigma_{\tau}$. Since
one has ${}^3g_{rs} = {\tilde \phi}^{2/3}\, {}^3{\hat g}_{rs}$ with
$det\, {}^3{\hat g}_{rs} = 1$, $\tilde \phi$ is also called the {\it
conformal factor} of the 3-metric.

\medskip

The two pairs of canonical variables $R_{\bar a}$, $\Pi_{\bar a}$,
$\bar a = 1,2$, describe the generalized {\it tidal effects}, namely
the independent physical degrees of freedom of the gravitational
field. They are 3-scalars on $\Sigma_{\tau}$ and the configuration
tidal variables $R_{\bar a}$ parametrize {\it the two eigenvalues of
the 3-metric ${}^3{\hat g}_{rs}$ with unit determinant}. They are
Dirac observables {\it only} with respect to the gauge
transformations generated by 10 of the 14 first class constraints.
Let us remark that, if one fixes completely the gauge and one goes
to Dirac brackets, then the only surviving dynamical variables
$R_{\bar a}$ and $\Pi_{\bar a}$ become two pairs of {\it non
canonical} Dirac observables for that gauge: the two pairs of
canonical Dirac observables have to be found as a Darboux basis of
the copy of the reduced phase space identified by the gauge and they
will be (in general non-local) functionals of the $R_{\bar a}$,
$\Pi_{\bar a}$ variables. \bigskip

Therefore, the 14 arbitrary gauge variables are $\varphi_{(a)}(\tau,
\sigma^u)$, $\alpha_{(a)}(\tau, \sigma^u)$, $n(\tau, \sigma^u)$,
${\bar n}_{(a)}(\tau, \sigma^u)$, $\theta^i(\tau, \sigma^u)$,
$\pi_{\tilde \phi}(\tau, \sigma^u)$: they describe the following
generalized {\it inertial effects} \cite{36}:\medskip

a) $\alpha_{(a)}(\tau , \sigma^u )$ and $\varphi_{(a)}(\tau ,
\sigma^u )$ are the 6 configuration variables parametrizing the
O(3,1) gauge freedom in the choice of the tetrads in the tangent
plane to each point of $\Sigma_{\tau}$ and describe the
arbitrariness in the choice of a tetrad to be associated to a
time-like observer, whose world-line goes through the point $(\tau
,\vec \sigma )$. They fix {\it the unit 4-velocity of the observer
and the conventions for the orientation of three gyroscopes and
their transport along the world-line of the observer}. The  {\it
Schwinger time gauges} are defined by the gauge fixings
$\alpha_{(a)}(\tau,  \sigma^u) \approx 0$, $\varphi_{(a)}(\tau,
\sigma^u) \approx 0$.
\medskip

b) $\theta^i(\tau , \sigma^u )$ (depending only on the 3-metric)
describe the arbitrariness in the choice of the 3-coordinates in the
instantaneous 3-spaces $\Sigma_{\tau}$ of the chosen non-inertial
frame  centered on an arbitrary time-like observer. Their choice
will induce a pattern of {\it relativistic inertial forces} for the
gravitational field, whose potentials are the functions
$V_{ra}(\theta^i)$ present in the weak ADM energy ${\hat E}_{ADM}$.
\medskip

c) ${\bar n}_{(a)}(\tau , \sigma^u )$, the shift functions, describe
which points on different instantaneous 3-spaces have the same
numerical value of the 3-coordinates. They are the inertial
potentials describing the effects of the non-vanishing off-diagonal
components ${}^4g_{\tau r}(\tau , \sigma^u )$ of the 4-metric,
namely they are the {\it gravito-magnetic potentials} \footnote{In
the post-Newtonian approximation in harmonic gauges they are the
counterpart of the electro-magnetic vector potentials describing
magnetic fields \cite{38}: A) $N = 1 + n$, $n\, {\buildrel
{def}\over =}\, - {{4\, \sgn}\over {c^2}}\, \Phi_G$ with $\Phi_G$
the {\it gravito-electric potential}; B) $n_r\, {\buildrel
{def}\over =}\, {{2\, \sgn}\over {c^2}}\, A_{G\, r}$ with $A_{G\,
r}$ the {\it gravito-magnetic} potential; C) $E_{G\, r} =
\partial_r\, \Phi_G - \partial_{\tau}\, ({1\over 2}\, A_{G\, r})$ (the {\it
gravito-electric field}) and $B_{G\, r} = \epsilon_{ruv}\,
\partial_u\, A_{G\, v} = c\, \Omega_{G\, r}$ (the {\it
gravito-magnetic field}). Let us remark that in arbitrary gauges the
analogy with electro-magnetism  breaks down.} responsible of effects
like the dragging of inertial frames (Lens-Thirring effect) in the
post-Newtonian approximation. The shift functions are determined by
the $\tau$-preservation of the gauge fixings determining the gauge
variables $\theta^i(\tau,  \sigma^u)$.
\medskip

d) $\pi_{\tilde \phi}(\tau , \sigma^u )$, i.e. the York time
${}^3K(\tau , \sigma^u )$, describes the non-dynamical arbitrariness
in the choice of the convention for the synchronization of distant
clocks which remains in the transition from SR to GR. Since the York
time is present in the Dirac Hamiltonian, it is a {\it new inertial
potential} connected to the problem of the relativistic freedom in
the choice of the {\it shape of the instantaneous 3-space}, which
has no Newtonian analogue (in Galilei space-time time is absolute
and there is an absolute notion of Euclidean 3-space). Its effects
are completely unexplored. Instead the other components of the
extrinsic curvature of $\Sigma_{\tau}$ are dynamically determined
once a 3-coordinate system has been chosen in the 3-space.

\medskip

e) $1 + n(\tau , \sigma^u )$, the lapse function appearing in the
Dirac Hamiltonian, describes the arbitrariness in the choice of the
unit of proper time in each point of the simultaneity surfaces
$\Sigma_{\tau}$, namely how these surfaces are packed in the 3+1
splitting. The lapse function is determined by the
$\tau$-preservation of the gauge fixing for the gauge variable
${}^3K(\tau,  \sigma^u)$.

\bigskip

As shown in Ref.\cite{35}, the dynamical nature of space-time
implies that each solution (i.e. an Einstein 4-geometry) of
Einstein's equations (or of the associated ADM Hamilton equations)
dynamically selects a preferred 3+1 splitting of the space-time,
namely in GR the instantaneous 3-spaces  are dynamically determined
modulo only one inertial gauge function (the gauge freedom in clock
synchronization in GR). In the York canonical basis this function is
the {\it York time}, namely the trace of the extrinsic curvature of
the 3-space. Instead in SR the gauge freedom in clock
synchronization depends on four basic gauge functions, the
embeddings $z^{\mu}(\tau, \sigma^r)$, and both the 4-metric and the
whole extrinsic curvature tensor were derived inertial potentials.
Instead in GR the extrinsic curvature tensor of the 3-spaces is a
mixture of dynamical (tidal) pieces and inertial gauge variables
playing the role of inertial potentials.

\subsection{3-Orthogonal Schwinger Time Gauges and Hamilton Equations}

As shown in the first paper in Refs.\cite{40}, in the York canonical
basis the Dirac Hamiltonian (\ref{34}) becomes (the $\lambda$'s are
arbitrary Dirac multipliers; the Dirac multiplier $\lambda_r(\tau)$
implements the rest frame condition ${\hat P}^r_{ADM} \approx 0$)

\bea
  H_D&=& {1\over c}\, {\hat E}_{ADM} + \int d^3\sigma\, \Big[ n\,
{\cal H} - n_{(a)}\, {\cal H}_{(a)}\Big](\tau , \sigma^u )
+ \lambda_r(\tau )\, {\hat P}^r_{ADM} +\nonumber \\
 &+&\int d^3\sigma\, \Big[\lambda_n\, \pi_n + \lambda_{
{\bar n}_{(a)}}\, \pi_{{\bar n}_{(a)}} + \lambda_{\varphi_{(a)}}\,
\pi_{ \varphi_{(a)}} + \lambda_{\alpha_(a)}\,
\pi^{(\alpha)}_{(a)}\Big](\tau , \sigma^u ),
 \label{40}
 \eea

\noindent with the following expression for the weak ADM energy

\bea
  {\hat E}_{ADM} &=& c\, \int d^3\sigma\, \Big[{\check {\cal M}} -
  {{c^3}\over {16\pi\, G}}\, {\cal S} +
    {{2\pi\, G}\over {c^3}}\, {\tilde \phi}^{-1}\, \Big(
 - 3\, (\tilde \phi\, \pi_{\tilde \phi})^2 + 2\, \sum_{\bar b}\,
 \Pi^2_{\bar b} +\nonumber \\
 &+& 2\, \sum_{abtwiuvj}\, {{\epsilon_{abt}\, \epsilon_{abu}\, V_{wt}\,
 B_{iw}\, V_{vu}\, B_{jv}\, \pi_i^{(\theta )}\,
 \pi_j^{(\theta )}}\over {\Big[Q_a\, Q^{-1}_b - Q_b\,
 Q^{-1}_a \Big]^2}}\Big)\,\, \Big](\tau , \sigma^u ).
 \label{41}
 \eea

In it ${\cal S}(\tau, \sigma^u)$ is a function of $\tilde \phi$,
$\theta^i$ and $R_{\bar a}$ (given in Eq.(B8) of the first paper in
Ref.\cite{40}), which play the role of an inertial potential
depending on the choice of the 3-coordinates in the 3-space (it is
the $\Gamma-\Gamma$ term in the scalar 3-curvature of the 3-space).
\medskip

Eq.(\ref{41}) shows that the kinetic term, quadratic in the momenta,
is {\it not positive definite}. While the kinetic energy of the
tidal variables and the last term \footnote{It describes
gravito-magnetic effects.} are positive definite, there is the
negative kinetic terms (vanishing only in the gauges ${}^3K(\tau,
\sigma^u) = 0$) $- {{c^4}\over {24 \pi G}}\, \int D^3\sigma\, \tilde
\phi(\tau, \sigma^u)\, {}^3K^2(\tau, \sigma^u)$. It is an inertial
potential associated with the inertial gauge variable York time,
which is a momentum due to the Lorentz signature of space-time. It
was known that this quadratic form is not definite positive, but
only in the York canonical basis this can be made explicit.

\bigskip

In the York canonical basis it is possible to follow the procedure
for the fixation of a gauge natural from the point of view of
constraint theory when there are chains of first-class constraints
\cite{9}. This procedure implies that one has to add six gauge
fixings to the primary constraints without without secondaries
($\pi_{\varphi_{(a)}}(\tau, \sigma^u) \approx 0$,
$\pi_{\alpha_{(a)}}(\tau, \sigma^u) \approx 0$) and four gauge
fixings to the secondary super-Hamiltonian and super-momentum
constraints. These ten gauge fixings must be preserved in time,
namely their Poisson brackets with the Dirac Hamiltonian must
vanish. The $\tau$-preservation of the six gauge fixings determining
the gauge variables $\alpha_{(a)}(\tau, \sigma^u)$ and
$\varphi_{(a)}(\tau, \sigma^u)$ produces the equations determining
the six Dirac multipliers $\lambda_{\varphi_{(a)}}(\tau, \sigma^u)$,
$\lambda_{\alpha_{(a)}}(\tau, \sigma^u)$. The $\tau$-preservation of
the other four gauge fixings, determining the gauge variables
$\theta^i(\tau, \sigma^u)$ and the York time ${}^3K(\tau,
\sigma^u)$, produces four secondary gauge fixing constraints for the
determination of the lapse and shift functions. Then the
$\tau$-preservation of these secondary gauge fixings determines the
four Dirac multipliers $\lambda_n(\tau, \sigma^u)$, $\lambda_{{\bar
n}_{(a)}}(\tau, \sigma^u)$. Instead in numerical gravity one gives
independent gauge fixings for both the primary and secondary gauge
variables in such a way to minimize the computer time.

\bigskip

In Section V of the first paper in Refs.\cite{40} there is a review
of the gauges usually used in canonical gravity. It is shown that
the commonly used family of the harmonic gauges is not natural
according to the above procedure. The harmonic gauge fixings imply
hyperbolic PDE for the lapse and shift functions, to be added to the
hyperbolic PDE for the tidal variables. Therefore in harmonic gauges
both the tidal variables and the lapse and shift functions depend
(in a retarded way) from the {\it no-incoming radiation} condition
on the Cauchy surface in the past (so that the knowledge of ${}^3K$
from the initial time till today is needed).

\medskip

Instead the natural gauge fixings in the York canonical basis of ADM
tetrad gravity are the family of {\it Schwinger time gauges}, where
the O(3,1) gauge freedom of the tetrads is eliminated with the gauge
fixings (implying $\lambda_{\varphi_{(a)}}(\tau, \sigma^u) =
\lambda_{\alpha_{(a)}}(\tau, \sigma^u) = 0$)

\beq
 \alpha_{(a)}(\tau, \sigma^u) \approx 0,\qquad
 \varphi_{(a)}(\tau, \sigma^u) \approx 0,
 \label{42}
 \eeq

\noindent and the subfamily of the {\it 3-orthogonal gauges}

\beq
 \theta^i(\tau, \sigma^u) \approx 0,\qquad
 {}^3K(\tau, \sigma^u) \approx  F(\tau, \sigma^u) = numerical\, function,
 \label{43}
 \eeq

\noindent in which the 3-coordinates are chosen in such a way the
the 3-metric in the 3-spaces $\Sigma_{\tau}$ is diagonal. The
$\tau$-preservation of Eqs.(\ref{43}) gives four coupled elliptic
PDE for the lapse and shift functions. Therefore in these gauges
only the tidal variables (the gravitational waves after
linearization), and therefore only the eigenvalues of the 3-metric
with unit determinant inside $\Sigma_{\tau}$, depend (in a retarded
way) on the no-incoming radiation condition. The solutions $\tilde
\phi$ and $\pi_i^{(\theta)}$  of the constraints and the lapse $1 +
n$ and shift ${\bar n}_{(a)}$ functions depend only on the 3-space
$\Sigma_{\tau}$ with fixed $\tau$. If the matter consists of
positive energy particles (with a Grassmann regularization of the
gravitational self-energies) \cite{40} these solutions will contain
action-at-a-distance gravitational potentials (replacing the Newton
ones) and gravito-magnetic potentials.

\bigskip

In the family of 3-orthogonal gauges the weak ADM energy and the
super-Hamiltonian and super-momentum constraints (they are coupled
{\it elliptic} PDE for their unknowns) have the expression (see
Eq.(3.47) of the first paper in Ref.\cite{40} for the other weak ADM
Poincar\'e generators)

  \begin{eqnarray*}
 {\hat E}_{ADM}{|}_{\theta^i = 0} &=& c\, \int d^3\sigma\,
 \Big[{\cal M} {|}_{\theta^i = 0} -
  {{c^3}\over {16\pi\, G}}\, {\cal S}{|}_{\theta^i = 0} +
  \nonumber \\
  &+&  {{2\pi\, G}\over {c^3}}\, {\tilde \phi}^{-1}\, \Big(
 - 3\, (\tilde \phi\, \pi_{\tilde \phi})^2 + 2\, \sum_{\bar b}\,
 \Pi^2_{\bar b} +\nonumber \\
 &+& 2\, \sum_{abij}\, {{\epsilon_{abi}\, \epsilon_{abj}\,
 \pi_i^{(\theta )}\, \pi_j^{(\theta )}}\over {\Big[Q_a\, Q^{-1}_b - Q_b\,
 Q^{-1}_a \Big]^2}}\Big)\,\, \Big](\tau , \sigma^u ),
 \end{eqnarray*}

\begin{eqnarray*}
 {\cal H}(\tau , \sigma^u ){|}_{\theta^i = 0} &=&
  {{c^3}\over {16\pi\, G}}\, {\tilde \phi}^{1/6} (\tau , \sigma^u ) \,
 [ 8\,   {\hat \triangle}\, {\tilde \phi}^{1/6} -
  {}^3{\hat R}{|}_{\theta^i = 0}\, {\tilde \phi}^{1/6}](\tau , \sigma^u )
 + {\cal M}{|}_{\theta^i = 0}(\tau , \sigma^u ) +\nonumber \\
 &&{}\nonumber \\
  &+&  {{2\pi\, G}\over {c^3}}\, {\tilde \phi}^{-1}\, \Big[
- 3\, (\tilde \phi\, \pi_{\tilde \phi})^2 + 2\, \sum_{\bar b}\,
 \Pi^2_{\bar b} +\nonumber \\
 &+& 2\, \sum_{abij}\, {{\epsilon_{abi}\, \epsilon_{abj}\,
 \pi_i^{(\theta )}\, \pi_j^{(\theta )}}\over {\Big[Q_a\, Q^{-1}_b - Q_b\,
 Q^{-1}_a \Big]^2}}\Big](\tau , \sigma^u ), \nonumber \\
 &&{}\nonumber \\
   {\tilde {\bar {\cal H}}}_{(a)}{|}_{\theta^i = 0}(\tau , \sigma^u) &=&
 \phi^{-2}(\tau, \vec \sigma)\, \Big[\sum_{b \not= a}\,
 \sum_i\, {{ \epsilon_{abi}\, Q_b^{-1}}
 \over {Q_b\, Q_a^{-1} - Q_a\, Q_b^{-1}}}\, \partial_b\, \pi_i^{(\theta)}
 +\nonumber \\
 &+&2\, \sum_{b \not= a}\, \sum_i\,  {{ \epsilon_{abi}\, Q_a^{-1}}
 \over {\Big(Q_b\, Q_a^{-1} - Q_a\, Q_b^{-1}\Big)^2}}\,
  \sum_{\bar c}\, (\gamma_{\bar ca} - \gamma_{\bar cb})\, \partial_b\,
  R_{\bar c}\,\, \pi_i^{(\theta)} +\nonumber \\
 &&{}\nonumber \\
 &+&Q_a^{-1}\,  \Big(\phi^6\, \partial_a\, \pi_{\tilde \phi}
 +  \sum_{\bar b}\, (\gamma_{\bar ba}\, \partial_a\, \Pi_{\bar b}
 - \partial_a\, R_{\bar b}\, \Pi_{\bar b}) + {\cal M}_{a}\Big)
 \Big](\tau , \sigma^u),
 \end{eqnarray*}

\bea
 \hat \triangle &=& \sum_r\, Q_r^{-2}\, \Big[ \partial^2_r + 2\,
 \sum_{\bar a}\, \gamma_{\bar ar}\, \partial_r\, R_{\bar a}(\tau, \sigma^u)\,
 \partial_r\Big],\nonumber \\
  {\cal S}_{\theta^i=0}(\tau, \sigma^u) &=& \Big({\tilde
  \phi}^{1/3}\, \sum_a\, Q_a^{-2}\, \Big[{2\over 9}\, ({\tilde \phi}^{-1}\,
  \partial_a\, \tilde \phi)^2 +\nonumber \\
  &+& \sum_{\bar b}\, \Big(\sum_{\bar c}\, (2\, \gamma_{\bar ba}\,
  \gamma_{\bar ca} - \delta_{\bar b\bar c} )\, \partial_a\, R_{\bar
  c} - {2\over 3}\, \gamma_{\bar ba}\, {\tilde \phi}^{-1}\,
  \partial_a\, \tilde \phi  \Big)\, \partial_a\, R_{\bar b}\Big]
  \Big)(\tau, \sigma^u).\nonumber \\
  &&{}
 \label{44}
 \eea

\bigskip

In the first paper in Refs.\cite{40} there is the explicit form of
the Hamilton equations for all the canonical variables of the
gravitational field and of the matter replacing the standard 12 ADM
equations and the matter equations ${}^4\nabla_A\, T^{AB} = 0$ in
the Schwinger time gauges and their restriction to the 3-orthogonal
gauges. They could also be obtained from the effective Dirac
Hamiltonian of the 3-orthogonal gauges, which is evaluated by means
of a $\tau$-dependent canonical transformation sending the gauge
momentum $\pi_{\tilde \phi}(\tau, \sigma^u)$ in the gauge-fixing
conditions $\pi_{\tilde \phi}^{'}(\tau, \sigma^u) =  {{c^3}\over
{12\pi\, G}}\, \Big({}^3K(\tau, \sigma^u) - F(\tau, \sigma^u)\Big)
\approx 0$ and which is given  in Eq.(4.39) of the second paper of
Ref.\cite{40}.
\medskip

These equations are divided in five groups:\medskip

A) The {\it contracted Bianchi identities}, namely the evolution
equations for the solutions $\tilde \phi(\tau, \sigma^u)$ and
$\pi_i^{(\theta)}(\tau, \sigma^u)$ of the super-Hamiltonian and
super-momentum constraints: they are identities saying that, given a
solution of the constraints on a Cauchy surface, it remains a
solution also at later times.\hfill\break

B) The evolution equation for the four basic gauge variables
$\theta^i(\tau , \sigma^u)$ and ${}^3K(\tau , \sigma^u)$ (the
equation for the York time is the Raychaudhuri equation
\footnote{This equation is relevant for studying the developments of
caustics in a congruence of time-like geodesics for converging
values of the expansion $\theta$ and of singularities in Einstein
space-times \cite{69}. However the boundary conditions of
asymptotically Minkowskian space-times without super-translations
should avoid the singularity theorems as it happens with their
subfamily without matter of Ref.\cite{16}.}): these equations
determine the lapse and the shift functions once four gauge-fixings
for the basic gauge variables are given.\hfill\break

C) The equations $\partial_{\tau}\, n(\tau , \sigma^u) =
\lambda_n(\tau, \sigma^u)$ and $\partial_{\tau}\, {\bar
n}_{(a)}(\tau , \sigma^u) = \lambda_{{\bar n}_{(a)}}(\tau,
\sigma^u)$. Once the lapse and shift functions of the chosen gauge
have been found, they determine the associated Dirac
multipliers.\hfill\break

D) The {\it hyperbolic} evolution PDE for the tidal variables
$R_{\bar a}(\tau , \sigma^u)$, $\Pi_{\bar a}(\tau , \sigma^u)$. When
the equations for $\partial_{\tau}\, R_{\bar a}(\tau, \sigma^u)$ is
inverted to get $\Pi_{\bar a}(\tau, \sigma^u)$ in terms of $R_{\bar
a}(\tau, \sigma^u)$ and its derivatives, then the Hamilton equations
for $\Pi_{\bar a}(\tau, \sigma^u)$ become hyperbolic PDE for the
evolution of the physical tidal variable $R_{\bar a}(\tau,
\sigma^u)$.\hfill\break

E) The Hamilton equations for matter, when present.\medskip

Given a solution of the super-momentum and super-Hamiltonian
constraints and the Cauchy data for the tidal variables on an
initial 3-space, one can find a solution of Einstein's equations in
radar 4-coordinates adapted to a time-like observer in the chosen
gauge.

\subsection{The Congruence of Eulerian Observers and the
  non-Hamiltonian First-Order ADM Equations of Cosmological Spacetimes}

Like in SR one can consider the congruence of the Eulerian observers
with zero vorticity associated with the 3+1 splitting of space-time,
whose properties are described by Eq.(\ref{5}). In the first paper
of Ref.\cite{40} it is shown that in ADM tetrad gravity the
congruence has the following properties in each point $(\tau;
\sigma^r)$  \footnote{See the "1+3 point of view" of Ref.\cite{70}
for a discussion of gravity in terms of the second non-surface-
forming congruence of time-like observers associated with a 3+1
splitting of space-time.}:\medskip

a) The {\it acceleration} ${}^3a^A = l^B\, {}^4\nabla_B\, l^A =
{}^4g^{AB}\, {}^3a_B$ has the components ${}^3a^{\tau} = 0$,
${}^3a^r =   \sgn\, {\tilde \phi}^{-2/3}\, Q_a^{-2}\, V_{ra}\,
V_{sa}\, \partial_s\, ln (1 + n)$,  ${}^3a_{\tau} = -
 {\tilde \phi}^{-1/3}\, Q_a^{-1}\, V_{ra}\, {\bar n}_{(a)}$
 $\partial_r\, ln (1 + n)$, ${}^3a_r = - \partial_r\, ln\, (1 + n)$.
\hfill\break

b) The {\it expansion} \footnote{It measures the average expansion
of the infinitesimally nearby world-lines surrounding a given
world-line in the congruence.} coincides with the {\it York time}:

\beq
 \theta = {}^4\nabla_A\,\, l^A = - \sgn\, {}^3K =   - \sgn\,
 {{12\pi\, G}\over {c^3}}\, \pi_{\tilde \phi}.
 \label{45}
 \eeq

\noindent In cosmology the expansion is proportional to the Hubble
constant and the dimensionless   {\it cosmological deceleration
parameter} is $q = 3\, l^A\, {}^4\nabla_A\, {1\over {\theta}} - 1 =
- 3\, \theta^{-2}\, l^A\, \partial_A\, \theta - 1$.\hfill\break

c) By using Eqs.(\ref{24}) it can be shown that the {\it shear}
\footnote{It measures how an initial sphere in the tangent space to
the given world-line, which is Lie-transported along the world-line
tangent $l^{\mu}$ (i.e. it has zero Lie derivative with respect to
$l^{\mu}\, \partial_{\mu}$), is distorted towards an ellipsoid with
principal axes given by the eigenvectors of $\sigma^{\mu}{}_{\nu}$,
with rate given by the eigenvalues of $\sigma^{\mu}{}_{\nu}$.}
$\sigma_{AB} = \sigma_{BA} = - {{\sgn}\over 2}\, ({}^3a_A\, l_B +
{}^3a_B\, l_A) + {{\sgn}\over 2}\, ({}^4\nabla_A\, l_B +
{}^4\nabla_B\, l_A) - {1\over 3}\, \theta\, {}^3h_{AB} =
\sigma_{(\alpha )(\beta )}\, {}^4{\buildrel \circ \over {\bar
E}}^{(\alpha )}_A\, {}^4{\buildrel \circ \over {\bar E}}^{(\beta
)}_B$ has the following components $\sigma_{(o)(o)} =
\sigma_{(o)(r)} = 0$, $\sigma_{(a)(b)} = \sigma_{(b)(a)} =
({}^3K_{rs} - {1\over 3}\, {}^3g_{rs}\, {}^3K)\, {}^3{\bar
e}^r_{(a)}\, {}^3{\bar e}^s_{(b)}$, $\sum_a\, \sigma_{(a)(a)} = 0$.
$\sigma_{(a)(b)}$ depends upon the canonical variables $\theta^r$,
$\tilde \phi$, $R_{\bar a}$, $\pi^{(\theta )}_i$ and $\Pi_{\bar a}$.
\bigskip

By using Eqs.(\ref{39}) for the extrinsic curvature tensor one finds
that the diagonal elements $\sigma_{(a)(a)}$ of the shear are also
connected with the tidal momenta $\Pi_{\bar a}$, while the
non-diagonal elements $\sigma_{(a)(b)}{|}_{a \not= b}$ are connected
with the momenta $\pi_i^{(\theta)}$ (the unknowns in the
super-momentum constraints)

\begin{eqnarray*}
 \Pi_{\bar a} &=& - {{c^3}\over {8\pi\, G}}\, \tilde \phi\,
 \sum_a\, \gamma_{\bar aa}\, \sigma_{(a)(a)},\nonumber \\
 &&{}\nonumber \\
 \pi_i^{(\theta )} &=&  {{c^3}\over {8\pi\, G}}\,
 \tilde \phi\,  \sum_{wtab}\, A_{wi}\, V_{wt}\, Q_a\, Q_b^{-1}\, \epsilon_{tab}\,
 \sigma_{(a)(b)}{|}_{a\not= b},
 \end{eqnarray*}

 \bea
 {}^3K_{rs} &=& {\tilde \phi}^{2/3}\, \sum_{ab}\, \Big(- {{\sgn}\over 3}\,
 \theta\, \delta_{ab} + \sigma_{(a)(b)}\Big)\, Q_a\, Q_b\,
 V_{ra}\, V_{sb} \rightarrow \nonumber \\
 &{\rightarrow_{\theta^i \rightarrow 0}}& {\tilde \phi}^{2/3}\,
 Q_r\, Q_s\, \Big( - {{\sgn}\over 3}\, \theta +
 \sigma_{(a)(b)}\Big).
 \label{46}
 \eea

\bigskip

Therefore the Eulerian observers associated to the 3+1 splitting of
space-time induce a geometrical interpretation of some of the
momenta of the York canonical basis:\medskip

1) {\it their expansion $\theta$  is the gauge variable York time
${}^3K = {{12 \pi G}\over {c^3}}\, \pi_{\tilde \phi}$ determining
the non-dynamical gauge part of the shape of the instantaneous
3-spaces $\Sigma_{\tau}$ as a sub-manifold of space-time};

2) {\it the diagonal elements of their shear describe the tidal
momenta $\Pi_{\bar a}$, while the non-diagonal elements are
connected to the variables $\pi_i^{(\theta )}$}, determined by the
super-momentum constraints.

\bigskip

In Eq.(\ref{44}), valid in the 3-orthogonal gauges, the term
quadratic in the momenta $\pi_i^{(\theta)}$ in the weak ADM energy
and in the super-Hamiltonian constraint can be written as
${{c^3}\over {16 \pi G}}\, \tilde \phi\, \sum_{ab, a\not= b}\,
\sigma^2_{(a)(b)}$, while the super-momentum constraints can be
written in the form of PDE for the non-diagonal elements of the
shear

\bea
 {\bar {\cal H}}_{(a)}{|}_{\theta^i = 0}(\tau , \sigma^u) &=&
 - {{c^3}\over {8\pi\, G}}\,  {\tilde \phi}^{2/3}(\tau,
 \sigma^u)\, \Big(\sum_{b \not= a}\, Q_b^{-1}\, \Big[\partial_b\,
 \sigma_{(a)(b)} +\nonumber \\
 &+& \Big({\tilde \phi}^{-1}\, \partial_b\, \tilde \phi +
 \sum_{\bar b}\, (\gamma_{\bar ba} - \gamma_{\bar bb})\, \partial_b\,
 R_{\bar b}\Big)\, \sigma_{(a)(b)}\Big] -\nonumber \\
 &-& {{8\pi\, G}\over {c^3}}\, {\tilde \phi}^{-1}\,
 Q_a^{-1}\,  \Big[\tilde \phi\, \partial_a\, \pi_{\tilde \phi}
 +  \sum_{\bar b}\, (\gamma_{\bar ba}\, \partial_a\, \Pi_{\bar b}
 - \partial_a\, R_{\bar b}\, \Pi_{\bar b}) +\nonumber \\
 &+& {\cal M}_{a}\Big] \Big)(\tau , \sigma^u) \quad \approx 0.
 \label{47}
 \eea

\bigskip

As a consequence, by using ${}^3g_{rs}$ of Eq.(\ref{38}) and
${}^3K_{rs}$ of Eq.(\ref{46}), the first-order non-Hamiltonian ADM
equations (\ref{33}) can be re-expressed in terms of the
configurational variables $n$, ${\bar n}_{(a)}$, $\tilde \phi$,
$\theta^i$, $R_{\bar a}$, and of the expansion $\theta$ and shear
$\sigma_{(a)(b)}$ of the Eulerian observers. Then the 12 equations
can be put in the form of equations determining $\partial_{\tau}\,
\tilde \phi$, $\partial_{\tau}\, R_{\bar a}$, $\partial_{\tau}\,
\theta^i$, $\partial{\tau}\, \theta$ and $\partial_{\tau}\,
\sigma_{(a)(b)}$. In Eqs.(2.17) of the first paper in Ref.\cite{40}
this manipulation is explicitly done for the first six equations
(\ref{33}).

 \bigskip

These results are important for extending the identification of the
inertial and tidal variables of the gravitational field, achieved
with the York canonical basis, to cosmological space-times. Since
these space-times are only conformally asymptotically flat, the
Hamiltonian formalism is not defined. However, they are globally
hyperbolic and admit 3+1 splittings with the associated congruence
of  Eulerian observers. As a consequence, in them Einstein's
equations are usually replaced with the non-Hamiltonian first-order
ADM equations plus the super-Hamiltonian and super-momentum
constraints. Our analysis implies that, since the 4-metric can
always be put in the form of Eqs.(\ref{38}), the inertial gauge
variables of the cosmological space-times are $n$, ${\bar n}_{(a)}$,
$\theta^i$ and the expansion $\theta = - \sgn\, {}^3K$, while the
physical tidal variables are $R_{\bar a}$ and the diagonal
components of the shear $\sigma_{(a)(a)}$ ($\sum_a\, \sigma_{(a)(a)}
= 0$). The unknown in the super-Hamiltonian constraint is the
conformal factor $\tilde \phi$ of the 3-metric in $\Sigma_{\tau}$,
while the unknowns in the super-Hamiltonian constraints are the
non-diagonal components of the shear $\sigma_{(a)(b)}{|}_{a\not=
b}$.

  \vfill\eject

\section{Post-Minkowskian Linearization  in Non-Harmonic 3-Orthogonal Gauges
and Post-Minkowskian Gravitational Waves}
  \label{sec:7}

In the second paper of Ref.\cite{40} it was shown that in the family
of non-harmonic 3-orthogonal Schwinger gauges it is possible to
define a consistent {\it linearization} of ADM canonical tetrad
gravity plus matter (N charged scalar particles of masses $m_i$,
Grassmann-valued signs of energy $\eta_i$, Grassmann-valued electric
charges $Q_i$, plus the electro-magnetic field in the radiation
gauge) in the weak field approximation, to obtain a formulation of
{\it Hamiltonian Post-Minkowskian (HPM) gravity with non-flat
Riemannian 3-spaces and asymptotic Minkowski background}.

\medskip

In the standard linearization one introduces a {\it fixed Minkowski
background space-time}, introduces the decomposition
${}^4g_{\mu\nu}(x) = {}^4\eta_{\mu\nu} + {}^4h_{\mu\nu}(x)$ in an
inertial frame and studies the linearized equations of motion for
the small Minkowskian fields ${}^4h_{\mu\nu}(x)$. The approximation
is assumed valid over a {\it big enough characteristic length $L$
interpretable as the reduced wavelength $\lambda / 2\pi$ of the
resulting gravitational waves (GW)} (only for distances higher of
$L$ the linearization breaks down and curved space-time effects
become relevant). For the Solar System there is a PN approximation
in harmonic gauges, which is adopted in the BCRS \cite{5} and whose
3-spaces $t_B = const.$ have deviations of order $c^{-2}$ from
Euclidean 3-spaces.
\medskip

See Refs.\cite{44,71} and Appendix A of the second paper in
Refs.\cite{40} for a review of all the results of the standard
approach and of the existing points of view  on the subject
\cite{72,73,74,75,76}.
\medskip

In the class of asymptotically Minkowskian space-times without
super-translations the 4-metric tends to an asymptotic Minkowski
metric at spatial infinity, ${}^4g_{AB}\, \rightarrow
{}^4\eta_{AB}$, which can be used as an {\it asymptotic background}.
The decomposition ${}^4g_{AB} = {}^4\eta_{AB} + {}^4h_{(1)AB}$, with
a first order perturbation ${}^4h_{(1)AB}$ vanishing at spatial
infinity, is defined in a global non-inertial rest frame of an
asymptotically Minkowskian space-time deviating for first order
effects from a global inertial rest frame of an abstract Minkowski
space-time $M_{(\infty)}$. The non-Euclidean 3-spaces
$\Sigma_{\tau}$ will deviate by first order effects from the
Euclidean 3-spaces $\Sigma_{\tau (\infty)}$ of the inertial rest
frame of $M_{(\infty)}$ coinciding with the limit of $\Sigma_{\tau}$
at spatial infinity. When needed differential operators like the
Laplacian in $\Sigma_{\tau}$ will be approximated with the flat
Laplacian in $\Sigma_{\tau (\infty)}$.

\bigskip

If $\zeta << 1$ is a small a-dimensional parameter, a consistent
Hamiltonian linearization implies the following restrictions on the
variables of the York canonical basis in the family of 3-orthogonal
gauges with ${}^3K(\tau, \vec \sigma) = F(\tau, \vec \sigma)$ $=\,
numerical\, function$ (in this Section one uses the notation $\vec
\sigma$ for the curvilinear 3-coordinates $\sigma^r$)

\begin{eqnarray*}
 &&R_{\bar a}(\tau ,\vec \sigma ) = R_{(1)\bar a}(\tau,
 \vec \sigma)   = O(\zeta) << 1,\nonumber \\
 &&\Pi_{\bar a}(\tau ,\vec \sigma ) = \Pi_{(1)\bar a}(\tau,
 \vec \sigma)   = {1\over {L\, G}}\, O(\zeta),\nonumber \\
 &&{}\nonumber \\
 &&\tilde \phi(\tau ,\vec \sigma ) = \sqrt{det\, {}^3g_{rs}(\tau ,\vec \sigma )}
  = 1 + 6\, \phi_{(1)}(\tau ,\vec \sigma ) + O(\zeta^2),\nonumber \\
 &&N(\tau ,\vec \sigma ) = 1 + n(\tau ,\vec \sigma ) =
 1 + n_{(1)}(\tau ,\vec \sigma ) + O(\zeta^2),\nonumber \\
 && \sgn\, {}^4g_{\tau\tau}(\tau ,\vec \sigma ) =
 1 + \sgn\, {}^4h_{(1)\tau\tau}(\tau ,\vec \sigma )
 = 1 + 2\, n_{(1)}(\tau ,\vec \sigma ) + O(\zeta^2),\nonumber \\
 &&{\bar n}_{(a)}(\tau ,\vec \sigma ) = - \sgn\, {}^4g_{\tau
 a}(\tau ,\vec \sigma ) = - \sgn\, {}^4h_{(1)\tau r}(\tau ,\vec \sigma )
 = {\bar n}_{(1)(a)}(\tau ,\vec \sigma ) + O(\zeta^2),
 \end{eqnarray*}

 \bea
 &&{}^3K(\tau ,\vec \sigma ) = {{12\pi\, G}\over {c^3}}\, \pi_{\tilde
 \phi}(\tau ,\vec \sigma ) = {}^3K_{(1)}(\tau ,\vec \sigma ) = {{12\pi\, G}\over
 {c^3}}\, \pi_{(1) \tilde \phi}(\tau ,\vec \sigma ) =
 {1\over L}\, O(\zeta),\nonumber \\
  &&\sigma_{(a)(b)}{|}_{a\not= b}(\tau ,\vec \sigma ) =
  \sigma_{(1)(a)(b)}{|}_{a\not= b}(\tau ,\vec \sigma )
 = {1\over L}\, O(\zeta),\nonumber \\
 &&{}\nonumber \\
 &&{}^3g_{rs}(\tau ,\vec \sigma ) = - \sgn\, {}^4g_{rs}(\tau ,\vec
 \sigma ) = \delta_{rs} - \sgn\, {}^4h_{(1)rs}(\tau ,\vec \sigma )
 =\nonumber \\
 &=& [1 + 2\, (\Gamma_r^{(1)}(\tau ,\vec \sigma ) +
 2\, \phi_{(1)}(\tau ,\vec \sigma ))]\, \delta_{rs} + O(\zeta^2),
 \nonumber \\
 && \Gamma^{(1)}_a(\tau ,\vec \sigma ) =
 \sum_{\bar ar = 1}^2\, \gamma_{\bar aa}\, R_{\bar a}(\tau ,\vec \sigma
 ),\qquad
 R_{\bar a}(\tau ,\vec \sigma ) = \sum_{a=1}^3\, \gamma_{\bar aa}\,
 \Gamma^{(1)}_a(\tau ,\vec \sigma ).
 \label{48}
 \eea

The tidal variables $R_{\bar a}(\tau ,\vec \sigma )$ are slowly
varying over the length $L$ and times $L/c$; one has $({L\over
{{}^4{\cal R}}})^2 = O(\zeta)$, where ${}^4{\cal R}$ is the mean
radius of curvature of space-time.

\medskip

The consistency of the Hamiltonian linearization requires the
introduction of a {\it ultra-violet cutoff $M$ for matter}. For the
particles, described by the canonical variables ${\vec
\eta}_i(\tau)$ and ${\vec \kappa}_i(\tau)$, this implies the
conditions ${{m_i}\over M}, {{ {\vec \kappa}_i}\over M} = O(\zeta)$.
With similar restrictions on the electro-magnetic field one gets
that the energy-momentum tensor of matter is $T^{AB}(\tau, \vec
\sigma) = T^{AB}_{(1)}(\tau, \vec \sigma) + O(\zeta^2)$. Therefore
also the mass and momentum densities have the behavior ${\cal
M}(\tau, \vec \sigma) = {\cal M}_{(1)}(\tau, \vec \sigma) +
O(\zeta^2)$, ${\cal M}_r(\tau, \vec \sigma) = {\cal M}_{(1)r}(\tau,
\vec \sigma) + O(\zeta^2)$. This approximation is not reliable at
distances from the point particles less than the gravitational
radius $R_M = {{M\, G}\over {c^2}} \approx 10^{- 29}\, M$ determined
by the cutoff mass. {\it The weak ADM Poincar\'e generators become
equal to the Poincar\'e generators of this matter in the inertial
rest frame of the Minkowski space-time $M_{(\infty)}$ plus terms of
order $O(\zeta^2)$ containing GW and matter}. Finally the GW
described by this linearization must have wavelengths satisfying
$\lambda/2\pi \approx L >> R_M$. If all the particles are contained
in a compact set of radius $l_c$ (the source), one must have $l_c >>
R_M$ for particles with relativistic velocities and $l_c \geq R_M$
for slow particles (like in binaries). See Ref.\cite{44} for more
details.

\medskip

{\it With this Hamiltonian linearization one can avoid to make PN
expansions: one gets fully relativistic expressions, i.e. a HPM
formulation of gravity}.

\bigskip

The effective Hamiltonian adapted to the 3-orthogonal gauges and
replacing the weak ADM energy is ${1\over c}\, \Big({\hat
E}_{ADM(1)} + {\hat E}_{ADM(2)}\Big) + {{c^3}\over {12 \pi G}}\,
\int d^3\sgn\, \Big(\partial_{\tau}\, {}^3K_{(1)}\, \Big[1 + {6\over
{\triangle}}\,\Big({1\over 4}\, \sum_a\,
\partial_a^2\, \Gamma_a^{(1)} - {{2 \pi G}\over {c^3}}\,
{\cal M}_{(1)}\Big)\Big]\Big)(\tau, \sigma^u) + O(\zeta^3)$ in the
PM linearized theory.
\medskip

In the second paper of Refs.\cite{40} one has found the solutions of
the super-momentum and super-Hamiltonian constraints and of the
equations for the lapse and shift functions with the Bianchi
identities satisfied. Therefore one knows the first order quantities
$\pi^{(\theta)}_{(1)i}(\tau ,\vec \sigma )$, $\tilde \phi(\tau ,\vec
\sigma ) = 1 + 6\, \phi_{(1)}(\tau ,\vec \sigma )$, $1 +
n_{(1)}(\tau ,\vec \sigma )$, ${\bar n}_{(1)(a)}(\tau ,\vec \sigma
)$ (the quantities containing the {\it action-at-a-distance part} of
the gravitational interaction in the 3-orthogonal gauges) with an
explicit expression for the {\it PM Newton and gravito-magnetic
potentials}. In absence of the electro-magnetic field they are (the
terms in $\Gamma^{(1)}_a(\tau ,\vec \sigma )$ describe the
contribution of GW) \footnote{Quantities like $|{\vec \eta}_i(\tau)
- {\vec \eta}_j(\tau)|$ are the Euclidean 3-distance between the two
particles in the asymptotic 3-space $\Sigma_{\tau (\infty)}$, which
differs by quantities of order $O(\zeta)$ from the real
non-Euclidean 3-distance in $\Sigma_{\tau}$ as shown in Eq.(3.3) of
the third paper in Ref.\cite{40}.}

\begin{eqnarray*}
 {\tilde \phi}(\tau, \vec \sigma) &=& 1 + 6\, \phi_{(1)}(\tau, \vec  \sigma)
 =\nonumber \\
 &=& 1 + {{3\, G}\over {c^3}}\, \sum_i\, \eta_i\, {{\sqrt{m_i^2\, c^2 +
 {\vec \kappa}^2_i(\tau)}}\over
 { |\vec \sigma - {\vec \eta}_i(\tau)|}} -\nonumber \\
 &-& {3\over {8\pi}}\, \int d^3\sigma_1\, {{\sum_a\, \partial_{1a}^2\,
 \Gamma_a^{(1)}(\tau, {\vec \sigma}_1)}\over { |\vec \sigma -
 {\vec \sigma}_1|  }},\nonumber \\
 &&{}\nonumber \\
 \sgn\, {}^4g_{\tau\tau}(\tau, \vec \sigma) &=& 1 + 2\,
 n_{(1)}(\tau,  \vec \sigma) = 1 -
  2\, \partial_{\tau}\, {}^3{\cal K}_{(1)}(\tau, \vec \sigma)
  -\nonumber \\
 &-& {{2\, G}\over {c^3}}\, \sum_i\, \eta_i\, {{\sqrt{m_i^2\, c^2 +
 {\vec \kappa}^2_i(\tau)}}\over { |\vec \sigma -
 {\vec \eta}_i(\tau)|}}\, \Big(1 +
 {{{\vec \kappa}^2_i}\over {m_i^2\, c^2 + {\vec \kappa}_i^2}}\Big),
 \end{eqnarray*}

 \bea
 - \sgn\, {}^4g_{\tau a}(\tau, \vec \sigma) &=& {\bar
 n}_{(1)(a)}(\tau, \vec \sigma) =
  \partial_a\, {}^3{\cal K}_{(1)}(\tau, \vec \sigma) -\nonumber \\
 &-& {{G}\over {c^3}}\, \sum_i\, {{\eta_i}\over {|\vec \sigma - {\vec \eta}_i(\tau)|}}\,
 \Big( \frac{7}{2}\kappa_{ia}(\tau) -\nonumber \\
 &-& \frac{1}{2} {{(\sigma^a - \eta^a_i(\tau))\, {\vec \kappa}_i(\tau) \cdot (\vec \sigma
 - {\vec \eta}_i(\tau))}\over {|\vec \sigma - {\vec \eta}_i(\tau)|^2}} \Big)
 -\nonumber \\
 &-& \int {{d^3\sigma_1}\over {4\pi\, |\vec \sigma - {\vec \sigma}_1|}}\,
 \partial_{1a}\, \partial_{\tau}\, \Big[ 2\,
 \Gamma_a^{(1)}(\tau, {\vec \sigma}_1) -\nonumber \\
 &-& \int d^3\sigma_2\, {{\sum_c\,  \partial_{2c}^2\,
 \Gamma_c^{(1)}(\tau, {\vec \sigma}_2)}\over {8\pi\, |{\vec \sigma}_1 -
 {\vec \sigma}_2|}}\Big],\nonumber \\
 &&{}\nonumber \\
  \sigma_{(1)(a)(b)}{|}_{a \not= b}(\tau, \vec  \sigma) &=& {1\over 2}\,
 \Big(\partial_a\, {\bar n}_{(1)(b)} + \partial_b\, {\bar
 n}_{(1)(a)}\Big){|}_{a \not= b}(\tau, \vec \sigma).
 \label{49}
 \eea

\medskip

Instead the linearization of the Hamilton equations for the tidal
variables $R_{\bar a}(\tau, \vec \sigma)$ implies that they satisfy
the following wave equation \footnote{For the tidal momenta one gets
${{8\pi\, G}\over {c^3}}\, \Pi_{\bar a}(\tau, \vec \sigma) =
[\partial_{\tau}\, R_{\bar a} - \sum_a\, \gamma_{\bar aa}\,
\partial_a\, {\bar n}_{(1)(a)}](\tau, \vec \sigma) + O(\zeta^2)$, so that
the diagonal elements of the shear are $\sigma_{(1)(a)(a)}(\tau,
\vec \sigma) = [- \sum_{\bar a}\, \gamma_{\bar aa}\,
\partial_{\tau}\, R_{\bar a} + {\bar n}_{(1)(a)} - {1\over 3}\,
\sum_b\, {\bar n}_{(1)(b)}](\tau, \vec \sigma) + O(\zeta^2)$.}
($\triangle$ and $\Box$ are the flat Laplacian and the flat
D'Alambertian on $\Sigma_{\tau (\infty)}$)

\bea
  \partial_{\tau}^2\, R_{\bar a}(\tau, \vec \sigma)\, &=& \triangle\,
 R_{\bar a}(\tau, \vec \sigma) + \sum_a\, \gamma_{\bar aa}\, \Big[
 \partial_{\tau}\, \partial_a\, {\bar n}_{(1)(a)} +\nonumber \\
 &+& \partial_a^2\, n_{(1)} + 2\, \partial_a^2\, \phi_{(1)} - 2\,
 \partial_a^2\, \Gamma_a^{(1)} + {{8\pi\, G}\over {c^3}}\,
 T_{(1)}^{aa} \Big](\tau, \vec \sigma).
 \label{50}
 \eea

\noindent

By using Eqs.(\ref{49}) this wave equation becomes

 \bea
 \Box\, \sum_{\bar b}\, M_{\bar a\bar b}\,
 R_{\bar b}(\tau, \vec \sigma)\,\,  &=&\,\,\,
 E_{\bar a}(\tau, \vec \sigma),\nonumber \\
 &&{}\nonumber \\
  M_{\bar a\bar b} &=& \delta_{\bar a\bar b} - \sum_a\, \gamma_{\bar
 aa}\, {{\partial_a^2}\over {\triangle}}\, \Big(2\, \gamma_{\bar
 ba} - {1\over 2}\, \sum_b\, \gamma_{\bar bb}\, {{\partial_b^2}\over
 {\triangle}}\Big),\nonumber \\
 E_{\bar a}(\tau, \vec \sigma) &=&
 {{4\pi\, G}\over {c^3}}\, \sum_a\, \gamma_{\bar
 aa}\, \Big[{{\partial_{\tau}\, \partial_a}\over {\triangle}}\,
 \Big(4\, {\cal M}_{(1)a}
 - {{\partial_a}\over {\triangle}}\, \sum_c\, \partial_c\,
 {\cal M}_{(1)c}\Big) +\nonumber \\
 &&\qquad + 2\, T_{(1)}^{aa}
 + {{\partial_a^2}\over {\triangle}}\, \sum_b\,
 T_{(1)}^{bb}\Big](\tau, \vec \sigma),
 \nonumber \\
 &&{}\nonumber \\
 &&\Downarrow\nonumber \\
 &&{}\nonumber \\
  \Box\, \sum_b\, {\tilde M}_{ab}\, \Gamma_b^{(1)}(\tau, \vec \sigma)\,\,
 &=&\,\,\, \sum_{\bar a}\, \gamma_{\bar aa}\,
 E_{\bar a}(\tau, \vec \sigma),\nonumber \\
 &&{}\nonumber \\
  {\tilde M}_{ab} &=& \sum_{\bar a\bar b}\, \gamma_{\bar aa}\,
 \gamma_{\bar bb}\, M_{\bar a\bar b} = \delta_{ab}\, \Big(1
 - 2\, {{\partial_a^2}\over {\triangle}}\Big) + {1\over 2}\,
 \Big(1 + {{\partial_a^2}\over {\triangle}}\Big)\,
 {{\partial_b^2}\over {\triangle}},\nonumber \\
 &&\qquad \sum_a\, {\tilde M}_{ab} = 0,
 \qquad M_{\bar a\bar b} = \sum_{ab}\, \gamma_{\bar aa}\,
 \gamma_{\bar bb}\, {\tilde M}_{ab}.
 \label{51}
 \eea

To understand the meaning of the spatial operators $M_{\bar a\bar
b}$ and ${\tilde M}_{ab}$, one must consider the perturbation
${}^4h_{(1)rs}(\tau, \vec \sigma) = - 2\, \sgn\, \delta_{rs}\,
(\Gamma_r^{(1)} + 2\, \phi_{(1)})(\tau, \vec \sigma)$ of
Eq.(\ref{48}) and apply to it the following decomposition, given in
Ref\cite{45},

\bea
 {}^4h_{(1)rs}(\tau, \vec \sigma) &=& \Big( {}^4h^{TT}_{(1)rs}
  + {1\over 3}\, \delta_{rs}\, H_{(1)} + {1\over 2}\, (\partial_r\, \epsilon_{(1)s} +
 \partial_s\, \epsilon_{(1)r}) +\nonumber \\
 &+& (\partial_r\, \partial_s - {1\over
 3}\, \delta_{rs}\, \triangle)\, \lambda_{(1)} \Big)(\tau, \vec \sigma),
 \label{52}
 \eea

\noindent with $\sum_r\, \partial_r\, \epsilon_{(1)r} = 0$ and
${}^4h_{(1)rs}^{TT}$ traceless and transverse (TT), i.e. $\sum_r\,
{}^4h^{TT}_{(1)rr} = 0$, $\sum_r\, \partial_r\, {}^4h^{TT}_{(1)rs} =
0$. Since one finds $H_{(1)}(\tau, \vec \sigma) = - 12\, \sgn\,
\phi_{(1)}(\tau, \vec \sigma)$, $\lambda_{(1)}(\tau, \vec \sigma) =
- 3\, \sgn\,  \sum_u\, {{\partial_u^2}\over {\triangle^2}}\,
\Gamma_u^{(1)}(\tau, \vec \sigma)$ and $\epsilon_{(1)r}(\tau, \vec
\sigma) = - 4\, \sgn\, {{\partial_r}\over {\triangle}}\, \Big(
\Gamma_r^{(1)} - \sum_u\, {{\partial_u^2}\over {\triangle}}\,
\Gamma_u^{(1)} \Big)(\tau, \vec \sigma)$, it turns out that the TT
part of the spatial metric is independent from $\phi_{(1)}$ and has
the expression

\bea
 {}^4h^{TT}_{(1)rs}(\tau, \vec \sigma) &=&
 - \sgn\, \Big[\Big(2\, \Gamma_r^{(1)} +
 \sum_u\, {{\partial_u^2}\over {\triangle}}\, \Gamma_u^{(1)}\Big)\,
 \delta_{rs} -\nonumber \\
 &-&2\, {{\partial_r\, \partial_s}\over {\triangle}}\,
 (\Gamma_r^{(1)} + \Gamma_s^{(1)}) + {{\partial_r\,
 \partial_s}\over {\triangle}}\, \sum_u\, {{\partial_u^2}\over
 {\triangle}}\, \Gamma_u^{(1)}\Big](\tau, \vec \sigma),\nonumber \\
 &&{}\nonumber \\
 \Rightarrow && {}^4h^{TT}_{(1)aa}(\tau, \vec \sigma) = - 2\,
 \sgn\, \sum_b\, {\tilde M}_{ab}\, \Gamma_b^{(1)}(\tau, \vec
 \sigma).
 \label{53}
 \eea

Therefore the spatial operator ${\tilde M}_{ab}$ connects the tidal
variables $R_{\bar a}(\tau, \vec \sigma)$ of the York canonical
basis to the TT components of the 3-metric. By applying the
decomposition (\ref{52}) to the spatial part $T_{(1)}^{rs}(\tau,
\vec \sigma)$ of the energy-momentum one verifies that like in the
harmonic gauges \cite{44} the TT part of the 3-metric satisfies the
wave equation $\Box\, {}^4h^{TT}_{rs}(\tau, \vec \sigma) = - \sgn\,
{{16 \pi G}\over {c^3}}\, T_{(1)rs}^{(TT)}(\tau, \vec \sigma)$.
\bigskip

The retarded solution of the wave equation with a no-incoming
radiation condition gives the following expression for the tidal
variables (the HPM-GW)

\bea
 R_{\bar a}(\tau, \vec \sigma) &=& - \sum_a\, \gamma_{\bar aa}\,
 \Gamma_a^{(1)}(\tau, \vec \sigma)
 \cir \sum_{ab}\, \gamma_{\bar aa}\, {\tilde M}^{-1}_{ab}(\tau, \vec \sigma)\,
 \nonumber \\
 && {{2\, G}\over {c^3}}\, \int d^3\sigma_1\, {{T^{(TT)bb}_{(1)}(\tau
 - |\vec \sigma - {\vec \sigma}_1|; {\vec \sigma}_1)}\over {
 |\vec \sigma - {\vec \sigma}_1|}},\nonumber \\
 &&{}\nonumber \\
  {{8\pi\, G}\over {c^3}}\, \Pi_{\bar a}(\tau, \vec \sigma) &=&
 \Big(\sum_{\bar b}\, M_{\bar a\bar b}\, \partial_{\tau} \, R_{\bar
 b} - \sum_a\, \gamma_{\bar aa}\, \Big[{{4\pi\, G}\over {c^3}}\,
 {1\over {\triangle}}\, (4\, \partial_a\,
 {\cal M}_{(1)a} -\nonumber \\
 &-& {{\partial_a^2}\over {\triangle}}\, \sum_c\,
 \partial_c\, {\cal M}_{(1)c}) +
   \partial_a^2\, {}^3{\cal K}_{(1)} \Big]\Big)(\tau, \vec
 \sigma).
 \label{54}
 \eea

The explicit form of the inverse operator is given in the second
paper of Ref.\cite{40}. By using  the multipolar expansion of the
energy-momentum $T_{(1)}^{AB}$ of Ref.\cite{58} in the  HPM version
adapted to the rest-frame instant form of dynamics of Ref.\cite{26},
one gets

\bea
 R_{\bar a}(\tau, \vec \sigma) = -
 {G\over {c^3}}\, \sum_{ab}\, \gamma_{\bar aa}\, {\tilde
 M}^{-1}_{ab}\, {{\partial^2_{\tau}\, q^{(TT) aa
 | \tau\tau}(\tau - |\vec \sigma|)}\over {|\vec \sigma|}}
 + (higher\, multipoles),\nonumber \\
 &&{}
  \label{55}
  \eea

\noindent where $q^{(TT) aa | \tau\tau}(\tau)$ is the {\it TT mass
quadrupole} with respect to the center of energy (put in the origin
of the radar 4-coordinates). An analogous result holds for
${}^4h^{TT}_{rs}(\tau, \vec \sigma)$ and this implies a {\it HPM
relativistic version of the standard mass quadrupole emission
formula}.\medskip

Moreover, notwithstanding there is no gravitational self-energy due
to the Grassmann regularization, the energy, 3-momentum and angular
momentum balance equations in HPM-GW emission are verified by {\it
using the conservation of the asymptotic ADM Poincar\'e generators}
(the same happens with the asymptotic Larmor formula of the
electro-magnetic case with Grassmann regularization as shown in the
last paper of Ref.\cite{27}). See Refs.\cite{44,76,77} for the use
of the self-energy in the standard derivation of this result by
means of PN expansions.

\bigskip

Eqs.(\ref{49}) and (\ref{54}) show that the HPM linearization with
no-incoming radiation condition and Grassmann regularization is a
theory with only dynamical matter interacting through suitable
action-at-a-distance and retarded effective potentials. Instead in
relativistic atomic physics in SR the no-incoming radiation
condition and the Grassmann regularization kill also the retardation
leaving only the action-at-a-distance inter-particle Coulomb plus
Darwin potentials. See Eq.(7.22) of the second paper of
Ref.\cite{40} for the expression of the weak ADM energy till order
$O(\zeta^3)$.

\bigskip

Moreover it can be shown that the coordinate transformation $\bar
\tau = \tau$,  ${\bar \sigma}^r = \sigma^r + {1\over 2}\,
{{\partial_r}\over {\triangle}}\, \Big(4\, \Gamma_r^{(1)} - \sum_c\,
{{\partial_c^2}\over {\triangle}}\, \Gamma_c^{(1)}\Big)(\tau, \vec
\sigma)$, introducing new $\tau$-dependent  radar 3-coordinates on
the 3-space $\Sigma_{\tau}$, allows one to make a transition from
the 3-orthogonal gauge with the 4-metric given by Eqs.(\ref{48}) and
(\ref{49}) to a {\it generalized non-3-orthogonal TT gauge}
containing the TT 3-metric (\ref{53})

\begin{eqnarray*}
 &&{}^4g_{(1)AB} = {}^4\eta_{AB} +\nonumber \\
 &&+ \sgn\,  \left(
 \begin{array}{ccc}
 - 2\,\frac{\partial_\tau\,}{\Delta}\, {}^3K_{(1)} + \alpha(matter)
 &{}& - \frac{\partial_r}{\Delta}\, {}^3K_{(1)} + A_r(\Gamma_a^{(1)}) + \beta_r(matter)\\
 &&\\
  - \frac{\partial_s}{\Delta}\, {}^3K_{(1)} + A_s(\Gamma_a^{(1)}) + \beta_s(matter)&{}&
 \Big[B_r(\Gamma_a^{(1)}) + \gamma(matter)\Big]\, \delta_{rs}
 \end{array}
 \right) +\nonumber \\
 &+& O(\zeta^2),\nonumber \\
 &&{}\nonumber \\
 &&\Downarrow
 \end{eqnarray*}

 \bea
 {}^4{\bar g}_{AB} &=& {}^4\eta_{AB} +
 \sgn\, \left(
 \begin{array}{ccc}
 -2\,\frac{\partial_\tau\,}{\Delta}\, {}^3K_{(1)} +
 \alpha(matter)
 &{}&- \frac{\partial_r}{\Delta}\, {}^3K_{(1)} +
 \beta_r(matter)\\
 &&\\
 - \frac{\partial_s}{\Delta}\, {}^3K_{(1)} +
 \beta_s(matter)&{}&
 \sgn\, {}^4h^{TT}_{(1) rs} + \delta_{rs}\,  \gamma(matter)
 \end{array}
 \right) +\nonumber \\
 &+& O(\zeta^2).
 \label{56}
 \eea

The functions appearing in Eqs.(\ref{56}) are: $A_r(\Gamma_a^{(1)})
= - {1\over 2}\, \partial_{\tau}\, {{\partial_r}\over {\triangle}}\,
\Big(4\, \Gamma_r^{(1)} - \sum_c\, {{\partial_c^2}\over
{\triangle}}\, \Gamma_c^{(1)}\Big)$, $B_r(\Gamma_a^{(1)}) = - 2\,
\Big(\Gamma_r^{(1)} + {1\over 2}\, \sum_c\, {{\partial_c^2}\over
{\triangle}}\, \Gamma_c^{(1)}\Big)$, $\alpha(matter) = {{8\pi\,
G}\over {c^3}}\,$ ${1\over {\triangle}}\, \Big({\cal M}_{(1)} +
\sum_c\, T^{cc}_{(1)}\Big)$, $\beta_r(matter) = - {{4\pi\, G}\over
{c^3}}\, {1\over {\triangle}}\, \Big(4\, {\cal M}_{(1)r} -
{{\partial_r}\over {\triangle}}\, \sum_c\, \partial_c\, {\cal
M}_{(1)c}\Big)$, $\gamma(matter) = {{8\pi\, G}\over {c^3}}\, {1\over
{\triangle}}\, {\cal M}_{(1)}$.

\bigskip

Also in absence of matter this TT gauge differs from the usual
harmonic ones for the non-spatial terms depending upon the inertial
gauge variable {\it non-local York time}

\beq
 {}^3{\cal K}_{(1)}(\tau, \vec \sigma) = {1\over {\triangle}}\, {}^3K_{(1)}(\tau, \vec
 \sigma),
 \label{57}
 \eeq

\noindent describing the HPM form of the gauge freedom in clock
synchronization.

If one uses the coordinate system of the generalized TT gauge, one
can introduce the standard polarization pattern of GW for
${}^4h^{TT}_{rs}$ (see Refs.\cite{44,45,78}) and then the inverse
transformation gives the polarization pattern of HPM-GW in the
family of 3-orthogonal gauges.

\bigskip

If the matter sources have a compact support and if  the matter
terms ${1\over {\triangle}}\, {\cal M}_{(1)}(\tau, \vec \sigma)$ and
${1\over {\triangle}}\, {\cal M}_{(1)r}(\tau, \vec \sigma)$ are
negligible in the radiation zone far away from the sources, then
Eq.(\ref{56}) gives a {\it spatial TT-gauge} with still the explicit
dependence on the inertial gauge variable ${}^3{\cal K}_{(1)}(\tau,
\vec \sigma)$ (non existing in Newtonian gravity), which determines
the non-Euclidean nature of the instantaneous 3-spaces. Then one can
study the {\it far field} of compact matter sources: the restriction
to the Solar System of the resulting HPM 4-metric \footnote{See
Eq.(7.20) of the second paper in Ref.\cite{40}, where
${}^4g_{\tau\tau}(\tau, \vec \sigma)$ and ${}^4g_{\tau r}(\tau, \vec
\sigma)$ are explicitly depending on the non-local York time.} is
compatible with the harmonic PN 4-metric of BCRS \cite{5} if the
non-local York time is of order $c^{-2}$. The resulting shift
function should be used for the HPM description of gravito-magnetism
(see Refs.\cite{38,79,80} for the Lense-Thirring and other
associated effects).

\bigskip

The TT gauge allows one to reproduce the various descriptions of the
GW detectors and of the reference frames used in GW detection in
terms of HPM-GW: this is done in Subsection VIID of the second paper
of Ref.\cite{40}, where the effect of a HPM-GW on a test mass is
given in terms of the proper distance between two nearby geodesics.

\bigskip

The HPM-GW propagate  in non-Euclidean instantaneous 3-spaces
$\Sigma_{\tau}$ differing from the inertial asymptotic Euclidean
3-spaces $\Sigma_{\tau (\infty)}$ at the first order. In the family
of 3-orthogonal gauges with York time ${}^3K_{(1)}(\tau, \vec
\sigma) \approx F_{(1)}(\tau, \vec \sigma) =\, numerical\,
function$, the dynamically determined 3-spaces $\Sigma_{\tau}$ have
an intrinsic 3-curvature ${}^3{\hat R}{|}_{\theta^i=0} = 2\,
\sum_a\,\partial_a^2\, \Gamma_a^{(1)}$ determined only by the HPM-GW
(and therefore by the matter energy-momentum tensor in the past as
shown by Eq.(\ref{54})). Their extrinsic curvature tensor as
sub-manifolds of space-time is

\bea
 {}^3K_{(1)rs}\, \approx\, \sigma_{(1)(r)(s)}{|}_{r \not= s} +
 \delta_{rs}\, \Big({1\over 3}\, F_{(1)} - \partial_{\tau}\, \Gamma_r^{(1)}
 + \partial_r\, {\bar n}_{(1)(r)} - \sum_a\, \partial_a\, {\bar n}_{(1)(a)}\Big),
 \nonumber \\
 &&{}
 \label{58}
 \eea

\noindent with ${\bar n}_{(1)(r)}$, $\sigma_{(1)(r)(s)}{|}_{r \not=
s}$ and $\Gamma_r^{(1)}$ given in Eqs. (\ref{49}) and (\ref{54}).
The York time appears only in Eq.(\ref{58}): all the other PM
quantities depend on the non-local York time ${}^3{\cal
K}_{(1)}(\tau, \vec \sigma) \approx {1\over {\triangle}}\,
F_{(1)}(\tau, \vec \sigma)$

\bigskip

In the third paper of Refs.\cite{40}, where the matter is restricted
only to the particles \footnote{The properties of HPM transverse
electro-magnetic fields have still to be explored.}, one evaluates
all the properties of these HPM space-times:\medskip

a) the 3-volume element, the 3-distance and the intrinsic and
extrinsic 3-curvature tensors of the 3-spaces
$\Sigma_{\tau}$;\medskip

b) the proper time of a time-like observer;\medskip

c) the time-like and null 4-geodesics (they are relevant for the
definition of the radial velocity of stars as shown in the IAU
conventions of Ref.\cite{81} and in study of {\it time delays}
\cite{82}, \cite{80});\medskip

d) the red-shift and luminosity distance. In particular one finds
that the recessional velocity of a star is proportional to its
luminosity distance from the Earth at least for small distances.
This is in accord with the Hubble old red-shift-distance relation
which is formalized in the Hubble law (velocity-distance relation)
when the standard cosmological model is used (see for instance
Ref.\cite{46} on these topics). These results have a dependence on
the non-local York time, which could play a role in giving a
different interpretation of the data from super-novae, which are
used as a support for dark energy \cite{1}.

\bigskip

Finally, in Subsection IIIB of the second paper in Refs.\cite{40} it
is shown that this HPM linearization can be interpreted as the first
term of a HPM expansion in powers of the Newton constant $G$ in the
family of 3-orthogonal gauges. This expansion has still to be
studied. In particular it will be useful to check whether in the HPM
formulation there are phenomena (appearing at high orders in the
standard PN expansions) like the hereditary tails starting from
1.5PN [$O(({v\over c})^3)$] and the non-linear (Christodoulou)
memory starting from 3PN (see Ref.\cite{83} for a review)
\footnote{They imply that GW propagate not only on the flat
light-cone but also inside it (i.e. with all possible speeds $0 \leq
v \leq c$): there is an instantaneous wavefront followed by a tail
traveling at lower speed (it arrives later and then fades away) and
a persistent zero-frequency non-linear memory.}. This would allow
one to make a comparison with all the results of the PN expansions,
in which today there is control on the GW solution and on the matter
equations of motion till order 3.5PN [$O(({v\over c})^7)$] (for
binaries see the review in chapter 4 of Ref.\cite{44}) and well
established connections with numerical relativity (see the review in
Ref.\cite{84}) especially for the binary black hole problem (see the
review in Ref.\cite{85}).

\vfill\eject

\section{Post-Minkowskian Hamilton  Equations for  Particles, their
   Post-Newtonian Limit  and Dark Matter as a Relativistic Inertial
   Effect}
  \label{sec:8}

The PM Hamilton equations and their PN limit in 3-orthogonal gauges
for a system of N scalar particles of mass $m_i$ and
Grassmann-valued signs of energy $\eta_i$ is discussed in this
Section by using the results of the third paper in Ref.\cite{40}.
See Refs.\cite{71,78,86,87} for classical texts on the motion of
particles in gravitational fields and Refs.\cite{74,88,89} for more
recent developments \footnote{In this approach point particles are
considered as independent matter degrees of freedom with a Grassmann
regularization of the self-energies to get well defined world-lines
(see also Ref.\cite{89}): they are not considered as point-like
singularities of solutions of Einstein's equations (the point of
view of Ref.\cite{86}). Solutions of this type have to be described
with distributions and, as shown in Ref.\cite{90}, the most general
class of such solutions under mathematical control includes
singularities simulating matter shells, but not either strings or
particles. See also Ref.\cite{91}.}.\medskip

The treatment in the 3-orthogonal gauges of the PM Hamilton
equations for the electro-magnetic field in the radiation gauge is
given in the second paper of Ref.\cite{40}, while the PM Hamilton
equations for perfect fluids are given in Ref.\cite{41}.
\bigskip

With only particles the PM approximation with the ultraviolet cutoff
M implies ${\vec \kappa}_i(\tau) = {{m_i c\, {\dot {\vec
\eta}}_i(\tau)}\over {\sqrt{1 - {\dot {\vec \eta}}_i^2(\tau)}}} +
Mc\, O(\zeta)$, ${\cal M}_{(1)}(\tau, \vec \sigma) = \sum_i\,
\delta^3(\vec \sigma, {\vec \eta}_i(\tau))\, \eta_i\,$ ${{m_i
c}\over {\sqrt{1 - {\dot {\vec \eta}}_i^2(\tau)}}} + O(\zeta^2)$,
${\cal M}_{(1) r}(\tau, \vec \sigma) = \sum_i\, \delta^3(\vec
\sigma, {\vec \eta}_i(\tau))\, \eta_i\, {{m_i c\, {\dot {\vec
\eta}}_{i r}(\tau)}\over {\sqrt{1 - {\dot {\vec \eta}}_i^2(\tau)}}}
+ O(\zeta^2)$. Moreover one has ${\ddot \eta}_i(\tau) = O(\zeta)$.
The notation $\dot a(\tau) = {{d a(\tau)}\over {d\tau}}$ is used.

\medskip

One can make  a equal time development of the retarded kernel in
Eq.(\ref{54}) like in Ref.\cite{27} for the extraction of the Darwin
potential from the Lienard-Wiechert solution (see Eqs. (5.1)-(5.21)
of Ref.\cite{27} with $\sum_s\, P_{\perp}^{rs}(\vec \sigma)\, {\dot
\eta}^s_i(\tau)\, \mapsto \sum_{uv}\, {\cal P}_{bbuv}(\vec \sigma)\,
{{{\dot \eta}^u_i(\tau)\, {\dot \eta}_i^v(\tau)}\over {\sqrt{1 -
{\dot \eta}_i^2(\tau)}}}$). In this way one gets the following
expression of the HPM GW from point masses

\bea
 \Gamma^{(1)}_a(\tau, \vec \sigma) &\cir& - {2\, G\over {c^2}}\, \sum_{b}\, {\tilde
 M}^{-1}_{ab}(\vec \sigma)\, \sum_i\, \eta_i\, m_i\, \sum_{uv}\,
 {\cal P}_{bbuv}(\vec \sigma)\, {{{\dot
 \eta}^u_i(\tau)\, {\dot \eta}_i^v(\tau)}\over {\sqrt{1 - {\dot
 \eta}_i^2(\tau)}}}\nonumber \\
 && \Big[|\vec \sigma - {\vec \eta}_i(\tau)|^{- 1}
  + \sum_{m=1}^{\infty}\, {{1}\over {(2m)!}}\,
 \Big({\dot {\vec \eta}}_i(\tau) \cdot {{\partial}\over {\partial\,
 \vec \sigma}}\Big)^{2m}\, |\vec \sigma - {\vec \eta}_i(\tau)|^{2m -
 1}\Big] +\nonumber \\
 &+& O(\zeta^2),\nonumber \\
  &&{\cal P}_{rsuv} = {1\over 2}\, (\delta_{ru}\, \delta_{sv} +
 \delta_{rv}\, \delta_{su}) -\nonumber \\
 &&- {1\over 2}\, \Big(\delta_{rs} -
 {{\partial_r\, \partial_s}\over {\triangle}}\Big)\, \delta_{uv}
 + {1\over 2}\, \Big(\delta_{rs} + {{\partial_r\, \partial_s}\over
 {\triangle}}\Big)\, {{\partial_u\, \partial_v}\over {\triangle}}
 -\nonumber \\
 &&- {1\over 2}\, \Big[{{\partial_u}\over {\triangle}}\,
 (\delta_{rv}\, \partial_s + \delta_{sv}\, \partial_r)
 + {{\partial_v}\over {\triangle}}\, (\delta_{ru}\, \partial_s
 + \delta_{su}\, \partial_r)\Big],
 \label{59}
 \eea

\noindent with the retardation effects pushed to order $O(\zeta^2)$.

\bigskip

If the lapse and shift functions are rewritten in the form $n_{(1)}
= {\check n}_{(1)} - \partial_{\tau}\, {}^3{\cal K}_{(1)}$, ${\bar
n}_{(1)(r)} = {\check {\bar n}}_{(1)(r)} + \partial_r\, {}^3{\cal
K}_{(1)}$, to display their dependence on the inertial gauge
variable non-local York time, it can be shown that the PM Hamilton
equations for the particles imply the following form of the PM
Grassmann regularized second-order equations of motion showing
explicitly the equality of the inertial and gravitational masses of
the particles

 \begin{eqnarray*}
 m_i\, \eta_i\, {\ddot \eta}_i^r(\tau) &\cir& \eta_i\, \sqrt{1 -
 {\dot {\vec \eta}}_i^2(\tau)}\, \Big({\cal F}^r_i - {\dot
 \eta}_i^r(\tau)\, {\dot {\vec \eta}}_i(\tau) \cdot {\vec {\cal
 F}}_i\Big)(\tau |{\vec \eta}_i(\tau)| {\vec
 \eta}_{k \not= i}(\tau)) =\nonumber \\
 &{\buildrel {def}\over =}& \eta_i\, F^r_i(\tau |{\vec \eta}_i(\tau)| {\vec
 \eta}_{k \not= i}(\tau)),
 \end{eqnarray*}

\bea
 &&\eta_i\,    {\cal F}^r_i(\tau |{\vec \eta}_i(\tau)| {\vec
 \eta}_{k \not= i}(\tau)) =
   {{m_i\, \eta_i}\over {\sqrt{1 - {\dot {\vec \eta}}^2_i(\tau)}}}\,
 \Big(  - {{\partial\, {\check n}_{(1)}(\tau,
 {\vec \eta}_i(\tau))}\over {\partial\, \eta^r_i}} +\nonumber \\
 &&+  {{{\dot \eta}^r_i(\tau)}\over {1 - {\dot {\vec
 \eta}}^2_i(\tau)}}\, \sum_u\, \Big[{\dot \eta}^u_i(\tau)\, {{\partial\,
 {\check n}_{(1)}}\over {\partial\,
 \eta^u_i}} + \sum_{j \not= i}\, {\dot \eta}^u_j(\tau)\, {{\partial\,
 {\check n}_{(1)} }\over {\partial\,
 \eta^u_j}} \Big](\tau, {\vec \eta}_i(\tau)) +\nonumber \\
 &&+ \Big(\sum_u\, {\dot \eta}^u_i(\tau)\, \Big[{{\partial\, {\check
 {\bar n}}_{(1)(u)}}\over {\partial\, \eta^r_i}} - {{\partial\, {\check
 {\bar n}}_{(1)(r)}}\over {\partial\, \eta^u_i}}\Big] -
  \sum_{j \not= i}\, \sum_u\, {\dot \eta}^u_j(\tau)\, {{\partial\, {\check
 {\bar n}}_{(1)(r)}}\over {\partial\, \eta^u_j}} -\nonumber \\
 &&-{{{\dot \eta}^r_i(\tau)}\over {1 - {\dot {\vec \eta}}^2_i(\tau)}}\,
 \sum_u\, {\dot \eta}_i^u(\tau)\, \sum_s\, \Big[{\dot
 \eta}_i^s(\tau)\, {{\partial\, {\check {\bar n}}_{(1)(u)}}\over
 {\partial\, \eta^s_i}} +\nonumber \\
 &&+ \sum_{j \not= i}\, {\dot \eta}_j^s(\tau)\,
 {{\partial\, {\check {\bar n}}_{(1)(u)}}\over {\partial\, \eta^s_j}}
 \Big]\Big)(\tau, {\vec \eta}_i(\tau)) +
  \Big(\sum_u\, ({\dot \eta}^u_i(\tau))^2\, {{\partial\, (\Gamma_u^{(1)} +
 2\, \phi_{(1)})}\over {\partial\, \eta^r_i}} -\nonumber \\
 &&- {\dot \eta}^r_i(\tau)\, \sum_u\, \Big[ {\dot \eta}^u_i(\tau)\,
 \Big(2\, {{\partial\, (\Gamma_r^{(1)} + 2\, \phi_{(1)})}\over {\partial\, \eta^u_i}}
 + \sum_c\, {{({\dot \eta}^c_i(\tau))^2}\over {1 - {\dot {\vec \eta}}^2_i(\tau)}}\,
 {{\partial\, (\Gamma_c^{(1)} + 2\, \phi_{(1)})}\over {\partial\, \eta^u_i}}
 \Big) +\nonumber \\
 &&+ \sum_{j \not= i}\, {\dot \eta}^u_j(\tau)\,
 \Big(2\, {{\partial\, (\Gamma_r^{(1)} + 2\, \phi_{(1)})}\over {\partial\, \eta^u_j}}
 + \nonumber \\
 &&+ \sum_c\, {{({\dot \eta}^c_i(\tau))^2}\over {1 - {\dot {\vec \eta}}^2_i(\tau)}}\,
 {{\partial\, (\Gamma_c^{(1)} + 2\, \phi_{(1)})}\over {\partial\, \eta^u_j}}
 \Big) \Big]\Big)(\tau, {\vec \eta}_i(\tau)) -\nonumber \\
 &-&  {{{\dot \eta}^r_i(\tau)}\over {1 - {\dot {\vec \eta}}^2_i(\tau)}}
 \Big[\partial_{\tau}^2{|}_{{\vec \eta}_i}\, {}^3{\cal K}_{(1)} +
 2\, \sum_s\, {\dot \eta}_i^s(\tau)\, {{\partial\, \partial_{\tau}{|}_{{\vec
 \eta}_i}\, {}^3{\cal K}_{(1)} }\over {\partial\, \eta_i^s}} +
 \nonumber \\
 &+& \sum_{su}\, {\dot \eta}_i^s(\tau)\,
 {\dot \eta}_i^u(\tau)\, {{\partial^2\, {}^3{\cal K}_{(1)}}\over
 {\partial\, \eta_i^u\, \partial\, \eta_i^s}}
 \Big](\tau, {\vec \eta}_i(\tau))
  \Big) + O(\zeta^2).
 \label{60}
 \eea

\medskip

The effective action-at-a-distance force ${\vec F}_i(\tau)$
contains\medskip

a) the contribution of the lapse function ${\check n}_{(1)}$, which
generalizes the Newton force;\medskip

b) the contribution of the shift functions ${\check {\bar
n}}_{(1)(r)}$, which gives the gravito-magnetic effects;\medskip

c) the retarded contribution of HPM GW, described by the functions
$\Gamma_r^{(1)}$ of Eq.(\ref{59});\medskip

d) the contribution of the volume element $\phi_{(1)}$ (${\tilde
\phi} = 1 + 6\, \phi_{(1)} + O(\zeta^2)$), always summed to the HPM
GW, giving forces of Newton type;\medskip

e) the contribution of the inertial gauge variable (the non-local
York time) ${}^3{\cal K}_{(1)} = {1\over {\triangle}}\,
{}^3K_{(1)}$.\medskip

In the electro-magnetic case in SR \cite{28} the regularized coupled
second-order equations of motion of the particles obtained by using
the Lienard-Wiechert solutions for the electro-magnetic field are
independent by the type of Green function (retarded or advanced or
symmetric) used. The electro-magnetic retardation effects, killed by
the Grassmann regularization, are connected with QED radiative
corrections to the one-photon exchange diagram. This is not strictly
true in the gravitational case. The effect of retardation is not
killed by the Grasmann regularization but only pushed to
$O(\zeta^2)$: at this order it should give extra contributions to
the second-order equations of motion. This shows that our
semi-classical approximation, obtained with our Grassmann
regularization, of a unspecified "quantum gravity" theory does not
take into account only a "one-graviton exchange diagram": in the
spin 2 case there is an extra retardation effect showing up only at
higher HPM orders \footnote{In the electro-magnetic case the
Grassmann regularization implies $Q_i\, {\dot \eta}_i^r(\tau - |\vec
\sigma|) = Q_i\, {\dot \eta}_i(\tau)$ and equations of motion of the
type ${\ddot \eta}^r_i(\tau) = Q_i\, ...$ with $Q^2_i = 0$. In the
gravitational case the equations of motion are of the type $\eta_i\,
{\ddot \eta}^r_i(\tau) = \eta_i\, ...$ with $\eta^2_i = 0$, but the
Grassmann regularized retardation in Eq.(\ref{54}) gives
Eq.(\ref{59}) only at the lowest order in $\zeta$ and has
contributions of every order $O(\zeta^k)$.}.

\subsection{The Center-of-Mass Problem in General Relativity and in
the HPM Linearization.}

As said in Section 5, the 3-universe is described in a non inertial
rest frame with non-Euclidean 3-spaces $\Sigma_{\tau}$ tending to
Euclidean inertial ones $\Sigma_{\tau (\infty)}$ at spatial
infinity. Both matter and gravitational degrees of freedom live
inside $\Sigma_{\tau}$ and their internal 3-center of mass is
eliminated by the rest-frame condition ${\hat P}^r_{ADM} \approx 0$
(implied by the absence of super-translations) if also the condition
${\hat K}^r_{ADM} = {\hat J}^{\tau r}_{ADM}\approx 0$ is added like
in SR. The 3-universe may be described as an external decoupled
center of mass carrying a pole-dipole structure: ${\hat E}_{ADM}$ is
the invariant mass and ${\hat J}^{rs}_{ADM}$ the rest spin. As in SR
the condition ${\hat K}^r_{ADM} \approx 0$ selects the Fokker=Pryce
center of inertia as the natural time-like observer origin of the
radar coordinates: it follows a non-geodetic straight world-line
like the asymptotic inertial observers existing in these
space-times.

\medskip

This is a way out from the the problem of the center of mass in
general relativity and of its world-line, a still  open problem in
generic space-times as can be seen from Refs. \cite{58,92} (and
Ref.\cite{74} for the PN approach). Usually, by means of some
supplementary condition, the center of mass is associated to the
monopole of a multipolar expansion of the energy-momentum of a small
body (see Ref.\cite{26} for the special relativistic case).

\medskip

In SR the elimination of the internal 3-center of mass leads to
describe the dynamics inside $\Sigma_{\tau}$ only in terms of
relative variables (see Eqs.(\ref{15}) in the case of particles).
However relative variables {\it do not exist} in the non-Euclidean
3-spaces of curved space-times, where flat objects like ${\vec
r}_{ij}(\tau) = {\vec \eta}_i(\tau) - {\vec \eta}_j(\tau)$ have to
be replaced with a quantity proportional to the tangent vector to
the space-like 3-geodesics joining the two particles in the
non-Euclidean 3-space $\Sigma_{\tau}$ (see Ref. \cite{93} for an
implementation of this idea). Quantities like ${\vec
r}^2_{ij}(\tau)$ have to be replaced with the Synge world function
\cite{82,87,89} \footnote{It is a bi-tensor, i.e. a scalar in both
the points ${\vec \eta}_i(\tau)$ and ${\vec \eta}_j(\tau)$, defined
in terms of the space-like geodesic connecting them in
$\Sigma_{\tau}$. See Eq.(3.13) of the third paper in
Ref.\cite{40}.}. This problem is another reason why extended objects
tend to be replaced with point-like multipoles, which, however, do
not span a canonical basis of phase space (see Refs.\cite{26} for
SR).

\medskip

However, at the level of the HPM approximation one can introduce
relative variables for the particles, like the SR ones of
Eq.(\ref{15}), defined as 3-vectors in the asymptotic inertial rest
frame $\Sigma_{\tau (\infty)}$ by putting ${\vec \eta}_i(\tau) =
{\vec \eta}_{(o) i}(\tau) + {\vec \eta}_{(1) i}(\tau)$ and ${\vec
\kappa}_i(\tau) = {\vec \kappa}_{(o) i}(\tau) + {\vec \kappa}_{(1)
i}(\tau)$ with ${\vec \eta}_{(o) i}(\tau),\, {\vec \kappa}_{(o)
i}(\tau)\, = O(\zeta)$. This allows one to define HPM collective and
relative canonical variables for the particles, with the collective
variables eliminated by the conditions ${\hat P}^r_{ADM} \approx 0$
and ${\hat K}^r_{ADM} \approx 0$ (at the lowest order they become
the SR conditions).

\bigskip

In the case of two particles (with total and reduced masses $M = m_1
+ m_2$ and $\mu = {{m_1\, m_2}\over M}$) one puts ${\vec
\eta}_1(\tau) = {\vec \eta}_{12}(\tau) + {{m_2}\over M}\, {\vec
\rho}_{12}(\tau)$, ${\vec \eta}_2(\tau) = {\vec \eta}_{12}(\tau) -
{{m_1}\over M}\, {\vec \rho}_{12}(\tau)$, ${\vec \kappa}_1(\tau) =
{{m_1}\over M}\, {\vec \kappa}_{12}(\tau) + {\vec \pi}_{12}(\tau)$,
${\vec \kappa}_2(\tau) = {{m_2}\over M}\, {\vec \kappa}_{12}(\tau) -
{\vec \pi}_{12}(\tau)$ and goes to the new canonical basis ${\vec
\eta}_{12}(\tau) = {{m_1\, {\vec \eta}_1(\tau) + m_2\, {\vec
\eta}_2}\over M}$,  ${\vec \rho}_{12}(\tau) = {\vec \eta}_1(\tau) -
{\vec \eta}_2(\tau)$, ${\vec \kappa}_{12}(\tau) = {\vec
\kappa}_1(\tau) + {\vec \kappa}_2(\tau)$, ${\vec \pi}_{12}(\tau) =
{{m_2\, {\vec \kappa}_1(\tau) - m_1\, {\vec \kappa}_2(\tau)}\over
M}$.
\medskip

It can be shown that the conditions ${\hat P}^r_{ADM} \approx 0$ and
${\hat K}^r_{ADM} \approx 0$  imply

\bea
 {\vec \eta}_1(\tau) &\approx& \Big({{m_2}\over M} - A_{(o)}(\tau)\Big)\, {\vec
 \rho}_{(o)12}(\tau) + {{m_2}\over M}\, {\vec \rho}_{(1)12}(\tau) +
 {\vec f}_{(1)}(\tau)[rel.var., GW],\nonumber \\
 {\vec \eta}_2(\tau) &\approx& - \Big({{m_1}\over M} + A_{(o)}(\tau)\Big)\, {\vec
 \rho}_{(o)12}(\tau) - {{m_1}\over M}\, {\vec \rho}_{(1)12}(\tau) +
 {\vec f}_{(1)}(\tau)[rel.var., GW],\nonumber \\
 &&{}\nonumber \\
 &&A_{(o)}(\tau) = {{ {{m_2}\over M}\, \sqrt{m_1^2\, c^2 +
 {\vec \pi}^2_{(1)12}(\tau)} -  {{m_1}\over M}\, \sqrt{m_2^2\, c^2 +
 {\vec \pi}^2_{(1)12}(\tau)}}\over {\sqrt{m_1^2\, c^2 +
 {\vec \pi}^2_{(1)12}(\tau)} + \sqrt{m_2^2\, c^2 +
 {\vec \pi}^2_{(1)12}(\tau)} }},\nonumber \\
  &&{}
 \label{61}
 \eea

\noindent for some function ${\vec f}_{(1)}[rel.var., GW](\tau)
\approx {\vec \eta}_{(1)12}(\tau)$ depending on the relative
variables and the HPM GW of Eq.(\ref{59}) in absence of incoming
radiation. Then the equations of motion (\ref{60}) imply

\bea
 \mu\, {\ddot \rho}^r_{(o)12}(\tau) &\cir& {{m_2}\over M}\,
 F^r_{1}(\tau| {\vec \eta}_{(o)1}(\tau)| {\vec \eta}_{(o)2}(\tau))
 - {{m_1}\over M}\,  F^r_{2}(\tau| {\vec \eta}_{(o)2}(\tau)| {\vec
 \eta}_{(o)1}(\tau)),\nonumber \\
 &&{}
 \label{62}
 \eea

\noindent for the relative configurational variable. The collective
configurational variable has ${\vec \eta}_{(o)12}(\tau) \approx -
A_{(o)}(\tau)\, {\vec \rho}_{(o)12}(\tau)$ at the lowest order,
while at the first order there is an equation of motion equivalent
to ${\ddot \eta}^r_{(1)12}(\tau) \approx {{d^2}\over {d\tau^2}}\,
{\vec f}_{(1)}[rel.var., GW]$ $(\tau)$.

\subsection{The Post-Newtonian Expansion at all Orders in the Slow Motion
Limit.}

If all the particles are contained in a compact set of radius $l_c$,
one can add a {\it slow motion condition} in the form
$\sqrt{\epsilon} = {v\over c} \approx \sqrt{{{R_{m_i}}\over
{l_c}}}$, $i=1,..N$ ($R_{m_i} = {{2\, G\, m_i}\over {c^2}}$ is the
gravitational radius of particle $i$) with $l_c \geq R_M$ and
$\lambda >> l_c$ (see the Introduction). In this case one can do the
PN expansion of Eqs.(\ref{60}).
\medskip

After having put $\tau = c\, t$, one makes the following change of
notation

\bea
 &&{\vec \eta}_i(\tau) = {\vec {\tilde \eta}}_i(t),\qquad {\vec
 v}_i(t) = {{d {\vec {\tilde \eta}}_i(t)}\over {dt}},\qquad {\vec
 a}_i(t) = {{d^2 {\vec {\tilde \eta}}_i(t)}\over {dt^2}},\nonumber \\
 &&\qquad {\dot {\vec \eta}}_i(\tau) = {{{\vec v}_i(t)}\over c},\qquad
 {\ddot {\vec \eta}}_i(\tau) = {{{\vec a}_i(t)}\over {c^2}}.
 \label{63}
 \eea

For the non-local York time one uses the notation ${}^3{\tilde {\cal
K}}_{(1)}(t, \vec \sigma) = {}^3{\cal K}_{(1)}(\tau, \vec \sigma)$.
\medskip

Then one studies the PN expansion of the equations of motion
(\ref{60}) with the result (kPN means of order $O(c^{- 2 k})$)

\bea
 m_i\, {{d^2\, {\tilde \eta}_i^r(t)}\over {dt^2}}
 &=& m_i\, \Big[-G\, {{\partial}\over {\partial\,
 {\tilde \eta}_i^r}}\,  \sum_{j \not= i}\,  {{m_j}\over
 {|{\vec {\tilde \eta}}_i(t) - {\vec {\tilde \eta}}_j(t)|}} -
  {1\over c}\, {{d {\tilde \eta}^r_i(t)}\over {dt}} \Big(
  \partial^2_t{|}_{{\vec {\tilde \eta}}_i}\, {}^3{\tilde
  {\cal K}}_{(1)} +\nonumber \\
  &+& 2\, \sum_u\, v_i^u(t)\, {{\partial\, \partial_t{|}_{{\vec {\tilde \eta}}_i}\,
 {}^3{\tilde {\cal K}}_{(1)} }\over {\partial\, {\tilde \eta}_i^u}}
 + \sum_{uv}\, v_i^u(t)\, v^v_i(t)\, {{\partial^2\, {}^3{\tilde
 {\cal K}}_{(1)}}\over {\partial\, {\tilde \eta}_i^u\, \partial\, {\tilde \eta}_i^v}}
 \Big) (t, {\vec {\tilde \eta}}_i(t)) \Big] +\nonumber \\
 &+& F^r_{i(1PN)}(t) + (higher\, PN\, orders).
 \label{64}
 \eea

\noindent At the lowest order one finds the standard Newton
gravitational force\hfill\break ${\vec F}_{i (Newton)}(t) = - m_i\,
G\, {{\partial}\over {\partial\, {\tilde \eta}_i^r}}\,  \sum_{j
\not= i}\,  {{m_j}\over {|{\vec {\tilde \eta}}_i(t) - {\vec {\tilde
\eta}}_j(t)|}}$.

\medskip

The unexpected result is a 0.5PN force term containing all the
dependence upon the non-local York time. The (arbitrary in these
gauges) double rate of change in time of the trace of the extrinsic
curvature creates a 0.5 PN damping (or anti-damping since the sign
of the inertial gauge variable ${}^3{\cal K}_{(1)}$ of Eq.(\ref{57})
is not fixed) effect on the motion of particles. This is a inertial
effect (hidden in the lapse function) not existing in Newton theory
where the Euclidean 3-space is absolute.\medskip

\medskip
Then there are all the other kPN terms with k = 1, 1.5, 2,... Since
these results have been obtained without introducing ad hoc
Lagrangians for the particles, are not in the harmonic gauge and do
not contain terms of order $O(\zeta^2)$ and higher, it is not
possible to make a comparison with the standard PN expansion (whose
terms are known till the order 3.5PN \cite{44}). Therefore only the
1PN and 0.5PN terms will be considered in the next two Subsections.

\subsection{ The HPM Binaries at the 1PN Order}

Since in the next Subsection the 0.5PN term depending on the
non-local York time will be connected with dark matter at the level
of galaxies and clusters of galaxies and since there is no
convincing evidence of dark matter in the Solar System and near the
galactic plane of the Milky Way \cite{96}, it is reasonable to
assume ${}^3{\cal K}_{(1)}(\tau, \vec \sigma) = {1\over
{\triangle}}\, F_{(1)}(\tau, \vec \sigma) \approx 0$ near  a star
with planets and near a binary.

\medskip

In the description of Subsection 8.1 of the 1PN two-body problem,
which is relevant for the treatment of binary systems \footnote{For
binaries one assumes ${v\over c} \approx \sqrt{{{R_m}\over {l_c}}}
<< 1$, where $l_c \approx |{\vec {\tilde r}}|$ with ${\vec {\tilde
r}}(t)$ being the relative separation after the decoupling of the
center of mass. Often one considers the case $m_1 \approx m_2$. See
chapter 4 of Ref. \cite{44} for a review of the emission of GW's
from circular and elliptic Keplerian orbits and of the induced
inspiral phase.} as shown in Chapter VI of Refs.\cite{44} based on
Ref.\cite{47,94,95}, it can be shown that the relative momentum in
the rest frame has the 1PN expression $\pi_{12}(\tau) =
\pi_{(1)12}(\tau) \approx \mu\, {\vec v}_{(rel)(o)12}(t)\, \Big[1 +
{{m_1^3 + m_2^3}\over {2\, M^3}}\, ({{{\vec v}_{(rel)(o)12}(t)}\over
c})^2\Big]$, where ${\vec v}_{(rel)(o)12}(t) = {{d\, {\vec
\rho}_{(o)12}(t)}\over {dt}}$ is the velocity of the lowest order
${\vec \rho}_{(o)12}(\tau)$ of the relative variable.

\medskip

If one ignores the York time and  considers only positive energy
particles ($\eta_1, \eta_2 \mapsto + 1$), the 1PN equations of
motion for the relative variable of the binary implied by
Eqs.(\ref{62}) and 1PN expression of the weak ADM energy ${\hat
E}_{ADM}$ and of the rest spin ${\hat J}^{rs}_{ADM}$ (being
determined by the boundary conditions they are constants of the
motion implying planar motion in the plane orthogonal to the rest
spin ) can be shown to be

\bea
 {{d\, {\vec v}_{(rel)(o)12}(t)}\over {dt}} &=&
   - G\, M\, {{{\vec \rho}_{(o)12}(t)}\over
 {|{\vec \rho}_{(o)12}(t)|^3}} \Big[1+ \Big(1+3\frac{\mu}{M}\Big)
 \frac{v^2_{(rel)(o)12}(t)}{c^2} -\nonumber \\
 &-& \frac{3\mu}{2M}\,\Big(\frac{{\vec v}_{(rel)(o)12}(t)\cdot
 {\vec \rho}_{(o)12}(t)\big)}{|{\vec \rho}_{(o)12}(t)|}\Big)^2\Big]+\nonumber\\
 &-& \frac{G\, M\,}{|{\vec \rho}_{(o)12}(t)|^3}\Big(
 4-\frac{2\mu}{M}
 \Big){\vec v}_{(rel)(o)12}(t)\,\frac{{\vec v}_{(rel)(o)12}(t)\cdot
 {\vec \rho}_{(o)12}(t)\big)}{|{\vec \rho}_{(o)12}(t)|}.\nonumber \\
 &&{}\nonumber \\
  {\hat E}_{ADM(1PN)} &=& \sum_i\, m_i\, c^2 + \mu\, \Big({1\over
 2}\, {\vec v}^2_{(rel)(o)12}(t) \,\Big[ 1 +
  {{m_1^3 + m_2^3}\over {M^3}}\, ({{{\vec v}_{(rel)(o)12}(t)}\over
 c})^2\Big] -\nonumber \\
 &-& {{G\, M}\over {|{\vec \rho}_{(o)12}(t)|}}\, \Big[ 1 + {1\over
 2}\, \Big((3 + {{\mu}\over {M}})\, {{{\vec v}^2_{(rel)(o)12}(t)}\over
 {c^2}} +\nonumber \\
 &+&  {{\mu}\over M}\, ({{{\vec v}_{(rel)(o)12}(t)}\over c} \cdot
 {{ {\vec \rho}_{(o)12}(t) }\over {|{\vec \rho}_{(o)12}(t)|}})^2
 \Big)\Big]\Big),\nonumber \\
 &&{}\nonumber \\
 {\hat J}^{rs}_{ADM(1PN)} &=& \Big(\rho^r_{(o)12}(t)\, v^s_{(rel)(o)12}(t) -
 \rho^s_{(o)12}(t)\, v^r_{(rel)(o)12}(t)\Big)\nonumber \\
 &&\Big[1 + {{m_1^3 + m_2^3}\over
 {2\, M^3}}\, ({{{\vec v}_{(rel)(o)12}(t)}\over c})^2\Big].
 \label{65}
 \eea

\bigskip

\medskip

Our 1PN equations (\ref{65}) in the 3-orthogonal gauges coincide
with Eqs. (2.5), (2.13) and (2.14) of the first paper in
Ref.\cite{47} (without $G^2$ terms since they are $O(\zeta^2)$),
which are obtained in the family of harmonic gauges starting from an
ad hoc 1PN Lagrangian for the relative motion of two test particles
(first derived by Infeld and Plebanski \cite{86}) \footnote{This is
also the starting point of the effective one body description of the
two-body problem of Refs. \cite{88}.}. These equations are the
starting point for studying the post-Keplerian parameters of the
binaries, which, together with the Roemer, Einstein and Shapiro time
delays (both near Earth and near the binary) in light propagation,
allow one to fit the experimental data from the binaries (see the
second paper in Ref.\cite{47} and Chapter VI of Ref.\cite{44}).
Therefore these results are reproduced also in our 3-orthogonal
gauge with ${}^3{\cal K}_{(1)}(\tau, \vec \sigma) = 0$.

 \subsection{From the  Three Signatures for Dark Matter Reinterpreted as
 Relativistic Inertial Effects Induced by the York Time to the Need
 of a PM ICRS}

To study the effects induced by the 0.5PN velocity-dependent
(friction or anti-friction) force term in Eqs.(\ref{64}), depending
on the inertial gauge variable non-local York time ${}^3{\tilde
{\cal K}}_{(1)}(t, \vec \sigma) = {1\over {\triangle}}\, {}^3{\tilde
K}_{(1)}(\tau, \vec \sigma) \approx {1\over {\triangle}}\,
F_{(1)}(\tau, \vec \sigma)$ with $F_{(1)}(\tau, \vec \sigma)$
arbitrary numerical function, it is convenient to rewrite such
equations in the form

 \bea
 \frac{d}{dt}\Big[ \,m_i\Big(1+\frac{1}{c}\,\frac{d}{dt}\,{}^3{\tilde {\cal K}}_{(1)}(t,
 {\vec {\tilde \eta}}_i(t))\,\Big)\, {{d\, {\tilde \eta}_i^r(t)}\over {dt}}\Big]
 &\cir &\,-G\, {{\partial}\over {\partial\,
 {\tilde \eta}_i^r}}\,  \sum_{j \not= i}\, \eta_j\, {{m_i\,m_j}\over
 {|{\vec {\tilde \eta}}_i(t) - {\vec {\tilde \eta}}_j(t)|}}+\nonumber\\
 &&\nonumber\\
 &+&{\cal O}(\zeta^2),
 \label{66}
\eea

\noindent because the damping or anti-damping factors in
Eq.(\ref{64}) are $\gamma_i(t, {\vec {\tilde \eta}}_i(t)) =
{{d^2}\over {dt^2}}\, {}^3{\tilde {\cal K}}_{(1)}(t, {\vec {\tilde
\eta}}_i(t))$ and ${\ddot {\vec {\tilde \eta}}}_i(t) = O(\zeta)$.

\medskip

As a consequence the velocity-dependent force can be {\it
reinterpreted} as the introduction of an {\it effective (time-,
velocity- and position-dependent) inertial mass term} for the
kinetic energy of each particle:

\bea
 m_i\, &\mapsto& m_i\,\Big(1+\frac{1}{c}\,\frac{d}{dt}\,{}^3{\tilde
 {\cal K}}_{(1)}(t, {\vec {\tilde \eta}}_i(t))\,\Big) =
  m_i + (\Delta\, m)_i(t, {\vec {\tilde \eta}}_i(t)),\nonumber \\
  &&{}
 \label{67}
 \eea

\noindent in each instantaneous 3-space. Instead in the Newton
potential there are the gravitational masses of the particles, equal
to the inertial ones in the 4-dimensional space-time due to the
equivalence principle. Therefore the effect is due to a modification
of the effective inertial mass in each non-Euclidean 3-space
depending on its shape as a 3-sub-manifold of space-time: {\it it is
the equality of the inertial and gravitational masses of Newtonian
gravity to be violated}! In Galilei space-time the Euclidean 3-space
is an absolute time-independent notion like Newtonian time: the
non-relativistic non-inertial frames live in this absolute 3-space
differently from what happens in SR and GR, where they are (in
general non-Euclidean) 3-sub-manifolds of the space-time.

\bigskip

Eqs. (\ref{64}), (\ref{66}) and (\ref{67}) can be applied to the
three main signatures of the existence of dark matter in the
observed masses of galaxies and clusters of galaxies, where the 1PN
forces are not important, namely the virial theorem \cite{49}, the
weak gravitational lensing \cite{50}, \cite{49} and the rotation
curves of spiral galaxies (see Ref.\cite{48} for a review), to give
a {\it reinterpretation of dark matter as a relativistic inertial
effect}.

\medskip

A) {\it Masses of clusters of galaxies from the virial theorem}. For
a bound system of N particles of mass $m$ (N equal mass galaxies) at
equilibrium, the virial theorem relates the average  kinetic energy
$< E_{kin} >$ in the system to the average  potential energy $<
U_{pot} >$ in the system: $< E_{kin} > = - {1\over 2}\, < U_{pot} >$
assuming Newton gravity. For the average kinetic energy of a galaxy
in the cluster one takes $< E_{kin} > \approx {1\over 2}\, m\, < v^2
>$, where $< v^2 >$ is the average of the square of the radial
velocity of single galaxies with respect to the center of the
cluster (measured with Doppler shift methods; the velocity
distribution is assumed isotropic). The average potential energy of
the galaxy is assumed of the form $< U_{pot} > \approx - G\, {{m\,
M}\over {\cal R}}$, where $M = N m$ is the total mass of the cluster
and ${\cal R}=\alpha\,R$ is a ''effective radius'' depending on the
cluster size $R$ (the angular diameter of the cluster and its
distance from Earth are needed to find $R$) and on the mass
distribution on the cluster (usually $\alpha\approx 1/2)$. Then the
virial theorem implies $M \approx {{\cal R}\over G}\, < v^2 >$. It
turns out that this mass $M$ of the cluster is usually at least an
order of magnitude bigger that the baryonic matter of the cluster
$M_{bar} = N\, m$ (spectroscopically determined). By applying
Eqs.(\ref{64}) to the equilibrium condition for a self-gravitating
system, i.e. $\frac{d^2}{dt^2} \sum_i\, m_i\, \mid
\vec{\tilde{\eta}}_i(t) \mid^2 = 0$ with $m_i = m$, one gets
$\sum_i\, m_i\, v_i^2(t) - G\, \sum_{i>j}\,  {{m_i\,m_j}\over
{|{\vec {\tilde \eta}}_i(t) - {\vec {\tilde \eta}}_j(t)|}}-
\frac{1}{c}\, \sum_i\, m_i\, \Big(\vec{\tilde{\eta}}_i(t)\cdot
\vec{v}_i(t)\Big)\,\gamma_i(t, {\vec {\tilde \eta}}_i(t)) = 0$ with
$m_i = m_j = m$. Therefore one can write $\langle U_{pot}\rangle = -
\frac{1}{N}\, \sum_{i>j}\, {{G\, m^2}\over {|{\vec {\tilde
\eta}}_i(t) - {\vec {\tilde \eta}}_j(t)|}} \approx G\, \frac{m\,
M_{bar}}{{\cal R}}$ (with ${\cal R}=R/2$) and $ \frac{1}{2}\, m\,
\langle v^2 \rangle = - \frac{1}{2}\, \langle U_{pot}\rangle +
\frac{m}{2c}\, \langle \Big(\vec{\tilde{\eta}} \cdot \vec{v}\Big)\,
\gamma(t, {\vec {\tilde \eta}})\rangle$ with the notation
$\langle\Big(\vec{\tilde{\eta}} \cdot \vec{v}\Big)\, \gamma(t, {\vec
{\tilde \eta}})\rangle = \frac{1}{N}\, \sum_i\,
\Big(\vec{\tilde{\eta}}_i(t) \cdot \vec{v}_i(t)\Big)\, \gamma_i(t,
{\vec {\tilde \eta}}_i(t))$ (it contains the non-local York time).
Therefore for the measured mass $M$ (the effective inertial mass in
3-space) one has

\bea
 M &=& {{\cal R}\over G}\, < v^2 > = M_{bar} + {{\cal R}\over {G\,c}}\,
 \langle\Big(\vec{\tilde{\eta}}\cdot\vec{v}\Big)\,\gamma(t,
 {\vec {\tilde \eta}})\rangle {\buildrel {def}\over =} M_{bar} + M_{DM},
 \nonumber \\
 &&{}
 \label{68}
 \eea

\noindent and one sees that the non-local York time can give rise to
a dark matter contribution $M_{DM} = M - M_{bar}$.

\medskip

B) {\it Masses of galaxies or clusters of galaxies from weak
gravitational lensing}. Usually one considers a galaxy (or a cluster
of galaxies) of big mass $M$ behind which a distant, bright object
(often a galaxy) is located. The light from the distant object is
bent by the massive one (the lens) and arrives on the Earth
deflected from the original propagation direction. As shown in
Ref.\cite{50} one has to evaluate Einstein deflection of light,
emitted by a source S at distance $d_S$ from the observer O on the
Earth, generated by the big mass  at a distance $d_D$ from the
observer O. The mass $M$, at distance $d_{DS}$ from the source S, is
considered as a point-like mass generating a 4-metric  of the
Schwarzschild type (Schwarzschild lens). The ray of light is assumed
to propagate in Minkowski space-time till near $M$, to be deflected
by an angle $\alpha$ by the local gravitational field of M and then
to propagate in Minkowski space-time till the observer O. The
distances $d_S$, $d_D$, $d_{DS}$, are evaluated by the observer O at
some reference time in some nearly-inertial Minkowski frame with
nearly Euclidean 3-spaces (in the Euclidean case $d_{DS} = d_S -
d_D$). If $\xi = \theta\, d_D$ is the impact parameter of the ray of
light at $M$ and if $\xi >> R_s = {{2\, G\, M}\over {c^2}}$ (the
gravitational radius), Einstein's deflection angle is $\alpha =
{{2\, R_s}\over {\xi}} = {{4\, G\, M}\over {c^2\, \xi}}$  and the
so-called Einstein radius (or characteristic angle) is $\alpha_o =
\sqrt{2\, R_s\, {{d_{DS}}\over {d_D\, d_S}}} = \sqrt{{{4\, G\,
M}\over {c^2}}\, {{d_{DS}}\over {d_D\, d_S}}}$. A measurement of the
deflection angle and of the three distances allows to get a value
for the mass $M$ of the lens, which usually turns out to be much
larger of its mass inferred from the luminosity of the lens. For the
calculation of the deflection angle one considers the propagation of
ray of light in a stationary 4-metric of the BCRS type and uses a
version of the Fermat principle containing an effective index of
refraction $n$. One has $n = {}^4g_{\tau\tau} = \sgn\, [1 - {{2
w}\over {c^2}} - 2\, \partial_{\tau}\, {}^3{\cal K}]$ in the PM
approximation. Since one has ${{2\, w}\over {c^2}} = - {{G\,
M_{bar}}\over {c^2\, |\vec \sigma|}}$, the definition $2\,
\partial_{\tau}\, {}^3{\cal K}_{(1)}\, {\buildrel {def}\over =}\, -
{{G\, M_{DM}}\over {c^2\, |\vec \sigma|}}$ leads to an Einstein
deflection angle

\beq
  \alpha = {{4\, G\, M}\over {c^2\, \xi}}\,
 \qquad with\quad M {\buildrel {def}\over =} M_{bar} + M_{DM}.
 \label{69}
 \eeq

Therefore also in this case the measured mass $M$ is the sum of a
baryonic mass $M_{bar}$ and of a dark matter mass $M_{DM}$ induced
by the non-local York time at the location of the lens.

\medskip

C) {\it Masses of spiral galaxy masses from their rotation curves}.
In this case one considers a two-body problem (a point-like galaxy
and a body circulating around it) described in terms of an internal
center of mass ${\tilde {\vec \eta}}_{12}(t) \approx {\tilde {\vec
\eta}}_{(1)12}(t)$ (${\tilde {\vec \eta}}_{(o)12}(t) = 0$ is the
origin of the 3-coordinates) and a relative variable ${\tilde {\vec
\rho}}_{12}(t)$. Then the sum and difference of Eqs.(\ref{64}) imply
the equations of motion for ${\tilde {\vec \eta}}_{(1)12}(t)$ and
${\tilde {\vec \rho}}_{12}(t)$. While the first equation implies a
small motion of the overall system, the second one has the form

\bea
   {{d^2\, {\tilde \rho}_{(1)12}^r(t)}\over {dt^2}}
  &\cir& - G\,  M\, {{{\tilde \rho}_{12}^r(t)}
 \over {|{\vec {\tilde \rho}_{12}}(t)|^3}} -
  {1\over  c}\, {{ d {\tilde \rho}_{12}^r(t)}\over {dt}}\,
 \gamma_{+}(t, {\vec {\tilde \rho}_{12}}(t),\vec{v}(t)),
 \nonumber \\
 &&{}\nonumber \\
  \gamma_{+}(t, {\vec {\tilde
 \rho}_{12}}(t),\vec{v}(t)) &=& {{m_1}\over M}\, \gamma_1(t, {{m_2}\over M}\, {\vec {\tilde
 \rho}_{12}}(t),\vec{v}(t)) +
 {{m_2}\over M}\,  \gamma_2(t, - {{m_1}\over M}\, {\vec {\tilde
 \rho}_{12}}(t),\vec{v}(t)),\nonumber \\
 &&{}
 \label{70}
 \eea

\noindent where $\gamma_i$ are the damping or anti-damping factors
defined after Eq.(\ref{66}). Eq.(\ref{70}) gives the two-body Kepler
problem with an extra perturbative force. Without it a Keplerian
solution with circular trajectory such that $\mid {\vec {\tilde
\rho}_{12}}(t)\mid=R=const.$ implies that the Keplerian velocity
$\vec{v}_o(t)=v_o\,\hat{n}(t)$ has the modulus vanishing at large
distances, $v_o = \sqrt{{{G\, M}\over {R}}} \rightarrow_{R
\rightarrow \infty}\, 0$. Instead the rotation curves of spiral
galaxies imply that the relative 3-velocity goes to constant for
large $R$, i.e. $v = \sqrt{{{G\, (M_{bar} + \Delta\, M(r))}\over r}}
\rightarrow_{R \rightarrow \infty}\, const.$ ($M_{bar}$ is the
spectroscopically determined baryon mass), so that the extra
required term $\Delta\, M(r)$ is interpreted as the mass $M_{DM}$ of
a dark matter halo.

The presence of the extra force term implies that the velocity must
be written as $\vec{v}(t) = \vec{v}_o(t) + \vec{v}_1(t)$ with
$v_1(t)$ a first order perturbative correction satisfying
$\frac{dv^r_1(t)}{dt} = - {v^r_o\over  c}\, \hat{n}(t)\,
 \gamma_{+}(t, {\vec {\tilde \rho}_{12}}(t), \vec{v}^r_o(t))$.
Therefore at the first order in the perturbation one gets $v^2(t) =
v_o^2\, \Big(\, 1 - \frac{2}{c}\, \hat{n}(t) \cdot\, \int_o^t dt_1\,
\hat{n}(t_1)\, \gamma_{+}(t_1, {\vec {\tilde \rho}_{12}}(t_1),$
$\vec{v}_o(t_1)) \,\Big)$. Therefore, after having taken a mean
value over a period $T$ (the time dependence of the mass of a galaxy
is not known) the effective mass of the two-body system is

\bea
 M_{eff} &=& \frac{\langle v^2\rangle\,R}{G}= M\, \left(\,1 -\left\langle
 \frac{2}{c}\,\hat{n}(t)\cdot\,\int^t dt_1\,\hat{n}(t_1)\,
 \gamma_{+}(t_1, {\vec {\tilde
 \rho}_{12}}(t_1),\vec{v}_o(t_1))\right\rangle
 \right)) =\nonumber\\
 &&\nonumber\\
 &=& M_{bar} + M_{DM}.
 \label{71}
 \eea

\noindent with a $\Delta\, M(r) = M_{DM}$ function only of the mean
value of the total time derivative of the non-local ${}^3{\cal
K}_{(1)}$ to be fitted to the experimental data. \bigskip

Therefore, the existence of the inertial gauge variable York time, a
property of the non-Euclidean  3-spaces as 3-sub-manifolds of
Einstein space-times (connected only to the general relativistic
remnant of the gauge freedom in clock synchronization, independently
from cosmological assumptions) implies the possibility of describing
part (or maybe all) dark matter as a {\it relativistic inertial
effect in Einstein gravity} without alternative explanations
using:\medskip

1) the non-relativistic MOND approach \cite{97} (where one modifies
Newton equations);\medskip

2)  modified gravity theories like the $f(R)$ ones (see for instance
Refs.\cite{98}; here one gets a modification of the Newton
potential);\medskip

3) the assumption of the existence of WIMP particles \cite{99}.
\medskip

Let us also remark that the 0.5PN effect has origin in the lapse
function and not in the shift one, as in the gravito-magnetic
elimination of dark matter proposed in Ref.\cite{100}.

\bigskip

The open problem with this explanation of dark matter is the
determination of the non-local York time from the data on dark
matter. From what is known  about dark matter in the Solar System
and inside the Milky Way near the galactic plane, it seems that
${}^3{\cal K}_{(1)}(\tau, \vec \sigma)$ is negligible near the stars
inside a galaxy. Instead the non-local York time (or better  a mean
value in time of its total time derivative) should be relevant
around the galaxies and the clusters of galaxies, where there are
big concentrations of mass and well defined signatures of dark
matter. Instead there is no indication on its value in the voids
existing among the clusters of galaxies.\medskip

Therefore the known data on dark matter do not allow one to get an
experimental determination of the York time ${}^3K_{(1)}(\tau, \vec
\sigma) = \triangle\, {}^3{\cal K}_{(1)}(\tau, \vec \sigma)$,
because to do it one needs {\it to know the non-local York time on
all the 3-universe} at a given $\tau$.

\medskip

Since, as said in the Introduction, at the experimental level {\it
the description of matter is intrinsically coordinate-dependent},
namely is connected with the conventions used by physicists,
engineers and astronomers for the modeling  of space-time, one has
to choose a gauge (i.e. a 4-coordinate system) in non-modified
Einstein gravity which is in agreement with the observational
conventions in astronomy. This way out from the gauge problem in GR
requires a choice of 3-coordinates on the instantaneous 3-spaces
identified by a choice of time and by a clock synchronization
convention, i.e. a fixation of the York time ${}^3K_{(1)}(\tau, \vec
\sigma)$. The convention resulting by one set of such choices would
give a {\it PM extension of ICRS}, with BCRS being its
quasi-Minkowskian approximation for the Solar System. Since the
existing ICRS \cite{3,5} has diagonal 3-metric,  3-orthogonal gauges
are a convenient choice.
\medskip

The real problem is the extraction of an indication of which kind of
function of time and 3-coordinates to use for the York time
${}^3K_{(1)}(\tau, \vec \sigma)$ from astrophysical data different
from the ones giving information about dark matter. Once one would
have a phenomenological parametrization of the York time, then the
data on dark matter would put restrictions on the induced
phenomenological parametrization of the non-local York time
${}^3{\cal K}_{(1)}(\tau, \vec \sigma) = {1\over {\triangle}}\,
{}^3K_{(1)}(\tau, \vec \sigma)$.  As it will be delineated in the
final Section, to implement this program one has to look at the
astrophysical data on dark energy after having succeeded to
interpret also it as a relativistic inertial effect in suitable
cosmological space-times in which one can induce the distinction
between inertial and tidal degrees of freedom of the gravitational
field from the previously discussed Hamiltonian framework.

\vfill\eject

\section{Dark Energy and Other Open Problems}
 \label{sec:9}

This Lecture contains a full review of an approach to SR and to
asymptotically Minkowskian classical canonical Einstein GR based on
a description of global non-inertial frames centered on a time-like
observer which is suggested by relativistic metrology. The {\it
gauge freedom in clock synchronization}, which does not exist in
Galilei space-time (Newton time and Euclidean 3-spaces are absolute)
and is not restricted in Minkowski space-time (it spans the class of
the admissible 3+1 splittings of this absolute space-time), is
restricted in GR to the gauge freedom connected with the inertial
gauge variable ${}^3K$, the York time, which determines the shape of
the instantaneous  non-Euclidean 3-spaces as 3-sub-manifolds of the
space-time.
\medskip

The study of canonical ADM tetrad gravity in asymptotically
Minkowskian space-times without super-translations (so that they
admit an asymptotic ADM Poincar\'e algebra at spatial infinity) in
the York canonical basis allowed one to disentangle the tidal
degrees of freedom of the gravitational field from the inertial
gauge ones (they include the York time), to find the family of
non-harmonic 3-orthogonal Schwinger time gauges and to define a HPM
linearization in them. The main properties of these non-harmonic
gauges are that only the HPM-GW (but not the lapse and shift
functions) are retarded quantities with a no-incoming radiation
condition and that one can naturally find which quantities depend
upon the York time.

\medskip

Relativistic particle mechanics, coupled to the electro-magnetic
field in the radiation gauge, has been studied both in SR and GR
with a suitable Grassmann regularization of the self-energies so to
get well defined equations of motion.\medskip

In SR, after a clarification of the problem of the relativistic
center of mass and the definition of inertial and non-inertial rest
frames of isolated systems, it was possible to develop a
formulation, the parametrized Minkowski theories, in which the
transitions among global non-inertial frames are gauge
transformations. Then isolated systems were described in the
rest-frame instant form of dynamics and the structure of their
Poincar\'e generators and of their relative variables in the
instantaneous Wigner 3-spaces was clarified. With this approach it
was possible to give a new formulation of the micro-canonical
ensemble in relativistic kinetic theory and to develop a formulation
of relativistic quantum mechanics and relativistic entanglement
taking into account the known results about relativistic bound
states and the spatial non-separability and non-locality induced by
the Lorentz signature of Minkowski space-time.\medskip

In GR it was possible to derive regularized equations of motion of
the particles in the non-inertial rest frame and to study their PM
limit in the HPM linearization in the 3-orthogonal gauges and the
emission of HPM GW (with the energy balance under control even in
absence of self-forces). Then the PN limit of these PM equations
allows one to recover the known 1PN results of harmonic gauges. The
more surprising result is that in the PN expansion of the PM
equations of motion there is a 0.5PN term in the forces depending
upon the York time. This opens the possibility to describe dark
matter as a relativistic inertial effect implying that the effective
inertial mass of particles in the 3-spaces is bigger of the
gravitational mass because it depends on the York time (i.e. on the
shape of the 3-space as a 3-sub-manifold of the space-time: this is
impossible in Newton gravity in Galilei space-time and leads to a
violation of the Newtonian equivalence principle).
\medskip

The proposed solution to the gauge problem in GR based on the
conventions of relativistic metrology for ICRS and the results of
the last Section on the re-interpretation of dark matter as a
relativistic inertial effect arising as a consequence of a
convention on the York time in an extended PM ICRS push toward the
necessity of similar re-interpretation also of dark energy in
cosmology \cite{1,101,102,103}. As it has been shown, the
identification of the tidal and inertial degrees of freedom of the
gravitational field can be reformulated in the framework of the
non-Hamiltonian first-order ADM equations by means of the
replacement of the Hamiltonian momenta with the expansion and the
shear of the Eulerian observers associated with the 3+1 splitting of
the space-time. Therefore this identification can also be applied to
the cosmological space-times which do not admit a Hamiltonian
formulation: also in them the identification of the instantaneous
3-spaces $\Sigma_{\tau}$, now labeled by a cosmic time, requires a
conventional choice of clock synchronization, i.e. a convention on
the York time ${}^3K$ defining the shape of the 3-spaces as
3-sub-manifolds of the space-time, and of 3-coordinates (the
3-orthogonal ones are acceptable also in cosmology).\medskip

In the standard $\Lambda$CDM cosmological model the class of
cosmological solutions of Einstein equations is restricted to
Friedmann-Robertson-Walker (FRW) space-times with nearly Euclidean
3-spaces (i.e. with a small internal 3-curvature). In them the
Killing symmetries connected with homogeneity and isotropy imply
($\tau$ is the cosmic time, $a(\tau)$ the scale factor) that the
York time is no more a gauge variable but coincides with the Hubble
constant: ${}^3K (\tau) = - {{\dot a(\tau)}\over {a(\tau)}} = -
H(\tau)$. However at the first order in cosmological perturbations
(see Ref.\cite{104} for a review) one has ${}^3K = - H +
{}^3K_{(1)}$ with ${}^3K_{(1)}$ being again an inertial gauge
variable to be fixed with a metrological convention. Therefore the
York time has a central role also in cosmology and one needs to know
the dependence on it of the main quantities, like the red-shift and
the luminosity distance from supernovae, which  require the
introduction of the notion of dark energy to explain the 3-universe
and its accelerated expansion in the framework of the standard
$\Lambda$CDM cosmological model.\medskip

Instead in inhomogeneous space-times without Killing symmetries like
the Szekeres ones \cite{105,106} the York time remains an arbitrary
inertial gauge variable. Therefore the main open problem of the
present approach is to see whether it is possible to find a
3-orthogonal gauge in a inhomogeneous Einstein space-time (at least
in a PM approximation) in which the convention on the inertial gauge
variable York time allows one to accomplish the following two tasks
simultaneously: a) to eliminate both dark matter and dark energy
through the choice of a 4-coordinate system (suggested by
astrophysical data) to be used in a consistent PM reformulation of
ICRS and b) to save the main good properties of the standard
$\Lambda$CDM cosmological model due to the inertial and dynamical
properties of the space-time. As matter one will take the dust,
whose description in the York canonical basis is given in
Ref.\cite{41}.

\bigskip

Also in the back-reaction approach \cite{107,108}  to cosmology,
according to which dark energy is a byproduct of the non-linearities
of GR when one considers spatial averages of 3-scalar quantities in
the 3-spaces on large scales to get a cosmological description of
the universe taking into account its observed inhomogeneity, one
gets that the spatial average of the product of the lapse function
and of the York time (a 3-scalar gauge variable) gives the effective
Hubble constant. Since this approach starts from the Hamiltonian
description of an asymptotically flat space-time and since all the
canonical variables in the York canonical basis, except the angles
$\theta^i$, are 3-scalars, the formalism presented in this Lecture
will allow to study the spatial average of nearly all the Hamilton
equations and not only of the super-Hamiltonian constraint and of
the Hamilton equation for the York time as in the existing
formulation of the approach. This will be done by using the perfect
fluids of Ref.\cite{41} as matter.
\medskip

Also the recent point of view of Ref.\cite{109}, taking into account
the relevance of the voids among the clusters of galaxies, has to be
reformulated in terms of the York time.\medskip

Finally one should find the dependence upon he York time of the
Landau-Lifschitz energy-momentum pseudo-tensor and re-express it as
the effective energy-momentum tensor of a viscous pseudo-fluid. One
will have to check whether  for certain choices of the York time the
resulting effective equation of state of the fluid has negative
pressure, realizing also in this way a simulation of dark energy.

\bigskip

Other open problems in GR under investigation are:\medskip

A) Find the second order of the HPM expansion to see whether in PM
space-times there is the emergence of hereditary terms \cite{44,83}
like the ones present in harmonic gauges.
\medskip

B) Study  the PM equations of motion of the transverse
electro-magnetic field trying to find Lienard-Wiechert-type
solutions in GR. Study astrophysical problems where the
electro-magnetic field is relevant.\medskip

C) Find the expression in the York canonical basis of the Weyl
scalars of the Newman-Penrose approach \cite{68} and then of the
four Weyl eigenvalues, which are tetrad-independent 4-scalar
invariants of the gravitational field. Is it possible to find a
canonical transformation replacing the 3-scalar tidal variables with
four 4-scalar functions of the Weyl eigenvalues? Are Weyl
eigenvalues Dirac observables?

\medskip

D) Try to make a multi-temporal quantization (see Refs,\cite{34,65})
of the linearized HPM theory over the asymptotic Minkowski
space-time, in which, after a Shanmugadhasan canonical
transformation to a new York canonical basis adapted to all the
constraints, only the tidal variables are quantized but not the
inertial gauge ones.  After this type of quantization, in which the
lapse and shift functions remain c-numbers, the space-time would
still be a classical 4-manifold: only the two eigenvalues of the
3-metric describing GW are quantized and therefore only 3-metric
properties like 3-distances, 3-areas, 3-volumes become quantum
properties. After having re-expressed the Ashtekar variables
\cite{110} for asymptotically Minkowskian space-times (see Appendix
B of Ref.\cite{4}) in this final York canonical basis it will be
possible to compare the outcomes of this new type of quantization
with loop quantum gravity.

\vfill\eject

\printindex

\end{document}